\documentclass[prb, reprint, 9pt, superscriptaddress,
notitlepage, nofootinbib, longbibliography,
floatfix]{revtex4-2}
\usepackage{natbib}
\usepackage{graphicx}
\usepackage{epsfig}
\usepackage{amsfonts} 
\usepackage{amsmath} 
\usepackage{amssymb}
\usepackage{diagbox}
\usepackage{dsfont}
\usepackage{bm}
\usepackage{braket}
\usepackage{color}
\usepackage{physics} 
\usepackage{bbm}
\usepackage{hyperref}
\usepackage{subcaption}
\usepackage{comment}
\usepackage[export]{adjustbox}

\renewcommand{\raggedright}{\leftskip=0pt \rightskip=0pt plus 0cm}
\newcommand{\be}{\begin{equation}}
	\newcommand{\ee}{\end{equation}}
\newcommand{\ba}{\begin{eqnarray}}
	\newcommand{\ea}{\end{eqnarray}}

\newsavebox{\foobox}

\definecolor{LinkColor}{rgb}{0,0,1}
\hypersetup{
	colorlinks=true,
	citecolor=LinkColor,
	linkcolor=LinkColor,
	urlcolor=LinkColor
}

\definecolor{gr}{rgb}{0,0,0}

\begin{document}
	
	\title{Mixed-state long-range order and criticality from measurement and feedback}

		\author{Tsung-Cheng Lu}\email{tlu@perimeterinstitute.ca}
			\affiliation{Perimeter Institute for Theoretical Physics, Waterloo, Ontario N2L 2Y5, Canada}
			
		\author{Zhehao Zhang}\email{zhehao@umail.ucsb.edu}
			\affiliation{Department of Physics, University of California, Santa Barbara, CA 93106, USA}
				
			\author{Sagar Vijay}\email{sagar@physics.ucsb.edu}
			\affiliation{Department of Physics, University of California, Santa Barbara, CA 93106, USA}				
		\author{Timothy H. Hsieh}\email{thsieh@perimeterinstitute.ca}
			\affiliation{Perimeter Institute for Theoretical Physics, Waterloo, Ontario N2L 2Y5, Canada}
	
	\begin{abstract}
 We propose a general framework for using local measurements, local unitaries, and non-local classical communication to construct quantum channels which can efficiently prepare mixed states with long-range quantum order or quantum criticality. As an illustration, symmetry-protected topological (SPT) phases can be universally converted into mixed-states with long-range entanglement, which can undergo phase transitions with quantum critical correlations of local operators and a  logarithmic scaling of the entanglement negativity, despite coexisting with volume-law  entropy. Within the same framework,  we present two applications using fermion occupation number measurement to convert (i) spinful free fermions in one dimension into a quantum-critical mixed state with enhanced algebraic correlations between spins and (ii) Chern insulators into a  mixed state with critical quantum correlations in the bulk.  The latter is an example where mixed-state quantum criticality can emerge  from a gapped state of matter in constant depth using local quantum operations and non-local classical communication.

	\end{abstract}
	
	\maketitle

	
	{
		\hypersetup{linkcolor=black}
		\tableofcontents
	}
	
	
	
	\section{Introduction}

Interacting quantum matter may exhibit long-range entanglement that is intimately connected to fascinating phenomena such as fractionalized quasiparticles and criticality. Typically the scope for exploring long-range entanglement has been limited to pure states, especially the ground states of many-body Hamiltonians. However, realistic physical systems require a mixed-state description due to constant exposure to an environment, thus motivating the consideration of long-range entanglement in many-body mixed states. In equilibrium, mixed states naturally arise as finite-temperature Gibbs states; however, long-range entanglement is typically fragile to thermal fluctuations \cite{bravyi2009no_go,hastings2011,yoshida2011,Lu_topo_nega_2020,lu2022_lre}. In contrast, out-of-equilibrium mixed states offer richer possibilities for stabilizing long-range entanglement. 
For example, there has been substantial recent progress in characterizing universal properties of various non-trivial states of matter, e.g. symmetry-protected topological (SPT) phases, quantum critical states, and topological order, subject to noise channels \cite{spt_Schuch_2022,lee2022_spt,Bi_2022_spt,wang_2022_spt,behrends2022surface,fan2023mixed,bao2023mixed,Lee_2023_criticality_decoherence,zou_2023_channel_criticality}. Signatures of the non-trivial nature of these mixed-states generally require probing non-linear observables in the mixed-state density matrix, e.g. measures of quantum entanglement.


In this work, we introduce a novel route for realizing long-range entangled mixed states. We provide a general framework for constructing quantum channels which allow for the efficient realization of a large class of mixed-state long-range entanglement, including GHZ, topological order, and quantum criticality. Specifically, our protocols generate pure states with certain probabilities, hence defining a mixed-state ensemble. Importantly, the long-range order/criticality can be efficiently probed through observables that are linear in the mixed-state density matrix.

\begin{figure}
		\centering
		\begin{subfigure}{0.46\textwidth}
			\includegraphics[width=\textwidth]{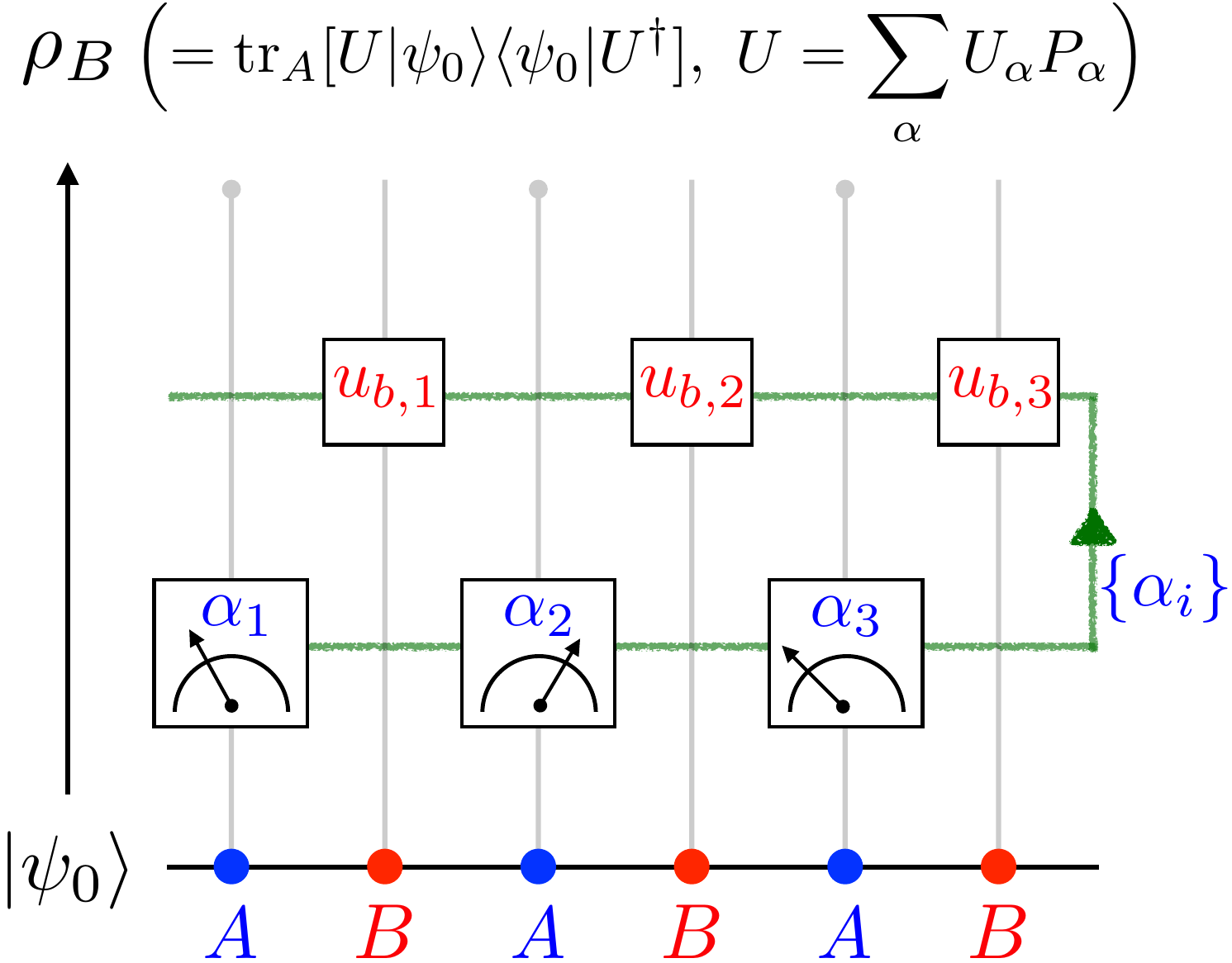}
		\end{subfigure}
\caption{We consider a system with $A, B$ degrees of freedom, which may correspond to sublattices or charge and spin of fermions, for example. We devise a depth-2 quantum channel that can effectively implement a controlled unitary, giving rise to various long-range quantum orders and criticality on the $B$ system. The first layer consists of single-site measurements on $A$, and the second layer consists of single-site unitaries on $B$ conditioned on the outcomes $\{\alpha\}$ in the first layer. This protocol outputs a density matrix $\rho_B$ on $B$ such that $\rho_B (=  \tr_A \ket{\psi}\bra{\psi})$ admits a purification $\ket{\psi}= U\ket{\psi_0}$ with $U$ being a controlled unitary $U= \sum_\alpha U_\alpha P_{\alpha}$ on an $AB$ composite system; $P_{\alpha}$  projects $A$ to a specific product state on $A$, and $U_{\alpha}$ is a product of on-site unitaries acting on $B$.    }
		\label{fig:main_fig}
	\end{figure}

	Our construction (Sec.\ref{sec:general}, see Fig.\ref{fig:main_fig} for schematic) relies on three ingredients, namely, local projective measurement, local unitary operations, and non-local classical communication. Starting with an input state, which we extensively bipartition in two disjoint Hilbert spaces, $A$ and $B$, we perform single-site measurements within $A$. Based on the measurement outcomes, recorded as classical data, we perform appropriate unitary feedback consisting of local unitary gates acting on $B$. We will allow the local unitary gates to depend on the {\it global} classical data, thus requiring non-local classical communication. Such a two-step protocol generically generates an ensemble of pure states associated with distinct measurement outcomes, hence defining a mixed state. The resulting mixed state on $B$ ($\rho_B$) may exhibit various long-range quantum orders and quantum criticality coexisting with extensive classical entropy.  We show that $\rho_B$ admits a purification that can be obtained by performing a controlled unitary on the composite $AB$ system. As such, our protocol may be regarded as implementing a controlled unitary that generically cannot be realized using finite-depth unitary circuits. This unitary description also provides a powerful handle allowing us to characterize $\rho_B$ by analyzing the parent Hamiltonian of its purification.

Our framework is  inspired by adaptive circuits,  where the choice of the applied local unitary gates depends on the global measurement outcomes in a way that post-measurement pure state trajectories associated with different measurement outcomes will be deterministically converted to the same target pure state. This architecture has provided a powerful framework that enables the  efficient preparation of a large class of non-trivial pure states, including gapped topological orders and  gapless conformal critical states in short times \cite{Raussendorf_2001_ghz,3d_cluster_state_2005,cirac_2008_optical,stace_2016_css,cirac_2021_locc,ashvin_2021_measurement,verresen2021_measurement_cold_atom,bravyi_2022_adaptive,lu2022measurement,ashvin_single_shot_2022,ashvin_hierarchy_2022}, which are impossible using any local unitary protocols. The operations required by adaptive circuits (mid-circuit measurements and feedback) are available in several quantum hardwares, and indeed the adaptive, finite-depth preparation of certain topological orders has been realized experimentally in recent works \cite{iqbal2023topological,foss2023experimental}. While our construction of quantum channels utilizes the same ingredients (local quantum operations and non-local classical communication) as in adaptive circuits, various post-measurement states generically do not converge in our protocol, thereby leading to a mixed-state ensemble that exhibits certain long-range quantum order and quantum criticality despite having extensive entropy.

 For the applications discussed in this work, the unitary feedback after local measurements is a product of onsite unitary operations, and hence, the long-range order in the resulting ensemble may equivalently be decoded from post-measurement pure-state trajectories via appropriate classical post-processing without unitary feedback (see e.g.  Ref.\cite{lee2022decoding,zhu2022nishimori}). However, our protocol can be generalized, e.g. by extending onsite-unitary feedback to a finite-depth local unitary circuit, or by considering multiple rounds of layers of measurement and unitary. In both cases, it is unclear if the properties of the resulting mixed state can be efficiently obtained by classical post-processing based on local measurement data. Furthermore, not all classical post-processing can be implemented efficiently as quantum channels.

In the rest of the Introduction, we provide an overview of two classes of applications. The first class of examples takes a symmetry-protected topological (SPT) order as an input state (Sec.\ref{sec:spt}). These are short-range entangled phases that can be prepared from product states using local unitary in finite time only when breaking the protecting symmetry \cite{spt_1d_2011,spt_2011}. We will focus on SPT phases characterized by decorated domain wall constructions \cite{dwSPT}. One class of examples are SPT phases in $d$-spatial dimensions which are protected by $\mathbb{Z}_2$ $p$-form $ \times \mathbb{Z}_2$ $q$-form symmetry with $p+q=d-1$. This type of SPT can be diagnosed by the long-range order in certain non-local operators, e.g. string operators in 1d or membrane operators in 2d. Based on these non-local operators, we show how to employ measurement and feedback to prepare a mixed state with $\mathbb{Z}_2$ long-range order coexisting with volume-law entropy. 	For instance, $\mathbb{Z}_2$ cat-state order and $\mathbb{Z}_2$ topological order can be prepared in one and two space dimensions in a mixed state, both of which cannot occur in equilibrium thermal states.

Notably, the convertibility to a long-range order is a universal property of the SPT: any pure state in the same SPT phase always leads to a long-range-ordered (and generically mixed) state characterized by the same universal properties that are related to the SPT order.  In addition, the output state is a reduced density matrix of a ground state of Hamiltonian $H$ that is ``dual'' to the parent  Hamiltonian $H_0$ of the input SPT.  Most interestingly, when $H_0$ is tuned to criticality, $H$ is critical as well, and the corresponding mixed state possesses a critical entanglement structure quantified by the entanglement negativity \cite{peres1996,horodecki1996,eisert99,vidal2002,plenio2005logarithmic}. For example, our protocol in 1D gives rise to a mixed state with volume-law scaling of von Neumann entropy, but logarithmic scaling of the entanglement negativity with subsystem size. Such mixed-state quantum criticality presents several unconventional features as we will discuss in this work.

Our application of converting a $\mathbb{Z}_2 \times\mathbb{Z}_2$ SPT to long-range order is directly inspired by Ref.\cite{Raussendorf_2001_ghz,3d_cluster_state_2005,ashvin_2021_measurement}. In particular, Ref.\cite{ashvin_2021_measurement} argues that measuring these SPTs will generically lead to certain long-range order in post-measurement pure state trajectories. While this can be shown analytically for fixed-point SPTs \cite{Raussendorf_2001_ghz,3d_cluster_state_2005}, the fate of non-fixed-point SPTs upon measurement was inconclusive\footnote{For non-fixed point SPTs, Ref.\cite{ashvin_2021_measurement} provides numerical evidence that measuring the Heisenberg spin-1 chain, an SPT protected by $\mathbb{Z}_2 \times\mathbb{Z}_2$ symmetry of $\pi$ rotations, leads a GHZ-type long-range order.}.  In contrast, our channel-based approach allows us to show that the emergence of long-range orders \textit{in a mixed state} is a universal property that persists throughout the entire SPT phase. Importantly, these orders can be efficiently probed through certain linear observables in the mixed-state ensemble that results from our quantum channel.

	The second class of examples focuses on spinful fermionic systems  (Sec.\ref{sec:fermions}). We consider extensive single-site fermion occupation number measurement followed by unitary feedback according to the measurement outcomes. In contrast to the single-site projective Pauli measurement, which trivializes the measured qubit, the occupation number for spinful fermions allows for richer phenomenology due to the possibility of a residual spin-1/2 degree of freedom in the subspace of a singly-occupied site. Indeed, it is known that starting with non-interacting spinful fermions, performing a Gutzwiller projection (projecting onto single-occupation subspace) effectively induces interactions that may lead to various exotic quantum spin liquids in two and higher spatial dimensions (see Ref.  \cite{Savary_2017_spin_liquid,Ng_2017_spin_liquid} for review). The Gutzwiller projection cannot be implemented as a fermion occupation-number measurement, without post-selecting the measurement outcome. In contrast, our scheme does not rely on post-selection; for each post-measurement trajectory, we apply a depth-1 local  unitary feedback, so that the mixed state, formed from the  ensemble of these trajectories, exhibits certain non-trivial features. We  also note that the fermion-occupation number measurement is readily experimentally available via quantum gas microscopes (see Ref.\cite{microscope_review_2021} for a review), which allows for the implementation of our protocols.  Below we briefly summarize  two applications,  in one and two space dimensions respectively.

	In 1d, we start from free spinful fermions (Sec.\ref{sec:1d_fermion}) with nearest neighbor hopping. In this case, it is known that spin-spin correlations decay algebraically with the separation between two sites with an exponent $2$.  We devise a protocol that outputs   a mixed state with an enhanced critical correlation characterized by a smaller exponent, namely, $1$. The essential idea is to notice that there is a long-range string order in the 1d free fermions \cite{squeeze_fermion_2004}, and measurement and feedback can convert this hidden string order into a truly long-range critical order. The resulting mixed state that describes correlations in the spin sector can be regarded as a reduced density matrix by tracing out the charge sector in a ground state of a parent Hamiltonian that we derive. Notably, this  Hamiltonian describes strongly interacting spinful fermions and therefore, our measurement-feedback channel may be viewed as a novel mean to effectively engineer interactions. We note that our protocol of boosting critical correlations is also applicable to interacting fermionic systems characterized by Luttinger liquids by exploiting their hidden string order.

	The 2d example (Sec.\ref{sec:chern}) takes Chern insulators  (see e.g.\cite{Bernevig_2013} for an introduction) as an input.  These are gapped invertible topological phases, which feature trivial bulks and non-trivial chiral edge modes on the boundary. Despite Chern insulators  having exponentially decaying correlations in the bulk, applying a  measurement-feedback channel leads to a mixed state with algebraic correlations. The essential property we use is a non-local membrane order parameter due to the topological response in Chern insulating states \cite{Girvin_1987_fqht,ryu_response_2020}. Such a non-local hidden order can be converted into algebraic decaying spin-spin correlations in the resulting mixed state using a non-local quantum channel. This therefore furnishes a remarkable example where quantum criticality  emerges from a gapped state of matter via a quantum channel involving local quantum operations but non-local classical communication.

\section{Non-local channel from local quantum operations and non-local classical communication}\label{sec:general}

Here we introduce the protocol for constructing quantum channels based on local quantum operations (measurement and unitary) and non-local classical communication (Fig.\ref{fig:main_fig}). Given a system with two types of degrees of freedom $A$ and $B$ initialized in the state $\rho_0= \ket{\psi_0}\bra{\psi_0}$, one performs simultaneous, extensive single-site measurement on every degree of freedom in $A$. This leads to  a particular pure state $\ket{\psi_\alpha} = \frac{P_\alpha\ket{\psi_0}}{  \sqrt{\bra{\psi_0} P_\alpha\ket{\psi_0}}   }$ with probability $\bra{\psi_0 } P_{\alpha} \ket{\psi_0}  $, where $\alpha$ labels the measurement outcome on $A$, and $P_\alpha$ is the projector associated with the measurement.  For each post-measurement state, we apply a unitary $U_\alpha$ acting on  $B$ based  on the outcome $\alpha$. Note that this requires non-local classical communication since the choice of a local unitary relies on distant  measurement outcomes recorded as classical data. The above measurement-feedback protocol leads to a mixed state   
\begin{equation}
	\rho  = \sum_{\alpha}  U_\alpha P_\alpha  \rho_0    P_\alpha U_\alpha^{\dagger}.
\end{equation}
With an appropriate choice of measurement and unitary feedback, the subsystem $B$ described by a reduced density matrix $\rho_B$ may exhibit various long-range quantum orders and criticality.

The aforementioned protocol may be viewed as a way to effectively implement a controlled unitary acting on the $AB$ composite system followed by tracing out $A$.  To see this, we notice that $\rho_B$ admits a purification $\ket{\psi}$ via $\rho_B= \tr_A \ket{\psi } \bra{\psi}$, where $\ket{\psi}$ is defined as

\begin{equation}\label{eq:purification}
\ket{\psi} = \sum_{\alpha} U_{\alpha}  P_{\alpha} \ket{ \psi_0}. 
\end{equation}

$U = \sum_{\alpha} U_{\alpha}  P_{\alpha} $ takes the form of a controlled unitary with $A$ being the control and $B$ being the target.  $U$ may not be realized as local unitary circuits, thereby enabling significant changes in the entanglement structure. In particular, $U$ provides a non-local transformation on operators according to Heisenberg evolution. For instance, the expectation of an operator $O_B$ supported on $B$ in the resulting mixed state $\rho_B$ amounts to the expectation of the operator $U^\dagger O_B U$ in the input pure state $\ket{\psi_0}$. As such, this non-local unitary transformation provides a powerful way to convert hidden orders in the input state into long-range order or criticality in the density matrix $\rho_B$, as we will illustrate using various examples. Moreover, this viewpoint of unitary transformation provides a useful way to describe the output $\rho_B$; since the initial state $\ket{\psi_0}$ and the purified state $\ket{\psi}$ are connected by a unitary $U$, the structure of $\rho_B $ can be characterized based on the Hamiltonian $H$ of $\ket{\psi}$ through  $H= U H_0 U^{\dagger}$,  where $H_0$ is the Hamiltonian of the input state $\ket{\psi_0}$.

\section{Entanglement structure of output mixed states}\label{sec:general_structure}
We first discuss general constraints on the entanglement properties of the mixed states generated from such protocols.

{\bf Bound on entanglement:}
The protocols presented in this work belong to LOCC, namely, local operations (onsite measurement and unitaries) and classical communication, and therefore, mixed-state entanglement cannot increase under these quantum channels (see 	Appendix.\ref{sec:constraint} for proof based on the entanglement of formation \cite{bennett1996}, a faithful entanglement measure for mixed states). This entanglement constraint provides a sharp distinction between the mixed-state orders that may and may not be realized within our protocols. For example, starting with an area-law entangled gapped state in 1d, our finite-depth channels cannot output a mixed-state with $\log L$ scaling entanglement. Namely, to output a quantum critical mixed state in 1d with $\log L $ scaling entanglement, the initial state must be critical with entanglement  $\gtrsim	O(\log L)$ as well, and   Sec.\ref{sec:1d_critical_spt} presents one such example. On the other hand, if one starts with a gapped, area-law state in 2d, the entanglement constraint does not rule out the possibility of realizing quantum critical mixed states (recall gapless conformal critical states obey an area law as well), and indeed this is realized in Sec.\ref{sec:chern}.

{\bf Sufficient conditions for a nontrivial mixed state:}
Despite the above constraint on the quantity of entanglement, our protocol can produce mixed states $\rho$ that are long-range entangled in the precise sense that they cannot be a mixture of trivial pure states \cite{hastings2011}. Namely, $\rho$ is long-range entangled if $\rho \neq  \sum_{n}p_n \ket{\phi_n } \bra{ \phi_n}$, where each $\ket{\phi_n}$ is a short-range entangled state that can be connected to a product state using a finite-time local unitary\footnote{This implies that the connected correlation between two distant operators $O_A, O_B$ is upper bounded by $\exp{-d/\xi}$ where $d$ is the separation between $O_A, O_B$, and $\xi$ is a finite correlation length.}. 

We prove that $\rho$ cannot be a mixture of short-range entangled states given the following two conditions (which will apply to all examples considered in this work). (i) {\it Global symmetry:} There exists a unitary operator $S$ with unit-norm expectation value, i.e.  $\tr(\rho S)=e^{i\theta}$ with $\theta \in \mathbb{R}$. (ii) {\it Long-range correlations for charged operators:} There exist charged operators $O_1, O_2$ (with respect to the symmetry $S$)\footnote{Namely, $SO_jS^{\dagger} = e^{iQ_j} O_j $ with $Q_j$ mod $2\pi \neq  0$ for $j=1,2$.}, whose correlation $\tr(\rho  O_1O_2)$  decay slower than $e^{-d/\xi}$, with $\xi >  0$, and $d$ being the spatial separation between $O_1$ and  $O_2$. For example, $\tr(\rho  O_1O_2)$ could be constant or follow a power-law decay with the separation $d$.  

The claim above can be proved by contradiction. Assume $\rho=  \sum_n p_n \ket{ \phi_n  } \bra{  \phi_n } $, i.e. a mixture of short-range entangled pure states $\ket{\phi_n}$. Since $\tr(\rho S) =e^{i\theta} $, it must be that $\langle \phi_n | S  |\phi_n \rangle=e^{i\theta}$ for each $\ket{\phi_n}$\footnote{This follows from the fact the eigenspectrum of the unitary $S$ is located on the unit circle in the complex plane.}. Further, $\ket{\phi_n}$ being short-range entangled means the connected correlation function decays exponentially:  $ \langle \phi_n | O_1 O_2 |  \phi_n \rangle  - \langle  \phi_n |O_1 | \phi_n\rangle \langle \phi_n |O_2|  \phi_n \rangle  \sim e^{-d/\xi_n}$ \cite{hastings_lrbound_2006} with $\xi_n$ upper bounded by a finite $\xi_{\text{max}}$. $\langle \phi_n | S  |\phi_n \rangle=e^{i\theta}$ and the assumption that each operator is charged with respect to $S$ imply that  $ \langle \phi_n  |  O_1  |  \phi_n \rangle =\langle \phi_n  |  O_2  |  \phi_n \rangle =    0$, and thus $\langle \phi_n  |    O_1 O_2   |  \phi_n \rangle \sim  e^{-d/\xi_n}$.  Since $ O_1 O_2 $ decays exponentially in each $\ket{\phi_n }$,  $  O_1 O_2 $ must decay exponentially in the mixed-state ensemble $\rho$, which contradicts the result that the decay of $\tr(\rho O_1 O_2)$ is slower than exponential. As a result, the initial assumption must be false.

As a demonstration, when the expectation value of a global $\mathbb{Z}_2$ symmetry generator follows $\expval{S} =\expval{\prod_i X_i }=1$ in the mixed state, $Z_iZ_j$ having non-zero expectation value as  the separation $d\to \infty$ implies a non-trivial mixed state with the GHZ-type order. Such an example will be presented in Sec.\ref{sec:1d_bspt}. If $Z_i Z_j$  instead decays algebraically, the corresponding non-trivial mixed state exhibits a quantum criticality with several unconventional properties, which will be discussed in \ref{sec:1d_critical_spt}. One may also generalize $S$ and charged operators $O_j$ to string-like operators, which can witness mixed-state topological order in two space dimensions (see Appendix.\ref{append:2d_spt} for details).

	\section{Mixed-state long-range order by measuring SPT phases}\label{sec:spt}

In this section, we apply the general framework discussed in Sec.\ref{sec:general} to convert SPT phases characterized by decorated domain-wall construction \cite{dwSPT} to mixed-state long-range orders. We will illustrate the main idea using 1d $\mathbb{Z}_2 \times \mathbb{Z}_2$ SPT. More general types of SPTs with distinct symmetries or higher-space dimensions and the corresponding emergent long-range order will be briefly outlined in Sec.\ref{sec:general_application} (with more details in Appendix).

While the measurement-feedback protocol for generating the long-range order from fixed-point SPTs has been known \cite{Raussendorf_2001_ghz,3d_cluster_state_2005,ashvin_2021_measurement}, the situation for SPTs away from fixed points remains elusive. Here we show that through the same measurement-feedback protocol, any state belonging to the same SPT phase can be universally converted to a long-range order (Sec.\ref{sec:1d_bspt}). Moreover, we develop a duality approach to better characterize the nature of the long-range order, which will be discussed in Sec.\ref{sec:duality}.

 We also note that Ref.\cite{lee2022decoding,zhu2022nishimori} discussed a protocol that can convert certain SPTs to a mixed-state ensemble with hidden long-range orders that can be decoded via non-linear observables or classical post-processing. However, the specific protocol is different from ours, and thus the type of long-range order that can be stabilized also differs. For example,  our setup  can realize  GHZ-type order in 1d and $\mathbb{Z}_2$ topological order in 2d, both of which are absent  in those works. More technically, 
 the protocol discussed in Ref.\cite{lee2022decoding,zhu2022nishimori} amounts to introducing thermal fluctuations, which is a qualitatively different perturbation than the ones we consider.

\subsection{Measuring 1d SPT}\label{sec:1d_bspt}

	\begin{figure*}
		\centering
		\begin{subfigure}{\textwidth}
			\includegraphics[width=\textwidth]{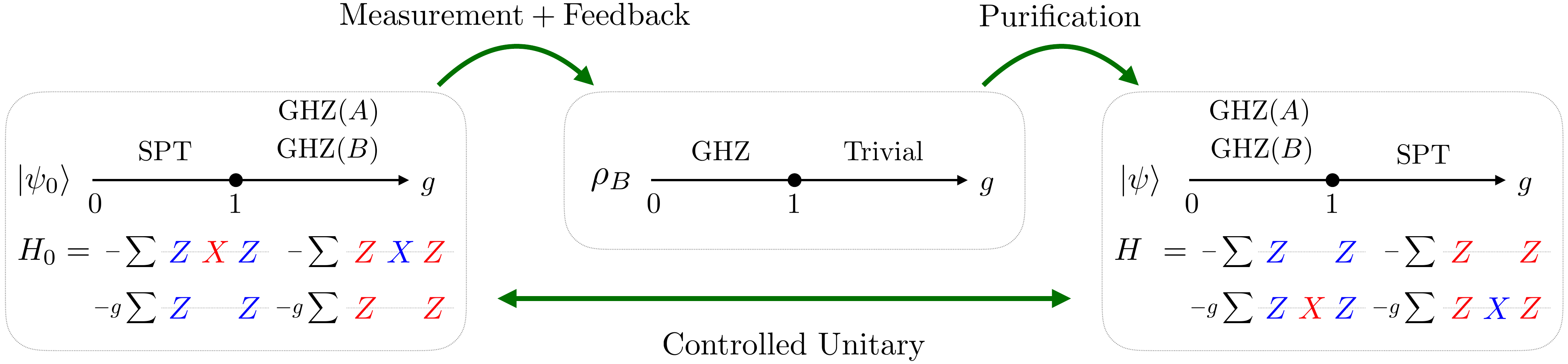}
		\end{subfigure}\caption{Starting with an input $\ket{\psi_0}$, which undergoes a transition from a $\mathbb{Z}_2\times \mathbb{Z}_2$ SPT order  in the $AB$ composite system to two independent GHZ-type symmetry breaking orders in $A, B$ respectively, a measurement-feedback protocol leads to a density matrix on $B$ (i.e. $\rho_B$), which undergoes a transition from a GHZ-type order to a trivial mixed state. $\rho_B$ admits a purification $\ket{\psi}$ that can be obtained from the input state $\ket{\psi_0}$ via a controlled unitary $U= \sum_{\alpha} U_{\alpha}P_\alpha$, and the Hamiltonian of $\ket{\psi}$ can be derived based on the transformation rules (Eq.\ref{eq:U_transform}). The purification therefore provides a useful handle to characterize the structure of the output density matrix $\rho_B$.}
		\label{fig:1dspt_main}
	\end{figure*}

We consider a 1d lattice with periodic boundary conditions, and the lattice sites are labeled with the ordering $(a,1), (b,1), (a,2), (b,2), ..., (a,L), (b,L)$. The cluster state Hamiltonian is defined as $- \sum_{i=1}^L  ( Z_{a,i}X_{b,i}Z_{a,i+1 }  +  Z_{b,i}X_{a,i+1}Z_{b,i+1 })$. The ground state exhibits an SPT order protected by the global $\mathbb{Z}_2\cross \mathbb{Z}_2$ symmetry generated by $\prod_{i=1}^LX_{a,i}$ on $A$ sublattice  and $\prod_{i=1}^L X_{b,i}$ on $B$ sublattice. Under symmetric local perturbations without gap closing, the SPT order is robust in the ground state $\ket{\psi_0}$, and can be diagnosed by the long-range string order (see e.g. Ref.\cite{Pollmann_2012_string_order})
\begin{equation}
 \bra{  \psi_0} Z_{b,i}  \left( \prod_{k=i+1 }^{j}   X_{a,k} \right)  Z_{b,j}  \ket{\psi_0} =c 
\end{equation}
with $0<c\leq 1$ as $\abs{i-j} \to \infty$ \footnote{$c=1$ in the ground state of cluster state Hamiltonian.}.  

Now we show that based on the long-range string order, one can prepare a mixed state with GHZ-like $\mathbb{Z}_2$  symmetry-breaking order using measurement and unitary feedback in constant depth.

To start, we measure Pauli-X for every site in $A$ sublattice and denote the measurement outcome of $X_{a,i}$ by  $\alpha_i$. Defining $\alpha= \{ \alpha_i    \}$ as the collection of outcomes, one obtains a post-measurement state \{$ \frac{P_\alpha\ket{\psi_0}   }{ \sqrt{  \bra{ \psi_0 } P_\alpha  \ket{ \psi_0  }    }     } $ 	with probability $p_\alpha   = \bra{ \psi_0 } P_\alpha  \ket{ \psi_0  }  $ and projector  $P_\alpha \equiv  \prod_{i} \frac{1+ \alpha_i X_{a,i}}{2}$. Without recording the outcome, the system is described by a mixed state, i.e. an ensemble of pure states corresponding to distinct measurement outcomes:	$\sum_{\alpha}  P_\alpha \rho _0  P_\alpha$ with $\rho_0=  \ket{\psi_0} \bra{\psi_0} $. This measurement-induced ensemble lacks any long-range order. For instance, the two-point function on $B$ sublattice with respect to the mixed state is $\bra{ \psi_0  } Z_{b,i} Z_{bj}   \ket{\psi_0}$, which is nothing but the two-point functions with respect to the initial SPT, therefore decaying exponentially with the separation.

In contrast, applying unitary feedback based on the measurement outcomes leads to a non-trivial mixed state. The essential idea is to choose a unitary such that the two-point $Z_bZ_b$ operator evaluated in the mixed state amounts to the string operator $Z_b X_a... X_aZ_b$  in the input SPT, thereby taking a non-zero expectation value.

First, for a post-measurement pure state with outcome $\alpha$, we apply a unitary $U_\alpha$ on $B$ sublattice:

	\begin{equation}\label{eq:1d_spt_correct}
		U_\alpha = \prod_{ i  }  X_{b,i}^{ \frac{1- \prod_{ j=1, 2, \cdots }^i \alpha_j  }{2}}. 
	\end{equation}
	In other words, $X_{b,i}$, a Pauli-X on $B$ sublattice, is applied when  there is an odd number of outcome $-1$ from the site $(a,1)$ to the site $(a,i)$. It follows that $Z_{b,i}$ conjugated by $U_\alpha$ will acquire a $1~ (-1)$ sign if there are even (odd) number of $-1$ measurement outcomes from the site $(a,1)$ to the site $(a,i)$. Correspondingly, one finds $U^{\dagger}_{\alpha}   Z_{b,i} Z_{b,j} U_{\alpha}  =  Z_{b,i} \left(\prod_{k=i+1 }^{j}   \alpha_{k}  \right)  Z_{b,j}$. The overall measurement and unitary operation lead to the  mixed state $\rho  = \sum_{\alpha}  U_\alpha P_\alpha  \rho_0    P_\alpha U_\alpha^{\dagger}$. The long-range order can be diagnosed by the two-point $ZZ$ correlation on $B$ sublattice:

	\begin{equation}\label{eq:1d_spt_derivation}
		\begin{split}
			\tr[   \rho  Z_{b,i} Z_{b,j}    ]  &= \sum_\alpha   \bra{\psi_0}     P_\alpha U_\alpha^{\dagger}  Z_{b,i} Z_{b,j } U_\alpha P_\alpha \ket{ \psi_0} \\
			&    =  \sum_\alpha   \bra{\psi_0}     P_\alpha    Z_{b,i} \left(\prod_{k=i+1 }^{j}   \alpha_{k}  \right)  Z_{b,j}   P_\alpha \ket{ \psi_0} \\
			&    = \bra{  \psi_0} Z_{b,i}  \left( \prod_{k=i+1 }^{j}   X_{a,k} \right)  Z_{b,j}  \ket{\psi_0}
		\end{split}
	\end{equation}
	where  we have used the fact that the measurement outcome $\alpha_k$ can be replaced with $X_{a,k}$ due to the projector $P_\alpha$, and $\sum_{\alpha} P_{\alpha}=1$. Therefore, the two-point function in $\rho$ is exactly the string order in the initial SPT order state, and hence  $\expval{ Z_{b,i} Z_{b,j}   } = c=O(1) >0$.  As discussed in Sec.\ref{sec:general_structure}, this non-decaying two-point function together with the  $\mathbb{Z}_2$ symmetry on $B$ sublattice, i.e. $\expval{\prod_i X_{b,i}}=1$, indicates that  $\rho_B (= \tr_A \rho)$ is a non-trivial mixed state with a GHZ-type quantum long-range order. Importantly, since the string order is a universal property 
of the input SPT, which is robust under any finite-time symmetry-preserving local unitary evolution, the convertibility to GHZ-type order is a universal property of the input SPT phase.

\subsection{Duality transformation}\label{sec:duality}
	
	As discussed  in our general framework (Sec.\ref{sec:general}), our protocol can be viewed as realizing a controlled unitary $U= \sum_{\alpha} P_{\alpha} U_{\alpha}$ acting on the $AB$ composite system  followed by tracing out $A$. Specifically, starting from the input $\ket{\psi_0}$ with the  Hamiltonian $H_0$, there exists a purification $\ket{\psi}$ of the output state $\rho_B  (=  \tr_A \ket{\psi} \bra{\psi})$ with  $ \ket{\psi} = U \ket{\psi_0}$. This allows us to  derive the parent Hamiltonian $H$ of $\ket{\psi}$ through $H= U H_0 U^{\dagger}$, so the structure of the output mixed state $\rho_B$ can be characterized. Fig.\ref{fig:1dspt_main} provides a summary for this subsection.

	With $U= \sum_{\alpha} P_{\alpha} U_{\alpha}$ ($U_{\alpha}$ defined in Eq.\ref{eq:1d_spt_correct}), one derives the transformation rule for operators under the conjugation of $U$ (see Appendix.\ref{append:1dspt_duality} for details).  
	
	\begin{equation} \label{eq:U_transform}
		\begin{split}
			&  X_{a,i}	  \to   X_{a,i},  \quad  X_{b,i}	  \to   X_{b,i},        \\
			&  Z_{a,i} Z_{a,i+1}	  \to   Z_{a,i}X_{b,i} Z_{a,i+1} , \\
			& Z_{b,i} Z_{b,i+1}	  \to   Z_{b,i}X_{a,i+1} Z_{b,i+1}   .
		\end{split}
	\end{equation}
Pauli-X is invariant since $U$ is diagonal in X basis.  On the other hand, neighboring $ZZ$ on one sublattice is attached with a Pauli-X on another sublattice in between two Pauli-Zs. This is quite intuitive since the unitary feedback is designed to transform the product of two Pauli-Zs on one sublattice with a sign that depends on the product of measurement outcomes (on another lattice) between these two Pauli-Zs. We also note that the above duality mapping can be understood as implementing a Kramers-Wannier (KW) duality conjugated by a unitary $U_{\text{CZ}}$  
 ($=\prod CZ_{(a,i), (b,i) }CZ_{(b,i),(a,i+1)  }$) that prepares a $\mathbb{Z}_2\times \mathbb{Z}_2 $ cluster SPT \footnote{For instance, sequentially applying $U_{\text{CZ}}$, KW duality, $U_{\text{CZ}}$, one finds $X_{b,i} \to  Z_{a,i} X_{b,i} Z_{a,i+1} \to Z_{a,i} X_{b,i} Z_{a,i+1} \to X_{b,i}$, and $Z_{a,i} Z_{a,i+1}\to Z_{a,i} Z_{a,i+1} \to  X_{b,i} \to Z_{a,i} X_{b,i}Z_{a,i+1}  $.}. This is dubbed twisted gauging in Ref.\cite{Oshikawa_2023_spt_ssb}, which is in contrast to the gauging via Kramers-Wannier duality.


The above duality mapping  provides a powerful tool to characterize the structure of the output mixed state; As an application, we consider the ground state of the following Hamiltonian as an input:

	\begin{equation}\label{eq:1d_spt_symmetric_deform}
		\begin{split}
			H_0  = & -\sum_{i}Z_{a,i} X_{b,i} Z_{a,i+1 }  -\sum_{i}Z_{b,i} X_{a,i+1} Z_{b,i +1 } \\
			&		    -g  \sum_{i}  Z_{a,i} Z_{a,i+1 }   -g  \sum_{i}  Z_{b,i}Z_{b,i+1 }.  
		\end{split}
	\end{equation}

	The phase diagram of $H_0$ can be completely determined. To see this, by conjugating a product of controlled-Z gate: $U_{\text{CZ}}=\prod CZ_{(a,i), (b,i) }CZ_{(b,i), (a,i+1)  }$, one obtains 
	
	\begin{equation}
		\begin{split}
			& -\sum_{i}X_{b,i} -\sum_{i} X_{a,i} \\
			&		    -g  \sum_{i}  Z_{a,i} Z_{a,i+1 }   -g  \sum_{i}  Z_{b,i}Z_{b,i+1 },
		\end{split}
	\end{equation} 
	i.e. two decoupled Ising chains on two sublattices, where  $\abs{g}<1$ and $\abs{g}>1$ correspond to trivial phase and spontaneous symmetry breaking (SSB) phase with GHZ $\mathbb{Z}_2$ orders on $A$ sublattice and $B$ sublattice. This implies the ground state of $H_0$ belongs to SPT and SSB phase with GHZ order for $\abs{g}<1$ and $\abs{g}>1$, respectively. 
	
	Using the transformation rule in Eq.\ref{eq:U_transform}, one finds  the measurement-feedback channel leads to a mixed state $\rho_B$ on $B$ sublattice, which is a reduced density matrix of the ground state $\ket{\psi}$ of the following Hamiltonian 
	
	\begin{equation}\label{eq:1d_spt_symmetric_measure}
		\begin{split}
			H  = & -\sum_{i}Z_{a,i}  Z_{a,i+1 }  -\sum_{i}Z_{b,i} Z_{b,i +1 } \\
			&		    -g  \sum_{i}  Z_{a,i}    X_{b,i}  Z_{a,i+1 }   -g  \sum_{i}  Z_{b,i}  X_{a,i+1}  Z_{b,i+1 }.  
		\end{split}
	\end{equation} 
	Comparing Eq.\ref{eq:1d_spt_symmetric_deform} and Eq.\ref{eq:1d_spt_symmetric_measure}, one finds they are dual to each other, and the phase of $H$ can be determined analogously: $A$ and $B$ sublattices  individually  exhibit  a GHZ order for  $\abs{g}<1$, and the $A\bigcup B$ together exhibits SPT order for $\abs{g}>1$. Consequently, $\rho_B$ possesses a mixed-state GHZ order for $\abs{g}<1$, and $\rho_B$ becomes a trivial mixed state for $\abs{g}>1$ (since the subsystem of the SPT gives a trivial mixed state).  In addition, since $\ket{\psi}$ can be obtained by applying extensive controlled-Z gates across $A, B$ sublattices (i.e. $U_{\text{CZ}}=\prod CZ_{(a,i), (b,i) }CZ_{(b,i), (a,i+1)  }$) on two decoupled Ising chains on A/B sublattice, one 
 expects $\rho_B  = \tr_A \ket{\psi} \bra{\psi}$ has volume-law entropy for any non-zero $g$ ($U_{\text{CZ}}$ acts trivially at $g=0$). This is indeed the case based on our Exact Diagonalization calculation (see  Appendix.\ref{sec:volume_law_entropy}).


While here we only consider one type of perturbation to illustrate the utility of the duality approach, in Appendix.\ref{sec:general_perturbation} we discuss other types of perturbation, including independently tunable perturbation strength in $Z_{a,i} Z_{a,i+1}, Z_{b,i} Z_{b,i+1}$ as well as onsite Pauli-X perturbation.  Interestingly, when perturbing the fixed-point SPT using onsite Pauli-Xs, the corresponding output  $\rho_B$ is exactly the (pure) ground state of the transverse-field Ising chain in the symmetry-broken phase in the subspace with $\prod_i X_{b,i}=1$.

	\subsection{Mixed-state quantum criticality}\label{sec:1d_critical_spt}
	Our protocol can also output a mixed state with volume-law entropy coexisting with critical (algebraic) long-range order. This occurs when applying our measurement-feedback channel to a critical state, i.e. the ground state of $H_0$ in Eq.\ref{eq:1d_spt_symmetric_deform} at $g=1$. In this case, the output will be a mixed state $\rho_B$ on $B$ sublattice by tracing out $A$ sublattice for the ground state $\ket{\psi}$ of $H$ in Eq.\ref{eq:1d_spt_symmetric_measure} at $g=1$ \footnote{In fact, at the critical point $g=1$, 
		$H=UH_0U^{\dagger}=H_0$, so $\ket{\psi_0} $ is the same as $ \ket{\psi} $  	defined in Eq.\ref{eq:psi_cft}.}, where $\ket{\psi}$ reads
	
	\begin{equation}\label{eq:psi_cft}
		\ket{\psi} = U_{\text{CZ}} \ket{\text{CFT}}_A \otimes \ket{\text{CFT}}_B. 
	\end{equation}  
 $\ket{\text{CFT}}_{A/B}$ denotes the ground state of a transverse-field Ising chain (on $A/B$ sublattice) at a critical point,  characterized by the 1+1D Ising CFT, and $U_{\text{CZ}}=\prod CZ_{(a,i), (b,i) }CZ_{(b,i), (a,i+1)  }$. The corresponding mixed state $\rho_B= \tr_A \ket{ \psi }\bra{\psi}$ exhibits quantum criticality diagnosed by certain operators. For example, since $U_{\text{CZ}}$ commutes with Pauli-Zs, the two-point $ZZ$ function is given by the single Ising critical chain, which exhibits an algebraic decay:  $\expval{Z_{b,i } Z_{b,j} } =   \bra{\text{CFT}}_B Z_{b,i } Z_{b,j} \ket{\text{CFT}_B } \sim  \frac{1}{\abs{i-j}^{\eta}  }   $ with $\eta=1/4 $ being a critical exponent in 1+1D Ising CFT. On the other hand, the disorder operator   $X_{b,i}X_{b,i+1}...X_{b,j}$ also exhibits an algebraic decay: 
	
	\begin{equation}\label{eq:1d_disorder}
		\begin{split}
			&\left\langle X _{b,i}...X_{b,j} \right\rangle\\
			&=  \bra{\text{CFT}}_A  \bra{\text{CFT}}_B    U_{CZ}^{\dagger}X _{b,i}...X_{b,j} U_{CZ}     \ket{\text{CFT}}_A  \ket{\text{CFT}}_B     \\
			&=  \bra{\text{CFT}}_A  \bra{\text{CFT}}_B    Z_{a,i}X _{b,i}...X_{b,j}Z_{a,j+1}     \ket{\text{CFT}}_A  \ket{\text{CFT}}_B     \\
			&= \left\langle Z_{a,i}Z_{a, j+1}  \right\rangle_{CFT,A} \left\langle X _{b,i}...X_{b,j}\right\rangle_{CFT,B}\\
			&\sim \frac{1}{\lvert i - j \rvert^{2\eta}},
		\end{split}
	\end{equation}
	where we  used the Kramers-Wannier duality to convert $X_{b,i}X_{b,i+1}...X_{b,j}$ to $Z_{b, i } Z_{b, j+1}$. Importantly, we note that the disorder operator is distinct from a single pure CFT, where $ X _{b,i}X_{b,i+1}...X_{b,j}  \sim \frac{1}{\lvert i - j \rvert^{\eta}}$.

	\begin{figure}
		\centering
		\begin{subfigure}{0.47\textwidth}
			\includegraphics[width=\textwidth]{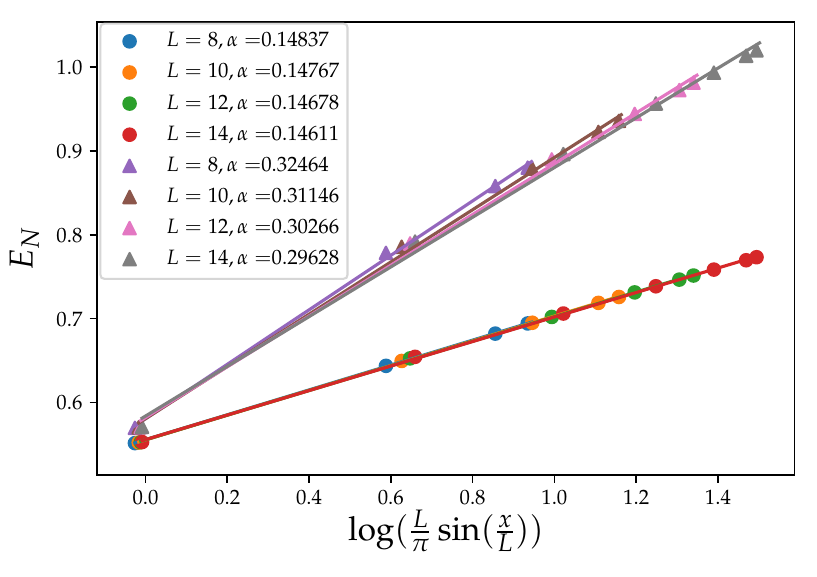}
		\end{subfigure}
		\caption{Entanglement negativity $E_N$ between two complementary segments of sizes $x$ and $L-x$ on $B$ sublattice. 
  Data denoted by circles are for the critical mixed state resulting from measurement and feedback on the SPT critical point (equivalently, $\rho_B =\tr_A \ket{\psi} \bra{\psi}  $ with $\ket{\psi}$ defined in Eq.\ref{eq:psi_cft}). The linear fit of the data indicates $\rho_B$ has an entanglement structure like that of a 1+1D CFT. As a comparison, triangles correspond to the pure 1+1D  Ising critical state $\ket{\text{CFT}}_B$ on $B$.}
		\label{fig:mixed_state_EN}
	\end{figure}

	The mixed state $\rho_B$ exhibits genuine quantum criticality. Specifically, based on the two conditions, i.e. algebraic decay in $ZZ$ two-point function and $\prod_i X_{b,i}=1$, $\rho_B$ cannot be an ensemble of short-range entangled pure states as discussed in Sec.\ref{sec:general_structure}.  To better characterize the entanglement signature for the mixed-state quantum criticality, we bipartition $B$ sublattice into two intervals $B_1$ and $B_2$, and quantify  their entanglement using  entanglement negativity $E_N$ \cite{peres1996,horodecki1996,eisert99,vidal2002,plenio2005logarithmic}, an entanglement measure of mixed states: $	E_{N}(\rho_B) = \ln \Big(\lvert\lvert \rho_B^{T_{B_1}} \rvert\rvert_1\Big)$
	, where the upper script $T_{B_1}$ denotes the partial transpose with respect to the subsystem $B_1$ and $\lvert\lvert ~\cdot ~\rvert \rvert_1$ denotes the trace norm. Using Exact Diagonalization (ED), we find (see Fig.\ref{fig:mixed_state_EN})  negativity follows a universal scaling form as in the 1+1D CFT \cite{wilczek_1994,calabrese_2004}: 
	\begin{equation}\label{eq:finite_EN}
		E_N(x, L) = \alpha \ln (\frac{L}{\pi}\sin(\frac{\pi x}{L})) + \beta
	\end{equation}
	where $x, L$ are the sizes of the subsystem $B_1$ and the entire $B$ sublattice, and $\alpha\approx 0.15$. As a comparison, the finite-size numerics for a pure critical Ising chain reports $\alpha\approx 0.30$\footnote{For a pure 1+1D CFT,  $\alpha$ is a universal number related to central charges, and $\alpha = \frac{1}{4}$ for Ising CFT \cite{Calabrese_2013_negativity}. The mismatch between $\frac{1}{4}$ and $0.30$ is attributed to finite-size effects.}. This indicates that the amount of long-range entanglement on $B$ sublattice decreases when coupling to  $A$ sublattice. Since the prefactor $\alpha$ is generically a universal number that relates to the number of low-energy degrees of freedom, one interpretation is that  coupling between $A, B$ sublattices diminishes the low-energy degrees of freedom that carry  long-distance entanglement  on $B$ sublattice.  This is also consistent with the fact that the disorder operator decays faster after coupling to $A$ sublattice (Eq.\ref{eq:1d_disorder}).

	Finally,  we remark that our protocol of converting hidden order is also applicable when the input state is mixed. One immediate application is that when the input pure SPT is subject to a symmetry-preserving noise channel, in which case the string order survives, the output of our measurement-feedback channel will remain  long-range ordered. In addition,  with the  critical state ($g=1$ in Eq.\ref{eq:1d_spt_symmetric_deform}) under a noise channel as an input, the output mixed state exhibits a log scaling of bipartite entanglement negativity (as in Eq.\ref{eq:finite_EN}) while the log-scaling prefactor decreases continuously as increasing the noise rate. See Appendix.\ref{append:decoherence} for a detailed discussion.

	\subsection{Generalization to other mixed-state long-range order}\label{sec:general_application}
	Here we briefly summarize several classes of mixed-state long-range order that can be realized based on our general framework. In all these examples, we derive an exact duality between the Hamiltonian of the input SPT and the Hamiltonian of a purified state whose sublattice encodes the long-range order, which allows us to establish the correspondence  between  phase diagram of the input state  and that of the output state. 

	\textbf{$\mathbb{Z}_2$ topological order in higher space dimensions}: our protocol can be  straightforwardly generalized by considering input states as  $\mathbb{Z}_2$ $p$-form $\times \mathbb{Z}_2 $ $q$-form SPTs in $d$-space dimension with  $p+q= d-1$. Using measurement-feedback channels, one can prepare a mixed state with  $\mathbb{Z}_2$ topological order at $d\geq 2$ space dimensions. In Appendix.\ref{append:2d_spt} we detail the realization of mixed-state $\mathbb{Z}_2$ topological order in $2d$.

	\textbf{$\mathbb{Z}_2$ fermionic topological order}: 
	Another straightforward  application is the preparation of mixed states with $\mathbb{Z}_2$ fermionic topological order. This can  be achieved by taking $\mathbb{Z}_2 \times \mathbb{Z}_2^f$ SPTs  \cite{fermion_spt_2014_wen,floquet_spt_2016_sondhi,Ashvin_2018_commuing,Verresen_kitaev_2021,ashvin_2021_measurement} as input, where the first $\mathbb{Z}_2$ symmetry acts on qubits and the second $\mathbb{Z}_2^f$ is the parity symmetry on fermions. Based on certain non-local hidden orders in these SPTs, we devise a protocol that outputs fermionic mixed states with  long-range order. In Appendix.\ref{append:kitaev_1d} we present the protocol that realizes the fermionic mixed state in 1d with  the same long-range  order as in the topological phase of the 1d Kitaev chain. In Appendix.\ref{append:fermion_2d_topo}  we present the protocol that realizes the fermionic mixed states in 2d with intrinsic topological order.

	\section{Mixed-state quantum criticality by measuring fermions}\label{sec:fermions}
	In this section, we construct quantum channels based on fermion occupation number measurement in spinful fermions. This   is naturally motivated  by Gutzwiller projection \cite{Gutzwiller_1963}, a standard approach to construct exotic states of matter by projecting  spinful non-interacting fermions into the subspace of single occupation number per lattice site, yielding a spin-1/2 wavefunction.  
 For example, Gutzwiller projecting free fermions at half-filling in 1d leads to a critical ground state of the Haldane-Shastry model \cite{Haldane_1988,Shastry_1988}, a long-range interacting Heisenberg antiferromagnet with  $1/r^2$ exchange coupling. In 2d, certain  spin liquids with topological order can  be constructed similarly (see Ref.  \cite{Savary_2017_spin_liquid,Ng_2017_spin_liquid} for review). More broadly, Gutzwiller projection may be viewed as an application of parton construction \cite{wen2004quantum}, a well-known approach to  constructing   non-trivial states by imposing certain   constraints on non-interacting particles.

	Gutzwiller projection/parton construction provides inspiration for  using measurement to  implement the projection. Indeed, Ref.\cite{lu2022measurement} has presented a protocol to prepare the ground state of the Kitaev honeycomb model \cite{kitaev_2006}, where notably, the desired  projection can be achieved by measuring fermions and applying a depth-1 local unitary feedback. However, beyond this specific model, starting with a generic state,  post-measurement states with unwanted measurement outcomes may not converge to the same target Gutzwiller projected state using finite-depth unitaries. This presents a difficulty in realizing Gutzwiller projection with measurement-based protocols.  Here our framework avoids this conundrum. We will show that occupation number measurement followed by appropriate unitary feedback enables the realization of non-trivial mixed states with certain long-distance quantum correlations.  A field-theoretic understanding of the resulting mixed-states and their correlations will be presented in forthcoming work \cite{Vijay_Forthcoming}.

\subsection{Measuring 1d fermion}\label{sec:1d_fermion}
Taking 1d spinful free fermions as input, below we will present a finite-depth protocol for realizing quantum-critical mixed states with critical correlations distinct from the input states.

Consider non-interacting, spinful fermions on a 1d lattice, where the annihilation fermion operators at site $i$ are denoted by $c_{i,s}, c_{i,s}^{\dagger}$ with site indices $i=1, 2, \cdots, L$ and spin indices $s=\uparrow, \downarrow$, we define a tight-binding Hamiltonian $H= - \sum_{i,s} (c_{i+1,s}^{\dagger}  c_{i,s}  + \text{h.c.} )$, whose ground state $\ket{\psi_0}$ takes the following form:  $   \prod_{k} c_{k,\uparrow}^{\dagger}  \prod_k c^{\dagger}_{k,\downarrow }\ket{0} $ with $k$ being the momentum indices. $\ket{\psi_0}$ exhibits an algebraic correlation between spins:
	\begin{equation}
		\expval{ S^{z}_i S^{z}_j  } \sim \frac{1}{(j-i)^2},
	\end{equation}
	with the spin operator $S^{z}_i=  \frac{1}{2} [  n_{i,\uparrow} - n_{i,\downarrow}   ]$, and $n_i=  n_{i,\uparrow} +n_{i,\downarrow} $ is the  number operator at site $i$. 
	
Interestingly, by decorating with a fermion parity string, the spin-spin correlation decays slower (see Ref.\cite{squeeze_fermion_2004} for details):
	
	\begin{equation}\label{eq:fermion_string}
		\expval{ S^{z}_i  \prod_{l=i+1}^{j-1} (-1)^{n_l} S^{z}_j  } \sim \frac{1}{j-i}
	\end{equation}
 
In other words, the fermion-parity string reveals a more ordered, long-range correlation that is hidden in free fermions. Conceptually the insertion of the fermion string has the effect of removing the charge fluctuations, thereby enhancing the order in the spin sector. The physics is akin to the spin-1 Affleck-Kennedy-Lieb-Tasaki (AKLT) chain \cite{aklt_1987} (belonging to the Haldane SPT phase \cite{haldane_spin_chain_1983}), where the ground state is a  superposition of a class of product states that have a staggered pattern of $+1$ and $-1$ with an arbitrary number of $0$s in between, e.g. $...+00-0+000-+...$. While the two-point function $S^z_iS^z_j$ decays exponentially due to the position disorder of $0$, inserting the string $\prod_{l=i+1}^{j-1}(-1)^ {S^z_l}$ in between $S^z_i$ and $S^z_j$ effectively removes the disorder, thereby revealing the hidden antiferromagnetic order \cite{hidden_order_Rommelse_1989}. The process of removing certain disorders by inserting string operators is dubbed the squeezed-space construction in Ref.\cite{squeeze_fermion_2004}.

More technically, the string order may be understood as the following: the free-fermion state consists of two decoupled fermion chains with opposite spin flavors: $\ket{\psi_0 } =\ket{\psi_0}_{\uparrow}\otimes \ket{\psi_0}_{\downarrow}$,  Each fermion chain exhibits a string order $c^{\dagger}_{i,s} \prod_{l=i+1}^{j-1} (-1)^{n_{l,s}}   c_{j,s} \sim  \frac{1}{ \sqrt{\abs{j-i}}  }$ with $s\in \{\uparrow,\downarrow \}$ \footnote{The non-local string order of a single fermion chain can be mapped the two-point function $ ( X_i+iY_i)( X_j-iY_j)$ of the spin-1/2 XY model $H=- \sum_{i} (X_i X_{i+1}   +  Y_i Y_{i+1}  ) $  via the Jordan-Wigner transformation. Since it is known that the two-point $\expval{X_i X_j} = \expval{Y_i  Y_j  }  \sim  \frac{1}{\sqrt{\abs{i-j}}}  $ in the XY model \cite{XY_McCoy_1968}, the string order of fermion chain obeys that same scaling as well.}. Considering a product of two string operators, one finds $S_i^{\dagger}\prod_{l=i+1}^{j-1} (-1)^{n_{l}} S_j^- \sim \frac{1}{j-i}$, where $S_i^{\pm}$ is the spin-raising/lowering operator. Using the spin rotational symmetry, one then recovers the string order in Eq.\ref{eq:fermion_string}.

	Given the above free-fermion state $\ket{\psi_0}$, measuring the fermion occupation number at each site gives the outcome $n = \{ n_1, n_2, \cdots, n_L  \} $ with each $n_i \in \{0,1,2\}$. Correspondingly, the post-measurement state becomes  $\frac{P_n\ket{\psi_0}   }{ \sqrt{  \bra{ \psi_0 } P_n  \ket{ \psi_0  }    }     }$	with probability $p_n   = \bra{ \psi_0 } P_n  \ket{ \psi_0  }     $.  The measurement-only protocol leads to a  mixed state $\rho_m=   \sum_{n}  P_n\rho_0  P_n$, i.e. an ensemble of pure states corresponding to distinct measurement outcomes. Since spin operators commute with fermion number operators, the  spin-spin correlation remains unchanged:  $		\tr \left[ \rho_m   S^{z}_i S^{z}_j  \right] = \bra{ \psi_0  }   S^{z}_i S^{z}_j  \ket{\psi_0}  \sim \frac{1}{(j-i)^2}.$

	Now we show that with  appropriate unitary feedback, one can obtain a mixed state with enhanced long-range correlation. Specifically,  the string order of the original input state $\ket{\psi_0}$ will be transformed into  spin-spin correlation with $S^{z}_i S^{z}_j \sim \frac{1}{j-i}$  in the resulting mixed state. This is in strong contrast to $S^{z}_i S^{z}_j  \sim \frac{1}{(j-i)^2}$ in the case without unitary feedback.

	The  protocol is as follows.  First, we measure the fermion occupation number on every site. For each post-measurement state with an outcome labeled by $n$, we  apply a local unitary $U_n$. This depth-2 protocol leads to a mixed state:	$\rho  = \sum_{n} U_n P_n\rho_0  P_n U^{\dagger}_n$. Now we look for the unitary $U_n$ that transforms $S_i^{z}$ to $S_i^{z}(-1)^{\sum_{j\leq i} n_j  }$ for all $i$. This can be achieved by choosing
	\begin{equation}\label{eq:U}		
		U_n = \prod_{i=1}^{L} \left( \mathcal{S}_i^x  \right)^{ \sum_{j\leq i} n_j },  
	\end{equation}
	where $\mathcal{S}_i^x$ is a spin-flip operator: $\mathcal{S}_i^x =2 S_i^{x}=c_{i,\uparrow}^{\dagger} c_{i,\downarrow} +  c_{i,\downarrow}^{\dagger} c_{i,\uparrow} $ in the subspace of $n_i=1$, and $\mathcal{S}_i^x=1$ (i.e. it acts trivially) in the subspace of $n_i=0,2$. In other words,  the operator $\mathcal{S}_i^{x}$ is applied at site $i$ when there is an odd number of fermions in the interval $[1,i]$.
	
	Under  the transformation by $U_n$, the spin-spin correlator becomes  $	U_n^{\dagger}S^{z}_i S^{z}_j   U_n  =  - S_i^{z} (-1)^{\sum_{l=i+1}^{j-1} n_l} S_j^{z}$, which acquires a sign that depends on the number of fermions between sites $i$ and $j$. A straightforward calculation shows that

	\begin{equation}
		\begin{split}
			\tr \left[ \rho   S^{z}_i S^{z}_j   \right]& = \sum_n  \bra{\psi_0}P_n U_n^{\dagger} S^{z}_i S^{z}_j     U_n P_n  \ket{\psi_0}\\
			&   =-  \bra{\psi_0}  S_i^{z} (-1)^{\sum_{l=i+1}^{j-1} n_j} S_j^{z}   \ket{\psi_0}  \\
			&\sim  \frac{1}{j-i}. 
		\end{split}
	\end{equation}
Therefore, the hidden string order of the input state $\ket{\psi_0}$ is converted to a critical correlation shared among spins in the output mixed state $\rho$. In particular, the reduced density matrix of $\rho$ that describes spins, obtained by tracing out charge fluctuations, can be purified into a ground state of a local Hamiltonian that describes \textit{interacting} spinful fermions (see Appendix.\ref{append:purification_1d_fermions}).

	One may also consider the spin-spin correlation in other orientations. The global $SU(2)$ spin-rotation symmetry of the input state $\ket{\psi_0}$ implies $\bra{\psi_0}S^{\nu}_i S^{\nu}_j  \ket{\psi_0}  \sim \frac{1}{(j-i)^2}$ and  $\bra{\psi_0} S^{\nu}_i  \prod_{l=i+1}^{j-1} (-1)^{n_l} S^{\nu}_j   \ket{\psi_0} \sim \frac{1}{j-i}$ for $\nu = x, y, z$. Since the feedback unitary consists of $\mathcal{S}_i^x$, which anticommutes with $S_i^y$, the channel also boosts the correlation in $y$ component, i.e. $S^{y}_i   S^{y}_j $ is enhanced from $ \frac{1}{(j-i)^2}$ to $\frac{1}{j-i}$ in the resulting mixed state.  On the other hand,  correlation in $x$ component  remains $\frac{1}{(j-i)^2}$ since the spin operator in $x$ component transforms trivially under the unitary feedback.

	The mixed state arising from our protocol is non-trivial in the sense that it cannot be written as a  mixture of short-range entangled states. To see this, we first note that due to the global spin-flip symmetry of the input state, i.e. $\prod_i \mathcal{S}_i^x  \ket{\psi_0 } = \ket{\psi_0 }$,  the output mixed state $\rho$ will have $\prod_i \mathcal{S}_i^x  =1$ in the expectation value as well  (since this operator commutes with both measurement operators and unitary correction).  Based on the algebraic decay $S_i^z S_j^z $ and $\prod_i \mathcal{S}_i^x =1$, one can again use the proof technique in Sec.\ref{sec:general_structure} to show that $\rho$ cannot be a mixture of short-range entangled states. This in turn suggests certain long-distance entanglement in the mixed state.  Specifically, in light of the well-known $\log L $ entanglement scaling of the input free fermions in 1d, the output state will likely exhibit a $\log L$ scaling mixed-state entanglement (e.g. quantified by entanglement negativity).

When considering interactions in the input state, our protocol continues to generate critical mixed states. As shown in Ref.\cite{squeeze_fermion_2004}, given interacting fermions described by a Luttinger liquid with the Luttinger parameters $K_s$ and $K_c$ corresponding to the stiffness of spins and charges, with repulsive charge interaction (implying $0<K_c<1$) and global $SU(2)$ spin-rotation symmetry (implying $K_s=1$),  the leading order spin-spin correlation behaves as $S^z_i S^z_j  \sim \frac{1}{(j-i)^{K_c+ K_s}}$ and the string operator behaves as 	$ S^{z}_i  \prod_{l=i+1}^{j-1} (-1)^{n_l} S^{z}_j \sim \frac{1}{(j-i)^{K_s}}$. Using the same measurement-feedback channel, one can then transform this string order to a critical spin-spin correlation $S^z_i S^z_j  \sim \frac{1}{(j-i)^{ K_s}}$ in the resulting mixed state, which is in contrast to the initial exponent $K_c+ K_s$.

	\subsection{Measuring  Chern insulators}\label{sec:chern}

Here, we show that starting with a two-dimensional gapped state with short-range correlations in the bulk,  measurement and feedback can remarkably lead to a mixed state with long-range {\it critical} correlations.

	Our starting point is the Chern insulator, a gapped state of matter which features integer-value quantized Hall conductance and gapless edge states.  Despite the short-range correlation in the bulk, i.e. exponential decay of two-point functions, there exists a non-local operator whose correlation decays algebraically. Denoting  a fermion creation operator by $c^{\dagger}(\vb{x}   )$ with $\vb{x}= (x,y)$ being the space coordinate, one defines a dressed fermion creation operator $ \tilde{c}^{\dagger}(\vb{x})  =  \eta(\vb{x}) c^{\dagger}(\vb{x})$,
where $\eta(\vb{x})$ is a non-local operator: 
	
	\begin{equation}
		\eta(\vb{x}) = e^{ i   \int d^2 x'  c^{\dagger}(  \vb{x}' ) c(  \vb{x}' )    \text{arg} (\vb{x}  -   \vb{x}')}, \label{statflux}
	\end{equation} 
	with  $\text{arg}(\vb{x})$ being the polar angle of $\vb{x}$. 
 
In a Chern  insulator $\ket{\nu=1}$ with associated Hall conductance $\sigma_H= \nu (e^2/h)$, the  dressed operators exhibit a critical correlation \cite{ryu_response_2020}:
	
	\begin{equation}
		\bra{\nu=1}    \tilde{c}(\vb{x})   \tilde{c}^{\dagger}(\vb{x'})   \ket {\nu=1}  \sim \frac{1}{\abs{\vb{x}  - \vb {x'}   }^{\alpha  }},  
	\end{equation}
where $\alpha$ is a non-universal constant related to the $U(1)$ response of Chern insulators.

	\begin{figure}
		\centering
		\begin{subfigure}{0.48\textwidth}
			\includegraphics[width=\textwidth]{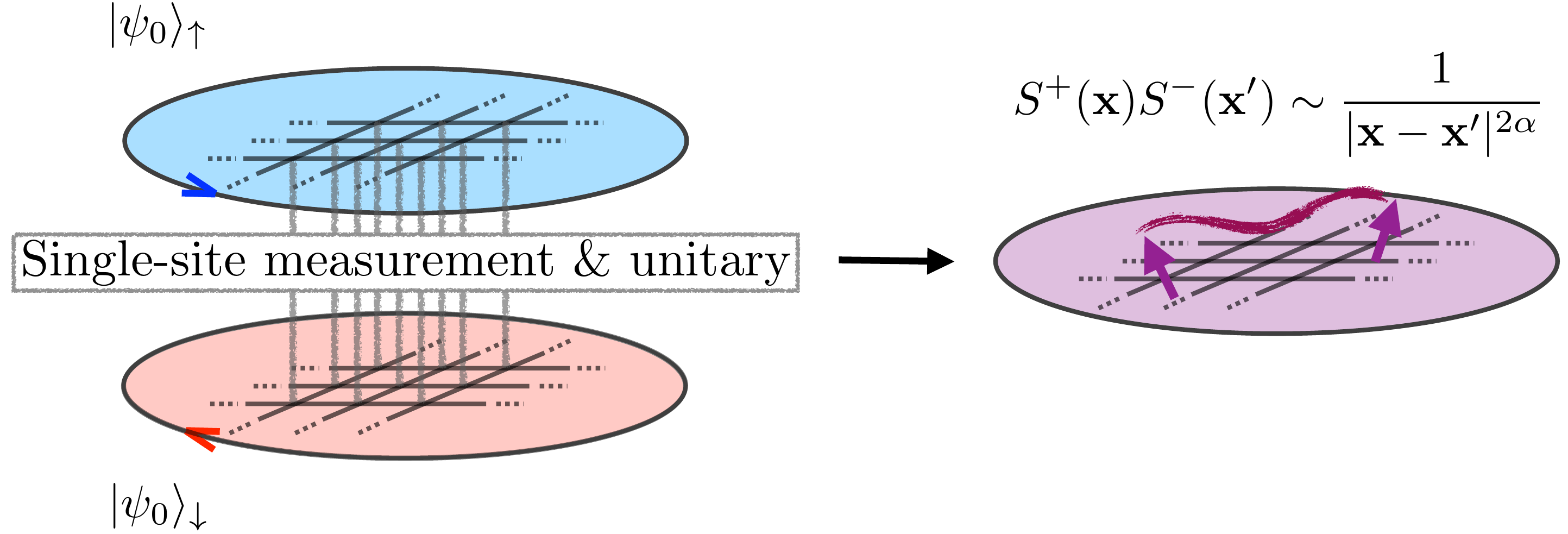}
		\end{subfigure}
		\caption{Spin-up and spin-down fermions are initialized in Chern insulating states with opposite Chern numbers. Such an initial state can be converted to a mixed state with algebraic correlations in the bulk through a two-step protocol consisting of single-site fermion occupation number measurement and unitary.} 
		\label{fig:chern}  
	\end{figure}
 
	As discussed in Ref.\cite{ryu_response_2020},  $\eta(\vb{x})$ can be understood  as the operator   associated  with the	infinitesimal time-evolution given by  inserting  a 2+1D $U(1)$ monopole, and the algebraic long-range order  results from the topological  response to such  monopole insertion in a Chern insulator. Note that the above critical order presents  an off-diagonal long-range order (ODLRO), which was first discovered by Girvin and MacDonald  \cite{Girvin_1987_fqht} in fractional quantum hall states at filling fraction $\nu=1/m$. In particular, these quantum hall states may be understood as a condensate of composite bosons by attaching fluxes to fermions, so the bare fermions need to be dressed by an appropriate non-local operator Eq. \ref{statflux} in order to reveal  the hidden ODLRO. For quantum hall states described  by Laughlin's wave function at $\nu=1$, $\alpha$ is  analytically found to be $1/2$ \cite{Girvin_1987_fqht}.

Below we will discuss a protocol that realizes  a critical mixed state with algebraic two-point correlations based on the hidden order in Chern insulators. To start, we consider an initial state $\ket{\psi_0}=\ket{\psi_0}_{\uparrow}  \otimes  \ket{\psi_0}_{\downarrow} $, where $\ket{\psi _0 }_\uparrow$ is the Chern insulator with $\nu=1$ consisting of spin-up fermions, and $\ket{\psi _0 }_\downarrow$ is the Chern insulator with $\nu= -1$  consisting of spin-down fermions. 	As a result, spin-up and spin-down fermions respectively exhibit an   algebraic hidden order:

	\begin{equation}
		\bra{\psi_0}_{s}    \tilde{c}_{s}(\vb{x})   \tilde{c}_{s}^{\dagger}(\vb{x'})   \ket {\psi_0}_{s}  \sim \frac{1}{\abs{\vb{x}  - \vb {x'}   }^{\alpha   }},  
	\end{equation}
	where $s=1, -1$ for spin-up and spin-down fermions, and the dressed fermion  operators are $\tilde{c}^{\dagger}_s(\vb{x})  =  \eta_s(\vb{x}) c^{\dagger}_s(\vb{x})$,
	with
	\begin{equation}
		\eta_s(\vb{x}) = e^{  i   s \int d^2 x'  c_s^{\dagger}(  \vb{x}' ) c_s(  \vb{x}' )    \text{arg} (\vb{x}  -   \vb{x}')}.
	\end{equation}
	Note the opposite sign in the exponent is due to the opposite sign of Chern numbers for spin-up and spin-down  fermions.

	The hidden critical  order  in the Chern insulators  can be detected through spin-raising/lowering operators   $S^{+}( \vb{x}  ) = c^{\dagger}_{\uparrow}(\vb{x})  c_{\downarrow}(\vb{x}) $ and  $S^{-}( \vb{x}  ) = c^{\dagger}_{\downarrow}(\vb{x})  c_{\uparrow}(\vb{x})$ by dressing them with the appropriate non-local operators. Specifically, let $\hat{n}_{\uparrow}(\vb{x}), \hat{n}_{\downarrow}(\vb{x})$ be the number operator of spin-up/down fermions, and define  $ \hat{\phi}(\vb{x})  =  \int d^2 x' \hat{n}(\vb{x})  \text{arg} (\vb{x}  -   \vb{x}')$ with $\hat{n}(\vb{x}) =\hat{n}_{\uparrow}(\vb{x})  +  \hat{n}_{\downarrow}(\vb{x}) $.  Then the   non-local operator $e^{i \hat{\phi}(\vb{x})}   S^{+}( \vb{x}  )  S^{-}( \vb{x}'  )  e^{-i \hat{\phi}(\vb{x}')}$ exhibits an algebraic decay:
	
	\begin{equation}
		\begin{split}
			&	 \bra{\psi_0}  e^{i \hat{\phi}(\vb{x})}   S^{+}( \vb{x}  )  S^{-}( \vb{x}'  )  e^{- i \hat{\phi}(\vb{x}')}  \ket{\psi_0}\\
			&  = \bra{\psi_0}   \tilde{c}^{\dagger}_{\uparrow}(  \vb{x})  \tilde{c}_{\downarrow}(  \vb{x})    \tilde{c}^{\dagger}_{\downarrow}(  \vb{x}')  \tilde{c}_{\uparrow}(  \vb{x}')     \ket{\psi_0} \\ 
			&  =  -   \bra{\psi_0}_{\uparrow}  \tilde{c}_{\uparrow}(\vb{x}')      \tilde{c}^{\dagger}_{\uparrow}(\vb{x})   \ket{\psi_0}_{\uparrow} \bra{\psi_0}_{\downarrow}  \tilde{c}_{\downarrow}(\vb{x})      \tilde{c}^{\dagger}_{\downarrow}(\vb{x}')   \ket{\psi_0}_{\downarrow} \\
			&\sim \frac{1}{ \abs{ \vb{x}-\vb{x}' }^{2\alpha }   }.
		\end{split}
	\end{equation}

	Now we show how to convert this hidden non-local order to a critical order in the two-point function  $ S^{+}( \vb{x}  )  S^{-}( \vb{x}'  )$ using measurement and feedback. In particular, since the non-local order in the input state is a universal property of the Chern insulator phase, our protocol will always give rise to a state with critical correlations as long as the input state is in the same phase of matter. Starting with the input state $\ket{\psi_0}$,  we measure the fermion occupation number on every site and then provide unitary feedback. The resulting density matrix takes the form: 
	\begin{equation}\label{eq:chern_rho}
		\rho  = \sum_{n} U_n P_n \ket{\psi_0} \bra{\psi_0}  P_n U^{\dagger}_n, 
	\end{equation}
	where $P_n$ is the projector to a definite fermion number configuration, and $U_n$ is the unitary feedback 
	\begin{equation}
		U_n = \prod_{\vb{x}} e^{  - i \phi_n(\vb{x} ) S^{z}(\vb{x } )}
	\end{equation}
	with   $S^z( \vb{x} )   =  \frac{1}{2} (  \hat{n}_{\uparrow}(\vb{x})  -  \hat{n}_{\downarrow}(\vb{x})   )$ and  the phase $ \phi_n(\vb{x} )  =  \int d^2 x'  n(\vb{x})  \text{arg} (\vb{x}  -   \vb{x}')$ depending on the measurement outcome. The resulting density matrix $\rho$ exhibits a critical two-point correlation inherited from the non-local hidden order of the input $\ket{\psi_0}$: 
	\begin{equation}\label{eq:chern_main}
		\begin{split}
			&	\tr \left[ \rho    S^{+}( \vb{x}  )  S^{-}( \vb{x}'  )      \right]  \\
			&= \bra{\psi_0}  e^{i \hat{\phi}(\vb{x})}   S^{+}( \vb{x}  )  S^{-}( \vb{x}'  )  e^{-i \hat{\phi}(\vb{x}')}  \ket{\psi_0}  \sim    \frac{1}{ \abs{ \vb{x}-\vb{x}' }^{2\alpha }   }.
		\end{split}
	\end{equation} 
	
	To see this, one first note that the spin operators under the transformation by  $U_n$ will be dressed by an appropriate phase, namely, 
	
	\begin{equation}
		\begin{split}
			&U_n^{\dagger} S^+( \vb{x} )  U_n  =   e^{i \phi_n(\vb{x})} S^+( \vb{x} )\\
			&U_n^{\dagger} S^-( \vb{x} )  U_n  =   e^{-i \phi_n(\vb{x})} S^-( \vb{x} ).
		\end{split}
	\end{equation}

	This implies  $P_n U_n^{\dagger}   S^{+}( \vb{x}  )  S^{-}( \vb{x}'  ) U_n P_n = P_n    e^{i \hat{\phi}(\vb{x})}   S^{+}( \vb{x}  )  S^{-}( \vb{x}'  )  e^{-i \hat{\phi}(\vb{x}')}$, where the phase $\phi_n(\vb{x})$ has been promoted to an operator $\hat{\phi}(\vb{x})$ under the projector $P_n$.  Using this result, it is then straightforward to derive Eq.\ref{eq:chern_main} with the density matrix in Eq.\ref{eq:chern_rho}.

	Similarly,  one may compute  $S^{-}(\vb{x})  S^+(\vb{x}')$, which turns out to be equal to  $S^{+}(\vb{x})  S^-(\vb{x}') $ in the operator expectation value, hence decaying algebraically. These two results together imply $S_x(\vb{x}) S_x(\vb{x}' ) +S_y(\vb{x}) S_y(\vb{x}' )    =  \frac{1}{2} (  S^{+}(\vb{x})  S^-(\vb{x}')    +    S^{-}(\vb{x})  S^+(\vb{x}')    )  \sim \frac{1}{\abs{ \vb{x}  - \vb{x}'  }^{2\alpha }}$, where  $S_x(\vb{x})  = \frac{1}{2}  (  S^+(\vb{x})   +  S^-(\vb{x})  )$, $S_y(\vb{x})  = \frac{1}{2i}  (  S^+(\vb{x})   -   S^-(\vb{x})  )$, i.e. the spin operator in $x$ and $y$ component. In particular, since the initial input state $\ket{\psi_0}$ satisfies $\prod_{\vb{x}} \mathcal{S}^z(\vb{x})  \ket{\psi_0} = \ket{\psi_0}  $,  where  $\mathcal{S}^z(\vb{x})$ is the $\pi$ rotation about $z$ axis, the expectation value of $\mathcal{S}^z(\vb{x})$ is one in the resulting mixed state. With the proof technique  used in Sec.\ref{sec:general_structure},  $\mathcal{S}^z(\vb{x}) = 1$ together with algebraic decay of $S_x(\vb{x}) S_x(\vb{x}' ) +S_y(\vb{x}) S_y(\vb{x}' )   $ indicate that the critical mixed state $\rho$ cannot be a mixture of short-range entangled state.

	\section{Summary and discussion}
	In this work, we present a general framework for realizing mixed-state quantum order and quantum criticality using finite-depth quantum channels consisting of local measurement, local unitary feedback, and non-local classical communication. As an illustration, our protocol universally converts certain SPT phases to mixed states with long-range (topological) orders. In addition, when the  input SPT is tuned to a critical point, our protocol outputs a quantum critical mixed state diagnosed by volume-law classical entropy, algebraic decay of correlations, as well as logarithmically scaling bipartite mixed-state entanglement. Within the same theoretical framework, we also show how to transform the correlation exponent in constant depth by considering 1d spinful free fermions. More interestingly, our protocol can convert Chern insulators to a mixed state with an algebraically decaying correlation in the bulk. This furnishes a notable example where mixed-state quantum criticality can emerge from gapped states of matter in finite depth.


Our work motivates several questions that are worth further exploration. First, while we provide several examples of realizing non-trivial mixed states with critical correlations, a deeper understanding  of the entanglement structure of these novel mixed-state quantum criticality is lacking. Second, since this work only focuses on the utility of depth-2 protocols with onsite measurement and onsite unitary feedback, it would be interesting to generalize the current protocol, which may enable the realization of more types of exotic mixed-state orders and criticality. For instance, the onsite unitary feedback may be generalized to finite-depth local unitary circuits, and the current depth-2 protocols may be extended to multiple rounds of measurement and unitary layers.

Finally, it remains unclear what are the limitations of local measurement, local unitary, and non-local classical communication for realizing non-trivial quantum orders and criticality. Answering this question would provide sharp distinctions between the states that can and cannot be realized in finite depth. While entanglement provides one such constraint as discussed in Sec.\ref{sec:general_structure}, it would be desirable to explore the limitations from various aspects in the future (see Ref.\cite{adaptivebounds,ashvin_hierarchy_2022} for  progress along this direction).


	\acknowledgements{The authors thank the Kavli Institute for Theoretical Physics (K.I.T.P.), where this research was initiated and partly performed. The KITP is supported, in part, by the National Science Foundation under Grant No. NSF PHY-1748958.  S.V. and Z.Z. thank Matthew Fisher, Ali Lavasani, Andreas Ludwig, and Cenke Xu for helpful discussions.  T.-C.L. and T.H. thank Tarun Grover, Shengqi Sang, Liujun Zou, and Yijian Zou for helpful discussions. T.-C.L. and T.H.'s research at Perimeter Institute is supported in part by the Government of Canada through the Department of Innovation, Science and Economic Development Canada and by the Province of Ontario through the Ministry of Colleges and Universities.
	}
	
	\bibliography{v1bib}

\begin{thebibliography}{79}%
\makeatletter
\providecommand \@ifxundefined [1]{%
 \@ifx{#1\undefined}
}%
\providecommand \@ifnum [1]{%
 \ifnum #1\expandafter \@firstoftwo
 \else \expandafter \@secondoftwo
 \fi
}%
\providecommand \@ifx [1]{%
 \ifx #1\expandafter \@firstoftwo
 \else \expandafter \@secondoftwo
 \fi
}%
\providecommand \natexlab [1]{#1}%
\providecommand \enquote  [1]{``#1''}%
\providecommand \bibnamefont  [1]{#1}%
\providecommand \bibfnamefont [1]{#1}%
\providecommand \citenamefont [1]{#1}%
\providecommand \href@noop [0]{\@secondoftwo}%
\providecommand \href [0]{\begingroup \@sanitize@url \@href}%
\providecommand \@href[1]{\@@startlink{#1}\@@href}%
\providecommand \@@href[1]{\endgroup#1\@@endlink}%
\providecommand \@sanitize@url [0]{\catcode `\\12\catcode `\$12\catcode
  `\&12\catcode `\#12\catcode `\^12\catcode `\_12\catcode `\%12\relax}%
\providecommand \@@startlink[1]{}%
\providecommand \@@endlink[0]{}%
\providecommand \url  [0]{\begingroup\@sanitize@url \@url }%
\providecommand \@url [1]{\endgroup\@href {#1}{\urlprefix }}%
\providecommand \urlprefix  [0]{URL }%
\providecommand \Eprint [0]{\href }%
\providecommand \doibase [0]{https://doi.org/}%
\providecommand \selectlanguage [0]{\@gobble}%
\providecommand \bibinfo  [0]{\@secondoftwo}%
\providecommand \bibfield  [0]{\@secondoftwo}%
\providecommand \translation [1]{[#1]}%
\providecommand \BibitemOpen [0]{}%
\providecommand \bibitemStop [0]{}%
\providecommand \bibitemNoStop [0]{.\EOS\space}%
\providecommand \EOS [0]{\spacefactor3000\relax}%
\providecommand \BibitemShut  [1]{\csname bibitem#1\endcsname}%
\let\auto@bib@innerbib\@empty
\bibitem [{\citenamefont {Bravyi}\ and\ \citenamefont
  {Terhal}(2009)}]{bravyi2009no_go}%
  \BibitemOpen
  \bibfield  {author} {\bibinfo {author} {\bibfnamefont {S.}~\bibnamefont
  {Bravyi}}\ and\ \bibinfo {author} {\bibfnamefont {B.}~\bibnamefont
  {Terhal}},\ }\bibfield  {title} {\bibinfo {title} {A no-go theorem for a
  two-dimensional self-correcting quantum memory based on stabilizer codes},\
  }\href {https://doi.org/10.1088/1367-2630/11/4/043029} {\bibfield  {journal}
  {\bibinfo  {journal} {New Journal of Physics}\ }\textbf {\bibinfo {volume}
  {11}},\ \bibinfo {pages} {043029} (\bibinfo {year} {2009})}\BibitemShut
  {NoStop}%
\bibitem [{\citenamefont {Hastings}(2011)}]{hastings2011}%
  \BibitemOpen
  \bibfield  {author} {\bibinfo {author} {\bibfnamefont {M.~B.}\ \bibnamefont
  {Hastings}},\ }\bibfield  {title} {\bibinfo {title} {Topological order at
  nonzero temperature},\ }\href
  {https://doi.org/10.1103/PhysRevLett.107.210501} {\bibfield  {journal}
  {\bibinfo  {journal} {Phys. Rev. Lett.}\ }\textbf {\bibinfo {volume} {107}},\
  \bibinfo {pages} {210501} (\bibinfo {year} {2011})}\BibitemShut {NoStop}%
\bibitem [{\citenamefont {Yoshida}(2011)}]{yoshida2011}%
  \BibitemOpen
  \bibfield  {author} {\bibinfo {author} {\bibfnamefont {B.}~\bibnamefont
  {Yoshida}},\ }\bibfield  {title} {\bibinfo {title} {Feasibility of
  self-correcting quantum memory and thermal stability of topological order},\
  }\href {https://doi.org/https://doi.org/10.1016/j.aop.2011.06.001} {\bibfield
   {journal} {\bibinfo  {journal} {Annals of Physics}\ }\textbf {\bibinfo
  {volume} {326}},\ \bibinfo {pages} {2566 } (\bibinfo {year}
  {2011})}\BibitemShut {NoStop}%
\bibitem [{\citenamefont {Lu}\ \emph {et~al.}(2020)\citenamefont {Lu},
  \citenamefont {Hsieh},\ and\ \citenamefont {Grover}}]{Lu_topo_nega_2020}%
  \BibitemOpen
  \bibfield  {author} {\bibinfo {author} {\bibfnamefont {T.-C.}\ \bibnamefont
  {Lu}}, \bibinfo {author} {\bibfnamefont {T.~H.}\ \bibnamefont {Hsieh}},\ and\
  \bibinfo {author} {\bibfnamefont {T.}~\bibnamefont {Grover}},\ }\bibfield
  {title} {\bibinfo {title} {Detecting topological order at finite temperature
  using entanglement negativity},\ }\href
  {https://doi.org/10.1103/PhysRevLett.125.116801} {\bibfield  {journal}
  {\bibinfo  {journal} {Phys. Rev. Lett.}\ }\textbf {\bibinfo {volume} {125}},\
  \bibinfo {pages} {116801} (\bibinfo {year} {2020})}\BibitemShut {NoStop}%
\bibitem [{\citenamefont {Lu}\ and\ \citenamefont {Vijay}(2022)}]{lu2022_lre}%
  \BibitemOpen
  \bibfield  {author} {\bibinfo {author} {\bibfnamefont {T.-C.}\ \bibnamefont
  {Lu}}\ and\ \bibinfo {author} {\bibfnamefont {S.}~\bibnamefont {Vijay}},\
  }\bibfield  {title} {\bibinfo {title} {Characterizing long-range entanglement
  in a mixed state through an emergent order on the entangling surface},\
  }\href@noop {} {\bibfield  {journal} {\bibinfo  {journal} {arXiv preprint
  arXiv:2201.07792}\ } (\bibinfo {year} {2022})}\BibitemShut {NoStop}%
\bibitem [{\citenamefont {de~Groot}\ \emph {et~al.}(2022)\citenamefont
  {de~Groot}, \citenamefont {Turzillo},\ and\ \citenamefont
  {Schuch}}]{spt_Schuch_2022}%
  \BibitemOpen
  \bibfield  {author} {\bibinfo {author} {\bibfnamefont {C.}~\bibnamefont
  {de~Groot}}, \bibinfo {author} {\bibfnamefont {A.}~\bibnamefont {Turzillo}},\
  and\ \bibinfo {author} {\bibfnamefont {N.}~\bibnamefont {Schuch}},\
  }\bibfield  {title} {\bibinfo {title} {Symmetry {P}rotected {T}opological
  {O}rder in {O}pen {Q}uantum {S}ystems},\ }\href
  {https://doi.org/10.22331/q-2022-11-10-856} {\bibfield  {journal} {\bibinfo
  {journal} {{Quantum}}\ }\textbf {\bibinfo {volume} {6}},\ \bibinfo {pages}
  {856} (\bibinfo {year} {2022})}\BibitemShut {NoStop}%
\bibitem [{\citenamefont {Lee}\ \emph {et~al.}(2022{\natexlab{a}})\citenamefont
  {Lee}, \citenamefont {You},\ and\ \citenamefont {Xu}}]{lee2022_spt}%
  \BibitemOpen
  \bibfield  {author} {\bibinfo {author} {\bibfnamefont {J.~Y.}\ \bibnamefont
  {Lee}}, \bibinfo {author} {\bibfnamefont {Y.-Z.}\ \bibnamefont {You}},\ and\
  \bibinfo {author} {\bibfnamefont {C.}~\bibnamefont {Xu}},\ }\bibfield
  {title} {\bibinfo {title} {Symmetry protected topological phases under
  decoherence},\ }\href@noop {} {\bibfield  {journal} {\bibinfo  {journal}
  {arXiv preprint arXiv:2210.16323}\ } (\bibinfo {year}
  {2022}{\natexlab{a}})}\BibitemShut {NoStop}%
\bibitem [{\citenamefont {Zhang}\ \emph {et~al.}(2022)\citenamefont {Zhang},
  \citenamefont {Qi},\ and\ \citenamefont {Bi}}]{Bi_2022_spt}%
  \BibitemOpen
  \bibfield  {author} {\bibinfo {author} {\bibfnamefont {J.-H.}\ \bibnamefont
  {Zhang}}, \bibinfo {author} {\bibfnamefont {Y.}~\bibnamefont {Qi}},\ and\
  \bibinfo {author} {\bibfnamefont {Z.}~\bibnamefont {Bi}},\ }\bibfield
  {title} {\bibinfo {title} {Strange correlation function for average
  symmetry-protected topological phases},\ }\href@noop {} {\bibfield  {journal}
  {\bibinfo  {journal} {arXiv preprint arXiv:2210.17485}\ } (\bibinfo {year}
  {2022})}\BibitemShut {NoStop}%
\bibitem [{\citenamefont {Ma}\ and\ \citenamefont
  {Wang}(2022)}]{wang_2022_spt}%
  \BibitemOpen
  \bibfield  {author} {\bibinfo {author} {\bibfnamefont {R.}~\bibnamefont
  {Ma}}\ and\ \bibinfo {author} {\bibfnamefont {C.}~\bibnamefont {Wang}},\
  }\bibfield  {title} {\bibinfo {title} {Average symmetry-protected topological
  phases s},\ }\href@noop {} {\bibfield  {journal} {\bibinfo  {journal} {arXiv
  preprint arXiv:2209.02723}\ } (\bibinfo {year} {2022})}\BibitemShut {NoStop}%
\bibitem [{\citenamefont {Behrends}\ \emph {et~al.}(2022)\citenamefont
  {Behrends}, \citenamefont {Venn},\ and\ \citenamefont
  {B{\'e}ri}}]{behrends2022surface}%
  \BibitemOpen
  \bibfield  {author} {\bibinfo {author} {\bibfnamefont {J.}~\bibnamefont
  {Behrends}}, \bibinfo {author} {\bibfnamefont {F.}~\bibnamefont {Venn}},\
  and\ \bibinfo {author} {\bibfnamefont {B.}~\bibnamefont {B{\'e}ri}},\
  }\bibfield  {title} {\bibinfo {title} {Surface codes, quantum circuits, and
  entanglement phases},\ }\href@noop {} {\bibfield  {journal} {\bibinfo
  {journal} {arXiv preprint arXiv:2212.08084}\ } (\bibinfo {year}
  {2022})}\BibitemShut {NoStop}%
\bibitem [{\citenamefont {Fan}\ \emph {et~al.}(2023)\citenamefont {Fan},
  \citenamefont {Bao}, \citenamefont {Altman},\ and\ \citenamefont
  {Vishwanath}}]{fan2023mixed}%
  \BibitemOpen
  \bibfield  {author} {\bibinfo {author} {\bibfnamefont {R.}~\bibnamefont
  {Fan}}, \bibinfo {author} {\bibfnamefont {Y.}~\bibnamefont {Bao}}, \bibinfo
  {author} {\bibfnamefont {E.}~\bibnamefont {Altman}},\ and\ \bibinfo {author}
  {\bibfnamefont {A.}~\bibnamefont {Vishwanath}},\ }\bibfield  {title}
  {\bibinfo {title} {Diagnostics of mixed-state topological order and breakdown
  of quantum memory},\ }\href@noop {} {\bibfield  {journal} {\bibinfo
  {journal} {arXiv preprint arXiv:2301.05689}\ } (\bibinfo {year}
  {2023})}\BibitemShut {NoStop}%
\bibitem [{\citenamefont {Bao}\ \emph {et~al.}(2023)\citenamefont {Bao},
  \citenamefont {Fan}, \citenamefont {Vishwanath},\ and\ \citenamefont
  {Altman}}]{bao2023mixed}%
  \BibitemOpen
  \bibfield  {author} {\bibinfo {author} {\bibfnamefont {Y.}~\bibnamefont
  {Bao}}, \bibinfo {author} {\bibfnamefont {R.}~\bibnamefont {Fan}}, \bibinfo
  {author} {\bibfnamefont {A.}~\bibnamefont {Vishwanath}},\ and\ \bibinfo
  {author} {\bibfnamefont {E.}~\bibnamefont {Altman}},\ }\bibfield  {title}
  {\bibinfo {title} {Mixed-state topological order and the errorfield double
  formulation of decoherence-induced transitions},\ }\href@noop {} {\bibfield
  {journal} {\bibinfo  {journal} {arXiv preprint arXiv:2301.05687}\ } (\bibinfo
  {year} {2023})}\BibitemShut {NoStop}%
\bibitem [{\citenamefont {Lee}\ \emph {et~al.}(2023)\citenamefont {Lee},
  \citenamefont {Jian},\ and\ \citenamefont
  {Xu}}]{Lee_2023_criticality_decoherence}%
  \BibitemOpen
  \bibfield  {author} {\bibinfo {author} {\bibfnamefont {J.~Y.}\ \bibnamefont
  {Lee}}, \bibinfo {author} {\bibfnamefont {C.-M.}\ \bibnamefont {Jian}},\ and\
  \bibinfo {author} {\bibfnamefont {C.}~\bibnamefont {Xu}},\ }\bibfield
  {title} {\bibinfo {title} {Quantum criticality under decoherence or weak
  measurement},\ }\href@noop {} {\bibfield  {journal} {\bibinfo  {journal}
  {arXiv preprint arXiv:2301.05238}\ } (\bibinfo {year} {2023})}\BibitemShut
  {NoStop}%
\bibitem [{\citenamefont {Zou}\ \emph {et~al.}(2023)\citenamefont {Zou},
  \citenamefont {Sang},\ and\ \citenamefont
  {Hsieh}}]{zou_2023_channel_criticality}%
  \BibitemOpen
  \bibfield  {author} {\bibinfo {author} {\bibfnamefont {Y.}~\bibnamefont
  {Zou}}, \bibinfo {author} {\bibfnamefont {S.}~\bibnamefont {Sang}},\ and\
  \bibinfo {author} {\bibfnamefont {T.~H.}\ \bibnamefont {Hsieh}},\ }\bibfield
  {title} {\bibinfo {title} {Channeling quantum criticality},\ }\href@noop {}
  {\bibfield  {journal} {\bibinfo  {journal} {arXiv preprint arXiv:2301.07141}\
  } (\bibinfo {year} {2023})}\BibitemShut {NoStop}%
\bibitem [{\citenamefont {Briegel}\ and\ \citenamefont
  {Raussendorf}(2001)}]{Raussendorf_2001_ghz}%
  \BibitemOpen
  \bibfield  {author} {\bibinfo {author} {\bibfnamefont {H.~J.}\ \bibnamefont
  {Briegel}}\ and\ \bibinfo {author} {\bibfnamefont {R.}~\bibnamefont
  {Raussendorf}},\ }\bibfield  {title} {\bibinfo {title} {Persistent
  entanglement in arrays of interacting particles},\ }\href
  {https://doi.org/10.1103/PhysRevLett.86.910} {\bibfield  {journal} {\bibinfo
  {journal} {Phys. Rev. Lett.}\ }\textbf {\bibinfo {volume} {86}},\ \bibinfo
  {pages} {910} (\bibinfo {year} {2001})}\BibitemShut {NoStop}%
\bibitem [{\citenamefont {Raussendorf}\ \emph {et~al.}(2005)\citenamefont
  {Raussendorf}, \citenamefont {Bravyi},\ and\ \citenamefont
  {Harrington}}]{3d_cluster_state_2005}%
  \BibitemOpen
  \bibfield  {author} {\bibinfo {author} {\bibfnamefont {R.}~\bibnamefont
  {Raussendorf}}, \bibinfo {author} {\bibfnamefont {S.}~\bibnamefont
  {Bravyi}},\ and\ \bibinfo {author} {\bibfnamefont {J.}~\bibnamefont
  {Harrington}},\ }\bibfield  {title} {\bibinfo {title} {Long-range quantum
  entanglement in noisy cluster states},\ }\href
  {https://doi.org/10.1103/PhysRevA.71.062313} {\bibfield  {journal} {\bibinfo
  {journal} {Phys. Rev. A}\ }\textbf {\bibinfo {volume} {71}},\ \bibinfo
  {pages} {062313} (\bibinfo {year} {2005})}\BibitemShut {NoStop}%
\bibitem [{\citenamefont {Aguado}\ \emph {et~al.}(2008)\citenamefont {Aguado},
  \citenamefont {Brennen}, \citenamefont {Verstraete},\ and\ \citenamefont
  {Cirac}}]{cirac_2008_optical}%
  \BibitemOpen
  \bibfield  {author} {\bibinfo {author} {\bibfnamefont {M.}~\bibnamefont
  {Aguado}}, \bibinfo {author} {\bibfnamefont {G.~K.}\ \bibnamefont {Brennen}},
  \bibinfo {author} {\bibfnamefont {F.}~\bibnamefont {Verstraete}},\ and\
  \bibinfo {author} {\bibfnamefont {J.~I.}\ \bibnamefont {Cirac}},\ }\bibfield
  {title} {\bibinfo {title} {Creation, manipulation, and detection of abelian
  and non-abelian anyons in optical lattices},\ }\href
  {https://doi.org/10.1103/PhysRevLett.101.260501} {\bibfield  {journal}
  {\bibinfo  {journal} {Phys. Rev. Lett.}\ }\textbf {\bibinfo {volume} {101}},\
  \bibinfo {pages} {260501} (\bibinfo {year} {2008})}\BibitemShut {NoStop}%
\bibitem [{\citenamefont {Bolt}\ \emph {et~al.}(2016)\citenamefont {Bolt},
  \citenamefont {Duclos-Cianci}, \citenamefont {Poulin},\ and\ \citenamefont
  {Stace}}]{stace_2016_css}%
  \BibitemOpen
  \bibfield  {author} {\bibinfo {author} {\bibfnamefont {A.}~\bibnamefont
  {Bolt}}, \bibinfo {author} {\bibfnamefont {G.}~\bibnamefont {Duclos-Cianci}},
  \bibinfo {author} {\bibfnamefont {D.}~\bibnamefont {Poulin}},\ and\ \bibinfo
  {author} {\bibfnamefont {T.~M.}\ \bibnamefont {Stace}},\ }\bibfield  {title}
  {\bibinfo {title} {Foliated quantum error-correcting codes},\ }\href
  {https://doi.org/10.1103/PhysRevLett.117.070501} {\bibfield  {journal}
  {\bibinfo  {journal} {Phys. Rev. Lett.}\ }\textbf {\bibinfo {volume} {117}},\
  \bibinfo {pages} {070501} (\bibinfo {year} {2016})}\BibitemShut {NoStop}%
\bibitem [{\citenamefont {Piroli}\ \emph {et~al.}(2021)\citenamefont {Piroli},
  \citenamefont {Styliaris},\ and\ \citenamefont {Cirac}}]{cirac_2021_locc}%
  \BibitemOpen
  \bibfield  {author} {\bibinfo {author} {\bibfnamefont {L.}~\bibnamefont
  {Piroli}}, \bibinfo {author} {\bibfnamefont {G.}~\bibnamefont {Styliaris}},\
  and\ \bibinfo {author} {\bibfnamefont {J.~I.}\ \bibnamefont {Cirac}},\
  }\bibfield  {title} {\bibinfo {title} {Quantum circuits assisted by local
  operations and classical communication: Transformations and phases of
  matter},\ }\href {https://doi.org/10.1103/PhysRevLett.127.220503} {\bibfield
  {journal} {\bibinfo  {journal} {Phys. Rev. Lett.}\ }\textbf {\bibinfo
  {volume} {127}},\ \bibinfo {pages} {220503} (\bibinfo {year}
  {2021})}\BibitemShut {NoStop}%
\bibitem [{\citenamefont {Tantivasadakarn}\ \emph {et~al.}(2021)\citenamefont
  {Tantivasadakarn}, \citenamefont {Thorngren}, \citenamefont {Vishwanath},\
  and\ \citenamefont {Verresen}}]{ashvin_2021_measurement}%
  \BibitemOpen
  \bibfield  {author} {\bibinfo {author} {\bibfnamefont {N.}~\bibnamefont
  {Tantivasadakarn}}, \bibinfo {author} {\bibfnamefont {R.}~\bibnamefont
  {Thorngren}}, \bibinfo {author} {\bibfnamefont {A.}~\bibnamefont
  {Vishwanath}},\ and\ \bibinfo {author} {\bibfnamefont {R.}~\bibnamefont
  {Verresen}},\ }\bibfield  {title} {\bibinfo {title} {Long-range entanglement
  from measuring symmetry-protected topological phases},\ }\href@noop {}
  {\bibfield  {journal} {\bibinfo  {journal} {arXiv preprint arXiv:2112.01519}\
  } (\bibinfo {year} {2021})}\BibitemShut {NoStop}%
\bibitem [{\citenamefont {Verresen}\ \emph {et~al.}(2021)\citenamefont
  {Verresen}, \citenamefont {Tantivasadakarn},\ and\ \citenamefont
  {Vishwanath}}]{verresen2021_measurement_cold_atom}%
  \BibitemOpen
  \bibfield  {author} {\bibinfo {author} {\bibfnamefont {R.}~\bibnamefont
  {Verresen}}, \bibinfo {author} {\bibfnamefont {N.}~\bibnamefont
  {Tantivasadakarn}},\ and\ \bibinfo {author} {\bibfnamefont {A.}~\bibnamefont
  {Vishwanath}},\ }\bibfield  {title} {\bibinfo {title} {Efficiently preparing
  ghz, topological and fracton states by measuring cold atoms},\ }\href@noop {}
  {\bibfield  {journal} {\bibinfo  {journal} {arXiv preprint arXiv:2112.03061}\
  } (\bibinfo {year} {2021})}\BibitemShut {NoStop}%
\bibitem [{\citenamefont {Bravyi}\ \emph {et~al.}(2022)\citenamefont {Bravyi},
  \citenamefont {Kim}, \citenamefont {Kliesch},\ and\ \citenamefont
  {Koenig}}]{bravyi_2022_adaptive}%
  \BibitemOpen
  \bibfield  {author} {\bibinfo {author} {\bibfnamefont {S.}~\bibnamefont
  {Bravyi}}, \bibinfo {author} {\bibfnamefont {I.}~\bibnamefont {Kim}},
  \bibinfo {author} {\bibfnamefont {A.}~\bibnamefont {Kliesch}},\ and\ \bibinfo
  {author} {\bibfnamefont {R.}~\bibnamefont {Koenig}},\ }\bibfield  {title}
  {\bibinfo {title} {Adaptive constant-depth circuits for manipulating
  non-abelian anyons},\ }\href@noop {} {\bibfield  {journal} {\bibinfo
  {journal} {arXiv preprint arXiv:2205.01933}\ } (\bibinfo {year}
  {2022})}\BibitemShut {NoStop}%
\bibitem [{\citenamefont {Lu}\ \emph {et~al.}(2022)\citenamefont {Lu},
  \citenamefont {Lessa}, \citenamefont {Kim},\ and\ \citenamefont
  {Hsieh}}]{lu2022measurement}%
  \BibitemOpen
  \bibfield  {author} {\bibinfo {author} {\bibfnamefont {T.-C.}\ \bibnamefont
  {Lu}}, \bibinfo {author} {\bibfnamefont {L.~A.}\ \bibnamefont {Lessa}},
  \bibinfo {author} {\bibfnamefont {I.~H.}\ \bibnamefont {Kim}},\ and\ \bibinfo
  {author} {\bibfnamefont {T.~H.}\ \bibnamefont {Hsieh}},\ }\bibfield  {title}
  {\bibinfo {title} {Measurement as a shortcut to long-range entangled quantum
  matter},\ }\href {https://doi.org/10.1103/PRXQuantum.3.040337} {\bibfield
  {journal} {\bibinfo  {journal} {PRX Quantum}\ }\textbf {\bibinfo {volume}
  {3}},\ \bibinfo {pages} {040337} (\bibinfo {year} {2022})}\BibitemShut
  {NoStop}%
\bibitem [{\citenamefont {Tantivasadakarn}\ \emph
  {et~al.}(2022{\natexlab{a}})\citenamefont {Tantivasadakarn}, \citenamefont
  {Verresen},\ and\ \citenamefont {Vishwanath}}]{ashvin_single_shot_2022}%
  \BibitemOpen
  \bibfield  {author} {\bibinfo {author} {\bibfnamefont {N.}~\bibnamefont
  {Tantivasadakarn}}, \bibinfo {author} {\bibfnamefont {R.}~\bibnamefont
  {Verresen}},\ and\ \bibinfo {author} {\bibfnamefont {A.}~\bibnamefont
  {Vishwanath}},\ }\bibfield  {title} {\bibinfo {title} {The shortest route to
  non-abelian topological order on a quantum processor},\ }\href@noop {}
  {\bibfield  {journal} {\bibinfo  {journal} {arXiv preprint arXiv:2209.03964}\
  } (\bibinfo {year} {2022}{\natexlab{a}})}\BibitemShut {NoStop}%
\bibitem [{\citenamefont {Tantivasadakarn}\ \emph
  {et~al.}(2022{\natexlab{b}})\citenamefont {Tantivasadakarn}, \citenamefont
  {Vishwanath},\ and\ \citenamefont {Verresen}}]{ashvin_hierarchy_2022}%
  \BibitemOpen
  \bibfield  {author} {\bibinfo {author} {\bibfnamefont {N.}~\bibnamefont
  {Tantivasadakarn}}, \bibinfo {author} {\bibfnamefont {A.}~\bibnamefont
  {Vishwanath}},\ and\ \bibinfo {author} {\bibfnamefont {R.}~\bibnamefont
  {Verresen}},\ }\bibfield  {title} {\bibinfo {title} {A hierarchy of
  topological order from finite-depth unitaries, measurement and feedforward},\
  }\href@noop {} {\bibfield  {journal} {\bibinfo  {journal} {arXiv preprint
  arXiv:2209.06202}\ } (\bibinfo {year} {2022}{\natexlab{b}})}\BibitemShut
  {NoStop}%
\bibitem [{\citenamefont {Iqbal}\ \emph {et~al.}(2023)\citenamefont {Iqbal},
  \citenamefont {Tantivasadakarn}, \citenamefont {Gatterman}, \citenamefont
  {Gerber}, \citenamefont {Gilmore}, \citenamefont {Gresh}, \citenamefont
  {Hankin}, \citenamefont {Hewitt}, \citenamefont {Horst}, \citenamefont
  {Matheny} \emph {et~al.}}]{iqbal2023topological}%
  \BibitemOpen
  \bibfield  {author} {\bibinfo {author} {\bibfnamefont {M.}~\bibnamefont
  {Iqbal}}, \bibinfo {author} {\bibfnamefont {N.}~\bibnamefont
  {Tantivasadakarn}}, \bibinfo {author} {\bibfnamefont {T.~M.}\ \bibnamefont
  {Gatterman}}, \bibinfo {author} {\bibfnamefont {J.~A.}\ \bibnamefont
  {Gerber}}, \bibinfo {author} {\bibfnamefont {K.}~\bibnamefont {Gilmore}},
  \bibinfo {author} {\bibfnamefont {D.}~\bibnamefont {Gresh}}, \bibinfo
  {author} {\bibfnamefont {A.}~\bibnamefont {Hankin}}, \bibinfo {author}
  {\bibfnamefont {N.}~\bibnamefont {Hewitt}}, \bibinfo {author} {\bibfnamefont
  {C.~V.}\ \bibnamefont {Horst}}, \bibinfo {author} {\bibfnamefont
  {M.}~\bibnamefont {Matheny}}, \emph {et~al.},\ }\bibfield  {title} {\bibinfo
  {title} {Topological order from measurements and feed-forward on a trapped
  ion quantum computer},\ }\href@noop {} {\bibfield  {journal} {\bibinfo
  {journal} {arXiv preprint arXiv:2302.01917}\ } (\bibinfo {year}
  {2023})}\BibitemShut {NoStop}%
\bibitem [{\citenamefont {Foss-Feig}\ \emph {et~al.}(2023)\citenamefont
  {Foss-Feig}, \citenamefont {Tikku}, \citenamefont {Lu}, \citenamefont
  {Mayer}, \citenamefont {Iqbal}, \citenamefont {Gatterman}, \citenamefont
  {Gerber}, \citenamefont {Gilmore}, \citenamefont {Gresh}, \citenamefont
  {Hankin} \emph {et~al.}}]{foss2023experimental}%
  \BibitemOpen
  \bibfield  {author} {\bibinfo {author} {\bibfnamefont {M.}~\bibnamefont
  {Foss-Feig}}, \bibinfo {author} {\bibfnamefont {A.}~\bibnamefont {Tikku}},
  \bibinfo {author} {\bibfnamefont {T.-C.}\ \bibnamefont {Lu}}, \bibinfo
  {author} {\bibfnamefont {K.}~\bibnamefont {Mayer}}, \bibinfo {author}
  {\bibfnamefont {M.}~\bibnamefont {Iqbal}}, \bibinfo {author} {\bibfnamefont
  {T.~M.}\ \bibnamefont {Gatterman}}, \bibinfo {author} {\bibfnamefont {J.~A.}\
  \bibnamefont {Gerber}}, \bibinfo {author} {\bibfnamefont {K.}~\bibnamefont
  {Gilmore}}, \bibinfo {author} {\bibfnamefont {D.}~\bibnamefont {Gresh}},
  \bibinfo {author} {\bibfnamefont {A.}~\bibnamefont {Hankin}}, \emph
  {et~al.},\ }\bibfield  {title} {\bibinfo {title} {Experimental demonstration
  of the advantage of adaptive quantum circuits},\ }\href@noop {} {\bibfield
  {journal} {\bibinfo  {journal} {arXiv preprint arXiv:2302.03029}\ } (\bibinfo
  {year} {2023})}\BibitemShut {NoStop}%
\bibitem [{\citenamefont {Lee}\ \emph {et~al.}(2022{\natexlab{b}})\citenamefont
  {Lee}, \citenamefont {Ji}, \citenamefont {Bi},\ and\ \citenamefont
  {Fisher}}]{lee2022decoding}%
  \BibitemOpen
  \bibfield  {author} {\bibinfo {author} {\bibfnamefont {J.~Y.}\ \bibnamefont
  {Lee}}, \bibinfo {author} {\bibfnamefont {W.}~\bibnamefont {Ji}}, \bibinfo
  {author} {\bibfnamefont {Z.}~\bibnamefont {Bi}},\ and\ \bibinfo {author}
  {\bibfnamefont {M.}~\bibnamefont {Fisher}},\ }\bibfield  {title} {\bibinfo
  {title} {Decoding measurement-prepared quantum phases and transitions: from
  ising model to gauge theory, and beyond},\ }\href@noop {} {\bibfield
  {journal} {\bibinfo  {journal} {arXiv preprint arXiv:2208.11699}\ } (\bibinfo
  {year} {2022}{\natexlab{b}})}\BibitemShut {NoStop}%
\bibitem [{\citenamefont {Zhu}\ \emph {et~al.}(2022)\citenamefont {Zhu},
  \citenamefont {Tantivasadakarn}, \citenamefont {Vishwanath}, \citenamefont
  {Trebst},\ and\ \citenamefont {Verresen}}]{zhu2022nishimori}%
  \BibitemOpen
  \bibfield  {author} {\bibinfo {author} {\bibfnamefont {G.-Y.}\ \bibnamefont
  {Zhu}}, \bibinfo {author} {\bibfnamefont {N.}~\bibnamefont
  {Tantivasadakarn}}, \bibinfo {author} {\bibfnamefont {A.}~\bibnamefont
  {Vishwanath}}, \bibinfo {author} {\bibfnamefont {S.}~\bibnamefont {Trebst}},\
  and\ \bibinfo {author} {\bibfnamefont {R.}~\bibnamefont {Verresen}},\
  }\bibfield  {title} {\bibinfo {title} {Nishimori's cat: stable long-range
  entanglement from finite-depth unitaries and weak measurements},\ }\href@noop
  {} {\bibfield  {journal} {\bibinfo  {journal} {arXiv preprint
  arXiv:2208.11136}\ } (\bibinfo {year} {2022})}\BibitemShut {NoStop}%
\bibitem [{\citenamefont {Chen}\ \emph
  {et~al.}(2011{\natexlab{a}})\citenamefont {Chen}, \citenamefont {Gu},\ and\
  \citenamefont {Wen}}]{spt_1d_2011}%
  \BibitemOpen
  \bibfield  {author} {\bibinfo {author} {\bibfnamefont {X.}~\bibnamefont
  {Chen}}, \bibinfo {author} {\bibfnamefont {Z.-C.}\ \bibnamefont {Gu}},\ and\
  \bibinfo {author} {\bibfnamefont {X.-G.}\ \bibnamefont {Wen}},\ }\bibfield
  {title} {\bibinfo {title} {Classification of gapped symmetric phases in
  one-dimensional spin systems},\ }\href
  {https://doi.org/10.1103/PhysRevB.83.035107} {\bibfield  {journal} {\bibinfo
  {journal} {Phys. Rev. B}\ }\textbf {\bibinfo {volume} {83}},\ \bibinfo
  {pages} {035107} (\bibinfo {year} {2011}{\natexlab{a}})}\BibitemShut
  {NoStop}%
\bibitem [{\citenamefont {Chen}\ \emph
  {et~al.}(2011{\natexlab{b}})\citenamefont {Chen}, \citenamefont {Gu},\ and\
  \citenamefont {Wen}}]{spt_2011}%
  \BibitemOpen
  \bibfield  {author} {\bibinfo {author} {\bibfnamefont {X.}~\bibnamefont
  {Chen}}, \bibinfo {author} {\bibfnamefont {Z.-C.}\ \bibnamefont {Gu}},\ and\
  \bibinfo {author} {\bibfnamefont {X.-G.}\ \bibnamefont {Wen}},\ }\bibfield
  {title} {\bibinfo {title} {Complete classification of one-dimensional gapped
  quantum phases in interacting spin systems},\ }\href
  {https://doi.org/10.1103/PhysRevB.84.235128} {\bibfield  {journal} {\bibinfo
  {journal} {Phys. Rev. B}\ }\textbf {\bibinfo {volume} {84}},\ \bibinfo
  {pages} {235128} (\bibinfo {year} {2011}{\natexlab{b}})}\BibitemShut
  {NoStop}%
\bibitem [{\citenamefont {Chen}\ \emph {et~al.}(2014)\citenamefont {Chen},
  \citenamefont {Lu},\ and\ \citenamefont {Vishwanath}}]{dwSPT}%
  \BibitemOpen
  \bibfield  {author} {\bibinfo {author} {\bibfnamefont {X.}~\bibnamefont
  {Chen}}, \bibinfo {author} {\bibfnamefont {Y.-M.}\ \bibnamefont {Lu}},\ and\
  \bibinfo {author} {\bibfnamefont {A.}~\bibnamefont {Vishwanath}},\ }\bibfield
   {title} {\bibinfo {title} {Symmetry-protected topological phases from
  decorated domain walls},\ }\href {https://doi.org/10.1038/ncomms4507}
  {\bibfield  {journal} {\bibinfo  {journal} {Nature Communications}\ }\textbf
  {\bibinfo {volume} {5}},\ \bibinfo {pages} {3507} (\bibinfo {year}
  {2014})}\BibitemShut {NoStop}%
\bibitem [{\citenamefont {Peres}(1996)}]{peres1996}%
  \BibitemOpen
  \bibfield  {author} {\bibinfo {author} {\bibfnamefont {A.}~\bibnamefont
  {Peres}},\ }\bibfield  {title} {\bibinfo {title} {Separability criterion for
  density matrices},\ }\href {https://doi.org/10.1103/PhysRevLett.77.1413}
  {\bibfield  {journal} {\bibinfo  {journal} {Phys. Rev. Lett.}\ }\textbf
  {\bibinfo {volume} {77}},\ \bibinfo {pages} {1413} (\bibinfo {year}
  {1996})}\BibitemShut {NoStop}%
\bibitem [{\citenamefont {Horodecki}\ \emph {et~al.}(1996)\citenamefont
  {Horodecki}, \citenamefont {Horodecki},\ and\ \citenamefont
  {Horodecki}}]{horodecki1996}%
  \BibitemOpen
  \bibfield  {author} {\bibinfo {author} {\bibfnamefont {M.}~\bibnamefont
  {Horodecki}}, \bibinfo {author} {\bibfnamefont {P.}~\bibnamefont
  {Horodecki}},\ and\ \bibinfo {author} {\bibfnamefont {R.}~\bibnamefont
  {Horodecki}},\ }\bibfield  {title} {\bibinfo {title} {Separability of mixed
  states: necessary and sufficient conditions},\ }\href
  {https://doi.org/https://doi.org/10.1016/S0375-9601(96)00706-2} {\bibfield
  {journal} {\bibinfo  {journal} {Physics Letters A}\ }\textbf {\bibinfo
  {volume} {223}},\ \bibinfo {pages} {1 } (\bibinfo {year} {1996})}\BibitemShut
  {NoStop}%
\bibitem [{\citenamefont {Eisert}\ and\ \citenamefont
  {Plenio}(1999)}]{eisert99}%
  \BibitemOpen
  \bibfield  {author} {\bibinfo {author} {\bibfnamefont {J.}~\bibnamefont
  {Eisert}}\ and\ \bibinfo {author} {\bibfnamefont {M.~B.}\ \bibnamefont
  {Plenio}},\ }\bibfield  {title} {\bibinfo {title} {A comparison of
  entanglement measures},\ }\href {https://doi.org/10.1080/09500349908231260}
  {\bibfield  {journal} {\bibinfo  {journal} {Journal of Modern Optics}\
  }\textbf {\bibinfo {volume} {46}},\ \bibinfo {pages} {145} (\bibinfo {year}
  {1999})}\BibitemShut {NoStop}%
\bibitem [{\citenamefont {Vidal}\ and\ \citenamefont
  {Werner}(2002)}]{vidal2002}%
  \BibitemOpen
  \bibfield  {author} {\bibinfo {author} {\bibfnamefont {G.}~\bibnamefont
  {Vidal}}\ and\ \bibinfo {author} {\bibfnamefont {R.~F.}\ \bibnamefont
  {Werner}},\ }\bibfield  {title} {\bibinfo {title} {Computable measure of
  entanglement},\ }\href {https://doi.org/10.1103/PhysRevA.65.032314}
  {\bibfield  {journal} {\bibinfo  {journal} {Phys. Rev. A}\ }\textbf {\bibinfo
  {volume} {65}},\ \bibinfo {pages} {032314} (\bibinfo {year}
  {2002})}\BibitemShut {NoStop}%
\bibitem [{\citenamefont {Plenio}(2005)}]{plenio2005logarithmic}%
  \BibitemOpen
  \bibfield  {author} {\bibinfo {author} {\bibfnamefont {M.~B.}\ \bibnamefont
  {Plenio}},\ }\bibfield  {title} {\bibinfo {title} {Logarithmic negativity: a
  full entanglement monotone that is not convex},\ }\href@noop {} {\bibfield
  {journal} {\bibinfo  {journal} {Physical review letters}\ }\textbf {\bibinfo
  {volume} {95}},\ \bibinfo {pages} {090503} (\bibinfo {year}
  {2005})}\BibitemShut {NoStop}%
\bibitem [{\citenamefont {Savary}\ and\ \citenamefont
  {Balents}(2016)}]{Savary_2017_spin_liquid}%
  \BibitemOpen
  \bibfield  {author} {\bibinfo {author} {\bibfnamefont {L.}~\bibnamefont
  {Savary}}\ and\ \bibinfo {author} {\bibfnamefont {L.}~\bibnamefont
  {Balents}},\ }\bibfield  {title} {\bibinfo {title} {Quantum spin liquids: a
  review},\ }\href {https://doi.org/10.1088/0034-4885/80/1/016502} {\bibfield
  {journal} {\bibinfo  {journal} {Reports on Progress in Physics}\ }\textbf
  {\bibinfo {volume} {80}},\ \bibinfo {pages} {016502} (\bibinfo {year}
  {2016})}\BibitemShut {NoStop}%
\bibitem [{\citenamefont {Zhou}\ \emph {et~al.}(2017)\citenamefont {Zhou},
  \citenamefont {Kanoda},\ and\ \citenamefont {Ng}}]{Ng_2017_spin_liquid}%
  \BibitemOpen
  \bibfield  {author} {\bibinfo {author} {\bibfnamefont {Y.}~\bibnamefont
  {Zhou}}, \bibinfo {author} {\bibfnamefont {K.}~\bibnamefont {Kanoda}},\ and\
  \bibinfo {author} {\bibfnamefont {T.-K.}\ \bibnamefont {Ng}},\ }\bibfield
  {title} {\bibinfo {title} {Quantum spin liquid states},\ }\href
  {https://doi.org/10.1103/RevModPhys.89.025003} {\bibfield  {journal}
  {\bibinfo  {journal} {Rev. Mod. Phys.}\ }\textbf {\bibinfo {volume} {89}},\
  \bibinfo {pages} {025003} (\bibinfo {year} {2017})}\BibitemShut {NoStop}%
\bibitem [{\citenamefont {Gross}\ and\ \citenamefont
  {Bakr}(2021)}]{microscope_review_2021}%
  \BibitemOpen
  \bibfield  {author} {\bibinfo {author} {\bibfnamefont {C.}~\bibnamefont
  {Gross}}\ and\ \bibinfo {author} {\bibfnamefont {W.~S.}\ \bibnamefont
  {Bakr}},\ }\bibfield  {title} {\bibinfo {title} {Quantum gas microscopy for
  single atom and spin detection},\ }\href
  {https://doi.org/10.1038/s41567-021-01370-5} {\bibfield  {journal} {\bibinfo
  {journal} {Nature Physics}\ }\textbf {\bibinfo {volume} {17}},\ \bibinfo
  {pages} {1316} (\bibinfo {year} {2021})}\BibitemShut {NoStop}%
\bibitem [{\citenamefont {Kruis}\ \emph {et~al.}(2004)\citenamefont {Kruis},
  \citenamefont {McCulloch}, \citenamefont {Nussinov},\ and\ \citenamefont
  {Zaanen}}]{squeeze_fermion_2004}%
  \BibitemOpen
  \bibfield  {author} {\bibinfo {author} {\bibfnamefont {H.~V.}\ \bibnamefont
  {Kruis}}, \bibinfo {author} {\bibfnamefont {I.~P.}\ \bibnamefont
  {McCulloch}}, \bibinfo {author} {\bibfnamefont {Z.}~\bibnamefont
  {Nussinov}},\ and\ \bibinfo {author} {\bibfnamefont {J.}~\bibnamefont
  {Zaanen}},\ }\bibfield  {title} {\bibinfo {title} {Geometry and the hidden
  order of luttinger liquids: The universality of squeezed space},\ }\href
  {https://doi.org/10.1103/PhysRevB.70.075109} {\bibfield  {journal} {\bibinfo
  {journal} {Phys. Rev. B}\ }\textbf {\bibinfo {volume} {70}},\ \bibinfo
  {pages} {075109} (\bibinfo {year} {2004})}\BibitemShut {NoStop}%
\bibitem [{\citenamefont {Bernevig}(2013)}]{Bernevig_2013}%
  \BibitemOpen
  \bibfield  {author} {\bibinfo {author} {\bibfnamefont {B.~A.}\ \bibnamefont
  {Bernevig}},\ }\href {https://doi.org/doi:10.1515/9781400846733} {\emph
  {\bibinfo {title} {Topological Insulators and Topological Superconductors}}}\
  (\bibinfo  {publisher} {Princeton University Press},\ \bibinfo {address}
  {Princeton},\ \bibinfo {year} {2013})\BibitemShut {NoStop}%
\bibitem [{\citenamefont {Girvin}\ and\ \citenamefont
  {MacDonald}(1987)}]{Girvin_1987_fqht}%
  \BibitemOpen
  \bibfield  {author} {\bibinfo {author} {\bibfnamefont {S.~M.}\ \bibnamefont
  {Girvin}}\ and\ \bibinfo {author} {\bibfnamefont {A.~H.}\ \bibnamefont
  {MacDonald}},\ }\bibfield  {title} {\bibinfo {title} {Off-diagonal long-range
  order, oblique confinement, and the fractional quantum hall effect},\ }\href
  {https://doi.org/10.1103/PhysRevLett.58.1252} {\bibfield  {journal} {\bibinfo
   {journal} {Phys. Rev. Lett.}\ }\textbf {\bibinfo {volume} {58}},\ \bibinfo
  {pages} {1252} (\bibinfo {year} {1987})}\BibitemShut {NoStop}%
\bibitem [{\citenamefont {Klein~Kvorning}\ \emph {et~al.}(2020)\citenamefont
  {Klein~Kvorning}, \citenamefont {Sp\aa{}nsl\"att}, \citenamefont {Chan},\
  and\ \citenamefont {Ryu}}]{ryu_response_2020}%
  \BibitemOpen
  \bibfield  {author} {\bibinfo {author} {\bibfnamefont {T.}~\bibnamefont
  {Klein~Kvorning}}, \bibinfo {author} {\bibfnamefont {C.}~\bibnamefont
  {Sp\aa{}nsl\"att}}, \bibinfo {author} {\bibfnamefont {A.~P.~O.}\ \bibnamefont
  {Chan}},\ and\ \bibinfo {author} {\bibfnamefont {S.}~\bibnamefont {Ryu}},\
  }\bibfield  {title} {\bibinfo {title} {Nonlocal order parameters for states
  with topological electromagnetic response},\ }\href
  {https://doi.org/10.1103/PhysRevB.101.205101} {\bibfield  {journal} {\bibinfo
   {journal} {Phys. Rev. B}\ }\textbf {\bibinfo {volume} {101}},\ \bibinfo
  {pages} {205101} (\bibinfo {year} {2020})}\BibitemShut {NoStop}%
\bibitem [{\citenamefont {Bennett}\ \emph {et~al.}(1996)\citenamefont
  {Bennett}, \citenamefont {DiVincenzo}, \citenamefont {Smolin},\ and\
  \citenamefont {Wootters}}]{bennett1996}%
  \BibitemOpen
  \bibfield  {author} {\bibinfo {author} {\bibfnamefont {C.~H.}\ \bibnamefont
  {Bennett}}, \bibinfo {author} {\bibfnamefont {D.~P.}\ \bibnamefont
  {DiVincenzo}}, \bibinfo {author} {\bibfnamefont {J.~A.}\ \bibnamefont
  {Smolin}},\ and\ \bibinfo {author} {\bibfnamefont {W.~K.}\ \bibnamefont
  {Wootters}},\ }\bibfield  {title} {\bibinfo {title} {Mixed-state entanglement
  and quantum error correction},\ }\href
  {https://doi.org/10.1103/PhysRevA.54.3824} {\bibfield  {journal} {\bibinfo
  {journal} {Phys. Rev. A}\ }\textbf {\bibinfo {volume} {54}},\ \bibinfo
  {pages} {3824} (\bibinfo {year} {1996})}\BibitemShut {NoStop}%
\bibitem [{\citenamefont {Bravyi}\ \emph {et~al.}(2006)\citenamefont {Bravyi},
  \citenamefont {Hastings},\ and\ \citenamefont
  {Verstraete}}]{hastings_lrbound_2006}%
  \BibitemOpen
  \bibfield  {author} {\bibinfo {author} {\bibfnamefont {S.}~\bibnamefont
  {Bravyi}}, \bibinfo {author} {\bibfnamefont {M.~B.}\ \bibnamefont
  {Hastings}},\ and\ \bibinfo {author} {\bibfnamefont {F.}~\bibnamefont
  {Verstraete}},\ }\bibfield  {title} {\bibinfo {title} {Lieb-robinson bounds
  and the generation of correlations and topological quantum order},\ }\href
  {https://doi.org/10.1103/PhysRevLett.97.050401} {\bibfield  {journal}
  {\bibinfo  {journal} {Phys. Rev. Lett.}\ }\textbf {\bibinfo {volume} {97}},\
  \bibinfo {pages} {050401} (\bibinfo {year} {2006})}\BibitemShut {NoStop}%
\bibitem [{\citenamefont {Pollmann}\ and\ \citenamefont
  {Turner}(2012)}]{Pollmann_2012_string_order}%
  \BibitemOpen
  \bibfield  {author} {\bibinfo {author} {\bibfnamefont {F.}~\bibnamefont
  {Pollmann}}\ and\ \bibinfo {author} {\bibfnamefont {A.~M.}\ \bibnamefont
  {Turner}},\ }\bibfield  {title} {\bibinfo {title} {Detection of
  symmetry-protected topological phases in one dimension},\ }\href
  {https://doi.org/10.1103/PhysRevB.86.125441} {\bibfield  {journal} {\bibinfo
  {journal} {Phys. Rev. B}\ }\textbf {\bibinfo {volume} {86}},\ \bibinfo
  {pages} {125441} (\bibinfo {year} {2012})}\BibitemShut {NoStop}%
\bibitem [{\citenamefont {Li}\ \emph {et~al.}(2023)\citenamefont {Li},
  \citenamefont {Oshikawa},\ and\ \citenamefont
  {Zheng}}]{Oshikawa_2023_spt_ssb}%
  \BibitemOpen
  \bibfield  {author} {\bibinfo {author} {\bibfnamefont {L.}~\bibnamefont
  {Li}}, \bibinfo {author} {\bibfnamefont {M.}~\bibnamefont {Oshikawa}},\ and\
  \bibinfo {author} {\bibfnamefont {Y.}~\bibnamefont {Zheng}},\ }\bibfield
  {title} {\bibinfo {title} {Non-invertible duality transformation between spt
  and ssb phases},\ }\href@noop {} {\bibfield  {journal} {\bibinfo  {journal}
  {arXiv preprint arXiv:2301.07899}\ } (\bibinfo {year} {2023})}\BibitemShut
  {NoStop}%
\bibitem [{\citenamefont {Holzhey}\ \emph {et~al.}(1994)\citenamefont
  {Holzhey}, \citenamefont {Larsen},\ and\ \citenamefont
  {Wilczek}}]{wilczek_1994}%
  \BibitemOpen
  \bibfield  {author} {\bibinfo {author} {\bibfnamefont {C.}~\bibnamefont
  {Holzhey}}, \bibinfo {author} {\bibfnamefont {F.}~\bibnamefont {Larsen}},\
  and\ \bibinfo {author} {\bibfnamefont {F.}~\bibnamefont {Wilczek}},\
  }\bibfield  {title} {\bibinfo {title} {Geometric and renormalized entropy in
  conformal field theory},\ }\href
  {https://doi.org/https://doi.org/10.1016/0550-3213(94)90402-2} {\bibfield
  {journal} {\bibinfo  {journal} {Nuclear Physics B}\ }\textbf {\bibinfo
  {volume} {424}},\ \bibinfo {pages} {443} (\bibinfo {year}
  {1994})}\BibitemShut {NoStop}%
\bibitem [{\citenamefont {Calabrese}\ and\ \citenamefont
  {Cardy}(2004)}]{calabrese_2004}%
  \BibitemOpen
  \bibfield  {author} {\bibinfo {author} {\bibfnamefont {P.}~\bibnamefont
  {Calabrese}}\ and\ \bibinfo {author} {\bibfnamefont {J.}~\bibnamefont
  {Cardy}},\ }\bibfield  {title} {\bibinfo {title} {Entanglement entropy and
  quantum field theory},\ }\href
  {https://doi.org/10.1088/1742-5468/2004/06/p06002} {\bibfield  {journal}
  {\bibinfo  {journal} {Journal of Statistical Mechanics: Theory and
  Experiment}\ }\textbf {\bibinfo {volume} {2004}},\ \bibinfo {pages} {P06002}
  (\bibinfo {year} {2004})}\BibitemShut {NoStop}%
\bibitem [{\citenamefont {Calabrese}\ \emph {et~al.}(2013)\citenamefont
  {Calabrese}, \citenamefont {Cardy},\ and\ \citenamefont
  {Tonni}}]{Calabrese_2013_negativity}%
  \BibitemOpen
  \bibfield  {author} {\bibinfo {author} {\bibfnamefont {P.}~\bibnamefont
  {Calabrese}}, \bibinfo {author} {\bibfnamefont {J.}~\bibnamefont {Cardy}},\
  and\ \bibinfo {author} {\bibfnamefont {E.}~\bibnamefont {Tonni}},\ }\bibfield
   {title} {\bibinfo {title} {Entanglement negativity in extended systems: a
  field theoretical approach},\ }\href
  {https://doi.org/10.1088/1742-5468/2013/02/p02008} {\bibfield  {journal}
  {\bibinfo  {journal} {Journal of Statistical Mechanics: Theory and
  Experiment}\ }\textbf {\bibinfo {volume} {2013}},\ \bibinfo {pages} {P02008}
  (\bibinfo {year} {2013})}\BibitemShut {NoStop}%
\bibitem [{\citenamefont {Gu}\ and\ \citenamefont
  {Wen}(2014)}]{fermion_spt_2014_wen}%
  \BibitemOpen
  \bibfield  {author} {\bibinfo {author} {\bibfnamefont {Z.-C.}\ \bibnamefont
  {Gu}}\ and\ \bibinfo {author} {\bibfnamefont {X.-G.}\ \bibnamefont {Wen}},\
  }\bibfield  {title} {\bibinfo {title} {Symmetry-protected topological orders
  for interacting fermions: Fermionic topological nonlinear
  $\ensuremath{\sigma}$ models and a special group supercohomology theory},\
  }\href {https://doi.org/10.1103/PhysRevB.90.115141} {\bibfield  {journal}
  {\bibinfo  {journal} {Phys. Rev. B}\ }\textbf {\bibinfo {volume} {90}},\
  \bibinfo {pages} {115141} (\bibinfo {year} {2014})}\BibitemShut {NoStop}%
\bibitem [{\citenamefont {von Keyserlingk}\ and\ \citenamefont
  {Sondhi}(2016)}]{floquet_spt_2016_sondhi}%
  \BibitemOpen
  \bibfield  {author} {\bibinfo {author} {\bibfnamefont {C.~W.}\ \bibnamefont
  {von Keyserlingk}}\ and\ \bibinfo {author} {\bibfnamefont {S.~L.}\
  \bibnamefont {Sondhi}},\ }\bibfield  {title} {\bibinfo {title} {Phase
  structure of one-dimensional interacting floquet systems. i. abelian
  symmetry-protected topological phases},\ }\href
  {https://doi.org/10.1103/PhysRevB.93.245145} {\bibfield  {journal} {\bibinfo
  {journal} {Phys. Rev. B}\ }\textbf {\bibinfo {volume} {93}},\ \bibinfo
  {pages} {245145} (\bibinfo {year} {2016})}\BibitemShut {NoStop}%
\bibitem [{\citenamefont {Tantivasadakarn}\ and\ \citenamefont
  {Vishwanath}(2018)}]{Ashvin_2018_commuing}%
  \BibitemOpen
  \bibfield  {author} {\bibinfo {author} {\bibfnamefont {N.}~\bibnamefont
  {Tantivasadakarn}}\ and\ \bibinfo {author} {\bibfnamefont {A.}~\bibnamefont
  {Vishwanath}},\ }\bibfield  {title} {\bibinfo {title} {Full commuting
  projector hamiltonians of interacting symmetry-protected topological phases
  of fermions},\ }\href {https://doi.org/10.1103/PhysRevB.98.165104} {\bibfield
   {journal} {\bibinfo  {journal} {Phys. Rev. B}\ }\textbf {\bibinfo {volume}
  {98}},\ \bibinfo {pages} {165104} (\bibinfo {year} {2018})}\BibitemShut
  {NoStop}%
\bibitem [{\citenamefont {Borla}\ \emph {et~al.}(2021)\citenamefont {Borla},
  \citenamefont {Verresen}, \citenamefont {Shah},\ and\ \citenamefont
  {Moroz}}]{Verresen_kitaev_2021}%
  \BibitemOpen
  \bibfield  {author} {\bibinfo {author} {\bibfnamefont {U.}~\bibnamefont
  {Borla}}, \bibinfo {author} {\bibfnamefont {R.}~\bibnamefont {Verresen}},
  \bibinfo {author} {\bibfnamefont {J.}~\bibnamefont {Shah}},\ and\ \bibinfo
  {author} {\bibfnamefont {S.}~\bibnamefont {Moroz}},\ }\bibfield  {title}
  {\bibinfo {title} {{Gauging the Kitaev chain}},\ }\href
  {https://doi.org/10.21468/SciPostPhys.10.6.148} {\bibfield  {journal}
  {\bibinfo  {journal} {SciPost Phys.}\ }\textbf {\bibinfo {volume} {10}},\
  \bibinfo {pages} {148} (\bibinfo {year} {2021})}\BibitemShut {NoStop}%
\bibitem [{\citenamefont {Gutzwiller}(1963)}]{Gutzwiller_1963}%
  \BibitemOpen
  \bibfield  {author} {\bibinfo {author} {\bibfnamefont {M.~C.}\ \bibnamefont
  {Gutzwiller}},\ }\bibfield  {title} {\bibinfo {title} {Effect of correlation
  on the ferromagnetism of transition metals},\ }\href
  {https://doi.org/10.1103/PhysRevLett.10.159} {\bibfield  {journal} {\bibinfo
  {journal} {Phys. Rev. Lett.}\ }\textbf {\bibinfo {volume} {10}},\ \bibinfo
  {pages} {159} (\bibinfo {year} {1963})}\BibitemShut {NoStop}%
\bibitem [{\citenamefont {Haldane}(1988)}]{Haldane_1988}%
  \BibitemOpen
  \bibfield  {author} {\bibinfo {author} {\bibfnamefont {F.~D.~M.}\
  \bibnamefont {Haldane}},\ }\bibfield  {title} {\bibinfo {title} {Exact
  jastrow-gutzwiller resonating-valence-bond ground state of the
  spin-$\frac{1}{2}$ antiferromagnetic heisenberg chain with
  1/${\mathrm{r}}^{2}$ exchange},\ }\href
  {https://doi.org/10.1103/PhysRevLett.60.635} {\bibfield  {journal} {\bibinfo
  {journal} {Phys. Rev. Lett.}\ }\textbf {\bibinfo {volume} {60}},\ \bibinfo
  {pages} {635} (\bibinfo {year} {1988})}\BibitemShut {NoStop}%
\bibitem [{\citenamefont {Shastry}(1988)}]{Shastry_1988}%
  \BibitemOpen
  \bibfield  {author} {\bibinfo {author} {\bibfnamefont {B.~S.}\ \bibnamefont
  {Shastry}},\ }\bibfield  {title} {\bibinfo {title} {Exact solution of an
  s=1/2 heisenberg antiferromagnetic chain with long-ranged interactions},\
  }\href {https://doi.org/10.1103/PhysRevLett.60.639} {\bibfield  {journal}
  {\bibinfo  {journal} {Phys. Rev. Lett.}\ }\textbf {\bibinfo {volume} {60}},\
  \bibinfo {pages} {639} (\bibinfo {year} {1988})}\BibitemShut {NoStop}%
\bibitem [{\citenamefont {Wen}(2004)}]{wen2004quantum}%
  \BibitemOpen
  \bibfield  {author} {\bibinfo {author} {\bibfnamefont {X.-G.}\ \bibnamefont
  {Wen}},\ }\bibfield  {title} {\bibinfo {title} {Quantum field theory of
  many-body systems: from the origin of sound to an origin of light and
  electrons},\ }\href
  {https://oxford.universitypressscholarship.com/view/10.1093/acprof:oso/9780199227259.001.0001/acprof-9780199227259}
  {\bibfield  {journal} {\bibinfo  {journal} {OUP Oxford}\ } (\bibinfo {year}
  {2004})}\BibitemShut {NoStop}%
\bibitem [{\citenamefont {Kitaev}(2006)}]{kitaev_2006}%
  \BibitemOpen
  \bibfield  {author} {\bibinfo {author} {\bibfnamefont {A.}~\bibnamefont
  {Kitaev}},\ }\bibfield  {title} {\bibinfo {title} {Anyons in an exactly
  solved model and beyond},\ }\href
  {https://doi.org/https://doi.org/10.1016/j.aop.2005.10.005} {\bibfield
  {journal} {\bibinfo  {journal} {Annals of Physics}\ }\textbf {\bibinfo
  {volume} {321}},\ \bibinfo {pages} {2} (\bibinfo {year} {2006})},\ \bibinfo
  {note} {january Special Issue}\BibitemShut {NoStop}%
\bibitem [{\citenamefont {Zhang}\ and\ \citenamefont
  {Vijay}()}]{Vijay_Forthcoming}%
  \BibitemOpen
  \bibfield  {author} {\bibinfo {author} {\bibfnamefont {Z.}~\bibnamefont
  {Zhang}}\ and\ \bibinfo {author} {\bibfnamefont {S.}~\bibnamefont {Vijay}},\
  }\bibfield  {title} {\bibinfo {title} {Forthcoming},\ }\href@noop {} {\
  }\BibitemShut {NoStop}%
\bibitem [{\citenamefont {Affleck}\ \emph {et~al.}(1987)\citenamefont
  {Affleck}, \citenamefont {Kennedy}, \citenamefont {Lieb},\ and\ \citenamefont
  {Tasaki}}]{aklt_1987}%
  \BibitemOpen
  \bibfield  {author} {\bibinfo {author} {\bibfnamefont {I.}~\bibnamefont
  {Affleck}}, \bibinfo {author} {\bibfnamefont {T.}~\bibnamefont {Kennedy}},
  \bibinfo {author} {\bibfnamefont {E.~H.}\ \bibnamefont {Lieb}},\ and\
  \bibinfo {author} {\bibfnamefont {H.}~\bibnamefont {Tasaki}},\ }\bibfield
  {title} {\bibinfo {title} {Rigorous results on valence-bond ground states in
  antiferromagnets},\ }\href {https://doi.org/10.1103/PhysRevLett.59.799}
  {\bibfield  {journal} {\bibinfo  {journal} {Phys. Rev. Lett.}\ }\textbf
  {\bibinfo {volume} {59}},\ \bibinfo {pages} {799} (\bibinfo {year}
  {1987})}\BibitemShut {NoStop}%
\bibitem [{\citenamefont {Haldane}(1983)}]{haldane_spin_chain_1983}%
  \BibitemOpen
  \bibfield  {author} {\bibinfo {author} {\bibfnamefont {F.~D.~M.}\
  \bibnamefont {Haldane}},\ }\bibfield  {title} {\bibinfo {title} {Nonlinear
  field theory of large-spin heisenberg antiferromagnets: Semiclassically
  quantized solitons of the one-dimensional easy-axis n\'eel state},\ }\href
  {https://doi.org/10.1103/PhysRevLett.50.1153} {\bibfield  {journal} {\bibinfo
   {journal} {Phys. Rev. Lett.}\ }\textbf {\bibinfo {volume} {50}},\ \bibinfo
  {pages} {1153} (\bibinfo {year} {1983})}\BibitemShut {NoStop}%
\bibitem [{\citenamefont {den Nijs}\ and\ \citenamefont
  {Rommelse}(1989)}]{hidden_order_Rommelse_1989}%
  \BibitemOpen
  \bibfield  {author} {\bibinfo {author} {\bibfnamefont {M.}~\bibnamefont {den
  Nijs}}\ and\ \bibinfo {author} {\bibfnamefont {K.}~\bibnamefont {Rommelse}},\
  }\bibfield  {title} {\bibinfo {title} {Preroughening transitions in crystal
  surfaces and valence-bond phases in quantum spin chains},\ }\href
  {https://doi.org/10.1103/PhysRevB.40.4709} {\bibfield  {journal} {\bibinfo
  {journal} {Phys. Rev. B}\ }\textbf {\bibinfo {volume} {40}},\ \bibinfo
  {pages} {4709} (\bibinfo {year} {1989})}\BibitemShut {NoStop}%
\bibitem [{\citenamefont {McCoy}(1968)}]{XY_McCoy_1968}%
  \BibitemOpen
  \bibfield  {author} {\bibinfo {author} {\bibfnamefont {B.~M.}\ \bibnamefont
  {McCoy}},\ }\bibfield  {title} {\bibinfo {title} {Spin correlation functions
  of the $x\ensuremath{-}y$ model},\ }\href
  {https://doi.org/10.1103/PhysRev.173.531} {\bibfield  {journal} {\bibinfo
  {journal} {Phys. Rev.}\ }\textbf {\bibinfo {volume} {173}},\ \bibinfo {pages}
  {531} (\bibinfo {year} {1968})}\BibitemShut {NoStop}%
\bibitem [{\citenamefont {Friedman}\ \emph {et~al.}(2022)\citenamefont
  {Friedman}, \citenamefont {Yin}, \citenamefont {Hong},\ and\ \citenamefont
  {Lucas}}]{adaptivebounds}%
  \BibitemOpen
  \bibfield  {author} {\bibinfo {author} {\bibfnamefont {A.~J.}\ \bibnamefont
  {Friedman}}, \bibinfo {author} {\bibfnamefont {C.}~\bibnamefont {Yin}},
  \bibinfo {author} {\bibfnamefont {Y.}~\bibnamefont {Hong}},\ and\ \bibinfo
  {author} {\bibfnamefont {A.}~\bibnamefont {Lucas}},\ }\bibfield  {title}
  {\bibinfo {title} {Locality and error correction in quantum dynamics with
  measurement},\ }\bibfield  {journal} {\bibinfo  {journal} {arXiv preprint
  arXiv:2206.09929}\ }\href {https://doi.org/10.48550/ARXIV.2206.09929}
  {10.48550/ARXIV.2206.09929} (\bibinfo {year} {2022})\BibitemShut {NoStop}%
\bibitem [{\citenamefont {Vidal}(2000)}]{vidal_monotone_2000}%
  \BibitemOpen
  \bibfield  {author} {\bibinfo {author} {\bibfnamefont {G.}~\bibnamefont
  {Vidal}},\ }\bibfield  {title} {\bibinfo {title} {Entanglement monotones},\
  }\href {https://doi.org/10.1080/09500340008244048} {\bibfield  {journal}
  {\bibinfo  {journal} {Journal of Modern Optics}\ }\textbf {\bibinfo {volume}
  {47}},\ \bibinfo {pages} {355} (\bibinfo {year} {2000})}\BibitemShut
  {NoStop}%
\bibitem [{\citenamefont {Wang}\ \emph {et~al.}(2016)\citenamefont {Wang},
  \citenamefont {Mu}, \citenamefont {Vedral},\ and\ \citenamefont
  {Fan}}]{renyi_eof_2016}%
  \BibitemOpen
  \bibfield  {author} {\bibinfo {author} {\bibfnamefont {Y.-X.}\ \bibnamefont
  {Wang}}, \bibinfo {author} {\bibfnamefont {L.-Z.}\ \bibnamefont {Mu}},
  \bibinfo {author} {\bibfnamefont {V.}~\bibnamefont {Vedral}},\ and\ \bibinfo
  {author} {\bibfnamefont {H.}~\bibnamefont {Fan}},\ }\bibfield  {title}
  {\bibinfo {title} {Entanglement r\'enyi $\ensuremath{\alpha}$ entropy},\
  }\href {https://doi.org/10.1103/PhysRevA.93.022324} {\bibfield  {journal}
  {\bibinfo  {journal} {Phys. Rev. A}\ }\textbf {\bibinfo {volume} {93}},\
  \bibinfo {pages} {022324} (\bibinfo {year} {2016})}\BibitemShut {NoStop}%
\bibitem [{\citenamefont {Kennedy}\ and\ \citenamefont
  {Tasaki}(1992{\natexlab{a}})}]{Kennedy_Tasaki_1992}%
  \BibitemOpen
  \bibfield  {author} {\bibinfo {author} {\bibfnamefont {T.}~\bibnamefont
  {Kennedy}}\ and\ \bibinfo {author} {\bibfnamefont {H.}~\bibnamefont
  {Tasaki}},\ }\bibfield  {title} {\bibinfo {title} {Hidden symmetry breaking
  and the haldane phase ins=1 quantum spin chains},\ }\href
  {https://doi.org/10.1007/BF02097239} {\bibfield  {journal} {\bibinfo
  {journal} {Communications in Mathematical Physics}\ }\textbf {\bibinfo
  {volume} {147}},\ \bibinfo {pages} {431} (\bibinfo {year}
  {1992}{\natexlab{a}})}\BibitemShut {NoStop}%
\bibitem [{\citenamefont {Kennedy}\ and\ \citenamefont
  {Tasaki}(1992{\natexlab{b}})}]{Kennedy_Tasaki_1992_prb}%
  \BibitemOpen
  \bibfield  {author} {\bibinfo {author} {\bibfnamefont {T.}~\bibnamefont
  {Kennedy}}\ and\ \bibinfo {author} {\bibfnamefont {H.}~\bibnamefont
  {Tasaki}},\ }\bibfield  {title} {\bibinfo {title} {Hidden
  ${\mathrm{z}}_{2}$\ifmmode\times\else\texttimes\fi{}${\mathrm{z}}_{2}$
  symmetry breaking in haldane-gap antiferromagnets},\ }\href
  {https://doi.org/10.1103/PhysRevB.45.304} {\bibfield  {journal} {\bibinfo
  {journal} {Phys. Rev. B}\ }\textbf {\bibinfo {volume} {45}},\ \bibinfo
  {pages} {304} (\bibinfo {year} {1992}{\natexlab{b}})}\BibitemShut {NoStop}%
\bibitem [{\citenamefont {Oshikawa}(1992)}]{Oshikawa_1992_KT}%
  \BibitemOpen
  \bibfield  {author} {\bibinfo {author} {\bibfnamefont {M.}~\bibnamefont
  {Oshikawa}},\ }\bibfield  {title} {\bibinfo {title} {Hidden z2*z2 symmetry in
  quantum spin chains with arbitrary integer spin},\ }\bibfield  {booktitle}
  {\emph {\bibinfo {booktitle} {Journal of Physics: Condensed Matter}},\ }\href
  {https://doi.org/10.1088/0953-8984/4/36/019} {\ \textbf {\bibinfo {volume}
  {4}},\ \bibinfo {pages} {7469} (\bibinfo {year} {1992})}\BibitemShut
  {NoStop}%
\bibitem [{\citenamefont {Ringel}\ and\ \citenamefont
  {Simon}(2015)}]{ringel_2015_hidden}%
  \BibitemOpen
  \bibfield  {author} {\bibinfo {author} {\bibfnamefont {Z.}~\bibnamefont
  {Ringel}}\ and\ \bibinfo {author} {\bibfnamefont {S.~H.}\ \bibnamefont
  {Simon}},\ }\bibfield  {title} {\bibinfo {title} {Hidden order and flux
  attachment in symmetry-protected topological phases: A laughlin-like
  approach},\ }\href {https://doi.org/10.1103/PhysRevB.91.195117} {\bibfield
  {journal} {\bibinfo  {journal} {Phys. Rev. B}\ }\textbf {\bibinfo {volume}
  {91}},\ \bibinfo {pages} {195117} (\bibinfo {year} {2015})}\BibitemShut
  {NoStop}%
\bibitem [{\citenamefont {Pachos}\ and\ \citenamefont
  {Plenio}(2004)}]{Plenio_2004_spt}%
  \BibitemOpen
  \bibfield  {author} {\bibinfo {author} {\bibfnamefont {J.~K.}\ \bibnamefont
  {Pachos}}\ and\ \bibinfo {author} {\bibfnamefont {M.~B.}\ \bibnamefont
  {Plenio}},\ }\bibfield  {title} {\bibinfo {title} {Three-spin interactions in
  optical lattices and criticality in cluster hamiltonians},\ }\href
  {https://doi.org/10.1103/PhysRevLett.93.056402} {\bibfield  {journal}
  {\bibinfo  {journal} {Phys. Rev. Lett.}\ }\textbf {\bibinfo {volume} {93}},\
  \bibinfo {pages} {056402} (\bibinfo {year} {2004})}\BibitemShut {NoStop}%
\bibitem [{\citenamefont {Doherty}\ and\ \citenamefont
  {Bartlett}(2009)}]{Bartlett_2009_spt}%
  \BibitemOpen
  \bibfield  {author} {\bibinfo {author} {\bibfnamefont {A.~C.}\ \bibnamefont
  {Doherty}}\ and\ \bibinfo {author} {\bibfnamefont {S.~D.}\ \bibnamefont
  {Bartlett}},\ }\bibfield  {title} {\bibinfo {title} {Identifying phases of
  quantum many-body systems that are universal for quantum computation},\
  }\href {https://doi.org/10.1103/PhysRevLett.103.020506} {\bibfield  {journal}
  {\bibinfo  {journal} {Phys. Rev. Lett.}\ }\textbf {\bibinfo {volume} {103}},\
  \bibinfo {pages} {020506} (\bibinfo {year} {2009})}\BibitemShut {NoStop}%
\bibitem [{\citenamefont {Scaffidi}\ \emph {et~al.}(2017)\citenamefont
  {Scaffidi}, \citenamefont {Parker},\ and\ \citenamefont
  {Vasseur}}]{gapless_spt_2017_vasseur}%
  \BibitemOpen
  \bibfield  {author} {\bibinfo {author} {\bibfnamefont {T.}~\bibnamefont
  {Scaffidi}}, \bibinfo {author} {\bibfnamefont {D.~E.}\ \bibnamefont
  {Parker}},\ and\ \bibinfo {author} {\bibfnamefont {R.}~\bibnamefont
  {Vasseur}},\ }\bibfield  {title} {\bibinfo {title} {Gapless
  symmetry-protected topological order},\ }\href
  {https://doi.org/10.1103/PhysRevX.7.041048} {\bibfield  {journal} {\bibinfo
  {journal} {Phys. Rev. X}\ }\textbf {\bibinfo {volume} {7}},\ \bibinfo {pages}
  {041048} (\bibinfo {year} {2017})}\BibitemShut {NoStop}%
\bibitem [{\citenamefont {Bl\"ote}\ and\ \citenamefont
  {Deng}(2002)}]{mc_ising_2002}%
  \BibitemOpen
  \bibfield  {author} {\bibinfo {author} {\bibfnamefont {H.~W.~J.}\
  \bibnamefont {Bl\"ote}}\ and\ \bibinfo {author} {\bibfnamefont
  {Y.}~\bibnamefont {Deng}},\ }\bibfield  {title} {\bibinfo {title} {Cluster
  monte carlo simulation of the transverse ising model},\ }\href
  {https://doi.org/10.1103/PhysRevE.66.066110} {\bibfield  {journal} {\bibinfo
  {journal} {Phys. Rev. E}\ }\textbf {\bibinfo {volume} {66}},\ \bibinfo
  {pages} {066110} (\bibinfo {year} {2002})}\BibitemShut {NoStop}%
\bibitem [{\citenamefont {Kitaev}(2001)}]{kitaev_chain_2001}%
  \BibitemOpen
  \bibfield  {author} {\bibinfo {author} {\bibfnamefont {A.~Y.}\ \bibnamefont
  {Kitaev}},\ }\bibfield  {title} {\bibinfo {title} {Unpaired majorana fermions
  in quantum wires},\ }\href {https://doi.org/10.1070/1063-7869/44/10S/S29}
  {\bibfield  {journal} {\bibinfo  {journal} {Physics-Uspekhi}\ }\textbf
  {\bibinfo {volume} {44}},\ \bibinfo {pages} {131} (\bibinfo {year}
  {2001})}\BibitemShut {NoStop}%
\bibitem [{\citenamefont {Alicea}(2010)}]{Alicea_2010_Majorana}%
  \BibitemOpen
  \bibfield  {author} {\bibinfo {author} {\bibfnamefont {J.}~\bibnamefont
  {Alicea}},\ }\bibfield  {title} {\bibinfo {title} {Majorana fermions in a
  tunable semiconductor device},\ }\href
  {https://doi.org/10.1103/PhysRevB.81.125318} {\bibfield  {journal} {\bibinfo
  {journal} {Phys. Rev. B}\ }\textbf {\bibinfo {volume} {81}},\ \bibinfo
  {pages} {125318} (\bibinfo {year} {2010})}\BibitemShut {NoStop}%
\bibitem [{\citenamefont {Vijay}\ and\ \citenamefont
  {Fu}(2016)}]{vijay_2016_surface_code}%
  \BibitemOpen
  \bibfield  {author} {\bibinfo {author} {\bibfnamefont {S.}~\bibnamefont
  {Vijay}}\ and\ \bibinfo {author} {\bibfnamefont {L.}~\bibnamefont {Fu}},\
  }\bibfield  {title} {\bibinfo {title} {Physical implementation of a majorana
  fermion surface code for fault-tolerant quantum computation},\ }\href
  {https://doi.org/10.1088/0031-8949/T168/1/014002} {\bibfield  {journal}
  {\bibinfo  {journal} {Physica Scripta}\ }\textbf {\bibinfo {volume} {2016}},\
  \bibinfo {pages} {014002} (\bibinfo {year} {2016})}\BibitemShut {NoStop}%
\end{thebibliography}%
	
	\newpage 
	\appendix


	\section{Constraints on mixed-state entanglement}\label{sec:constraint}

	Here we first review the basic notion of entanglement of formation, which we will use  to constrain the entanglement structure of output states from finite-depth quantum channels.\\ 
	
	\noindent \underline{\textit{Entanglement of formation}}: given a density matrix $\rho_{C\overline{C}}$ acting on the bipartite Hilbert space $\mathcal{H}_C \otimes \mathcal{H}_{\overline{C}}$, the entanglement between $C$ and $\overline{C}$ can be quantified via	entanglement of formation $E_f$ \cite{bennett1996}, which is defined as follows: one may  decompose a density matrix into  a convex sum of pure states, i.e.  $\rho_{C\overline{C}} =  \sum_i p_i  \ket{\psi_i} \bra{\psi_i}$ with $p_i \geq  0$ and $\sum_i p_i =1$. Then  $		E_f \equiv  \text{Min}   \left\{  \sum_i p_i S_C ( \tr_{\overline{C}} \ket{\psi_i }  \bra{\psi_i}    ) \right\}$,  where $S_C(\rho_C)$ is the von-Neumann entropy of $\rho_C$. Namely, $E_f$ is the average of bipartite entanglement entropy minimized over all possible ways of decomposing $\rho$ as a mixture of pure states. When $\rho_{C\overline{C}}$ is pure, $E_f $ reduces to the von-Neumann entanglement entropy. To demonstrate the idea of entanglement of formation, one may consider two qubits in a maximally mixed state, i.e. $\rho = \mathbb{I}/4$ with  $\mathbb{I}$ being the identity operator.  Certainly, the mixed state can be written as a uniform mixture of four Bell states $\frac{1}{\sqrt{2}}  (\ket{00}\pm \ket{11}  ),    \frac{1}{\sqrt{2}}  (\ket{01}\pm \ket{10}  )$, so the average entanglement is $\log 2$. However, the mixed state can also be written as a uniform mixture of four product states,  $\ket{00}, \ket{01}, \ket{10}, \ket{11}$, in which case one achieves the minimal average entanglement entropy, i.e. zero. Therefore $E_f=0$ for this maximally mixed state.\\

	\noindent \underline{\textit{Bounds on entanglement}}: in the main text, we construct quantum channels from LOCC, i.e. single-site measurement, single-site unitaries, and classical communication. As such, the mixed-state entanglement quantified by $E_f$ must be non-increasing under such quantum channels. To see this, starting with a state $\rho_0 = \ket{\psi_0}\bra{ \psi_0 }$,  performing single-site measurement followed by single-site unitaries leads to a mixed state $\rho$, a mixture of pure states  $ \ket{\psi_\alpha} =  \frac{U_{\alpha} P_\alpha  \ket{\psi_0} }{  \sqrt{     \bra{\psi_0 }P_{\alpha}   \ket{\psi_0}  }    }   $  with corresponding probability $p_\alpha= \bra{\psi_0} P_{\alpha }  \ket{\psi} $. Being an entanglement measure, $E_f$ cannot increase on average under LOCC \cite{bennett1996,vidal_monotone_2000}, indicating $   E_f(\rho_0)\geq  \sum_\alpha p_\alpha  E_f ( \ket{ \psi_\alpha }   ) =  \sum_\alpha p_\alpha S_C(   \tr_{\overline{C}} \ket{ \psi_\alpha } \bra{\psi_\alpha}   )$. Meanwhile, one has $\sum_\alpha p_\alpha S_C(   \tr_{\overline{C}} \ket{ \psi_\alpha } \bra{\psi_\alpha}   ) \geq E_f(\rho = \sum_\alpha p_\alpha  \ket{\psi_\alpha }\bra{\psi_\alpha}   ) $ because $E_f$ is the average entanglement entropy minimized over all possible pure state realizations. As a result, one finds $   E_f(\rho_0)\geq E_f(\rho)$.

When generalizing onsite unitary operations to multi-site, geometrically local unitary gates, the protocol no longer belongs to LOCC, and mixed-state entanglement may increase. To bound the entanglement growth, one may consider the Renyi-$n$ entanglement of formation \cite{renyi_eof_2016} defined as $ R_n  \equiv  \text{Min}   \left\{  \sum_i p_i S^{(n)}_C ( \tr_{\overline{C}} \ket{\psi_i }  \bra{\psi_i}    ) \right\}$,  where $S^{(n)}_C$ is the Renyi-$n$ entropy. Starting with a pure state, after a finite-depth protocol with local measurement and unitary gates (acting on $O(1)$ number of neighboring sites), one obtains an ensemble $\rho$ of pure state trajectories $\ket{\psi_{\alpha}}$ with probability $p_{\alpha}$. For each pure state trajectory $\ket{\psi_\alpha}$, the increase of Renyi-0 entanglement $S^{(0)}_C (\tr_{\overline{C}} \ket{\psi_\alpha}\bra{\psi_\alpha}   )  $,  also dubbed max entropy or Hartley entropy, is upper bounded by $D\abs{\partial A}$, where $D$ is the depth of the channel and $\abs{\partial A}$ is the bipartition boundary area. This is because Hartley entropy is simply bounded by the number of local measurements and unitary gates that act on the boundary. Therefore, the increase of  Renyi-0 entanglement of formation $R_0$ is also bounded by the area law. One immediate consequence is that if one starts with a pure product state (i.e. with  $R_0=0$),  then after a finite-depth channel, the  Renyi entanglement of formation of any Renyi index will follow an area law since  $R_n \leq R_0$ for any $n>0$.

	\section{Measuring 1d SPT}	
	\subsection{Derivation of operator duality}\label{append:1dspt_duality}

	Here we derive the transformation rules for operators under the controlled unitary  $U= \sum_{\alpha} P_{\alpha} U_{\alpha} $.  $U_{\alpha}$ is the unitary feedback defined as $U_\alpha = \prod_{ i  }  X_{b,i}^{ \frac{1- \prod_{ j=1, 2, \cdots }^i \alpha_j  }{2}}$.  $P_{\alpha}$ is a projector defined as $P_{\alpha} = \ket{\alpha} \bra{\alpha}$, which projects $A$ sublattice to a product state in Pauli-X basis. Below we provide two distinct approaches for derivation.\\ 
	
	\noindent \underline{\textit{Operator-based approach}}: 		
	First,  the  controlled unitary can be written as  $U=\prod_i X_{b,i}^{ \frac{1  -  \prod_{j=1}^i{ X_{a,j }   } }{2}  }$.   Since $U$ is diagonal in the Pauli-X basis, Pauli-X operators are invariant under the conjugation by $U$.  Next, we consider   $UZ_{b,i} Z_{b,i+1} U^{\dagger}$ (note that Pauli-Zs come in pairs due to the global $\mathbb{Z}_2$ symmetry). This operator can be simplified as $ uZ_{b,i} Z_{b,i+1}u^{\dagger}$, where 
	\begin{equation}
		u=X_{b,i}^{ \frac{1  -  \prod_{j=1}^i{ X_{a,j }   } }{2}  } X_{b,i+1}^{ \frac{1  -  \prod_{j=1}^{i+1}{ X_{a,j }   } }{2}  }. 
	\end{equation}
	In the subspace that $\prod_{j=1}^i X_{a,j}=1$, one finds $UZ_{b,i} Z_{b,i+1} U^{\dagger}  =  X_{b,i+1}^{   \frac{1 -X_{a,i+1}  }{2}  } Z_{b,i} Z_{b,i+1}  X_{b,i+1}^{   \frac{1 -X_{a,i+1}  }{2}  }   $. For  $X_{a,i+1}=1$, one has $UZ_{b,i} Z_{b,i+1} U^{\dagger}  =Z_{b,i} Z_{b,i+1}$ while for $X_{a,i+1}=-1$ one has $UZ_{b,i} Z_{b,i+1} U^{\dagger}  =-Z_{b,i} Z_{b,i+1}$. Combining the cases of $X_{a,i+1}=\pm 1$  leads to the result  $UZ_{b,i} Z_{b,i+1} U^{\dagger}  =Z_{b,i} X_{a,i+1}  Z_{b,i+1}$. Similarly, one finds $  U Z_{a,i} Z_{a,i+1} U^{\dagger}	  = Z_{a,i}X_{b,i} Z_{a,i+1}$.

Here we remark that as we see in the main text, the controlled unitary $U=\prod_i X_{b,i}^{ \frac{1  -  \prod_{j=1}^i{ X_{a,j }   } }{2}  }$ maps a $\mathbb{Z}_2 \times  \mathbb{Z}_2$ SPT to two $\mathbb{Z}_2$ symmetry-breaking (GHZ) ordered states on $A$ and $B$ sublattices. This is akin to the Kennedy-Tasaki transformation \cite{Kennedy_Tasaki_1992,Kennedy_Tasaki_1992_prb}, which transforms a Haldane spin-1 chain \cite{haldane_spin_chain_1983} with $\mathbb{Z}_2 \times  \mathbb{Z}_2$ SPT order to two spontaneous $\mathbb{Z}_2$ symmetry-breaking orders. Indeed, Kennedy-Tasaki transformation can also be realized as a controlled unitary \cite{Oshikawa_1992_KT}, which therefore fits into our general framework. In particular, we can take a Haldane spin-1 chain in the $\mathbb{Z}_2 \times  \mathbb{Z}_2$ SPT phase, and perform a layer of onsite measurement followed by a layer of onsite unitaries to realize  a mixed state with $\mathbb{Z}_2$ GHZ-like order. \\

	\noindent \underline{\textit{Wave-function based approach}}:  starting with a global $\mathbb{Z}_2 \times \mathbb{Z}_2$ symmetric input state $\ket{\psi_0}$,  in Pauli-X basis $\ket{\psi_0}$ can be written as 
	\begin{equation}\label{eq:spt_psi_0}
		\ket{\psi_0}   =\sum'_{\alpha,\beta} (-1)^{\chi(\alpha,\beta)} \psi(\alpha, \beta) \ket{\alpha,\beta}.
	\end{equation}
	$\alpha = \{ \alpha_i  =\pm1  \}, \beta =\{  \beta_i =\pm 1 \}$ denote the product state in Pauli-X basis in the sublattice $A$ and sublattice $B$. The $\mathbb{Z}_2$ symmetry on each sublattice implies that the allowed $\alpha, \beta$  must satisfy $\prod_{i } \alpha_i = \prod_{i} \beta_i =1$. We use $\Sigma'$ to denote that only those $\alpha, \beta$ that satisfy this constraint are considered.

\begin{figure}
		\centering
		\begin{subfigure}{0.47\textwidth}
			\includegraphics[width=\textwidth]{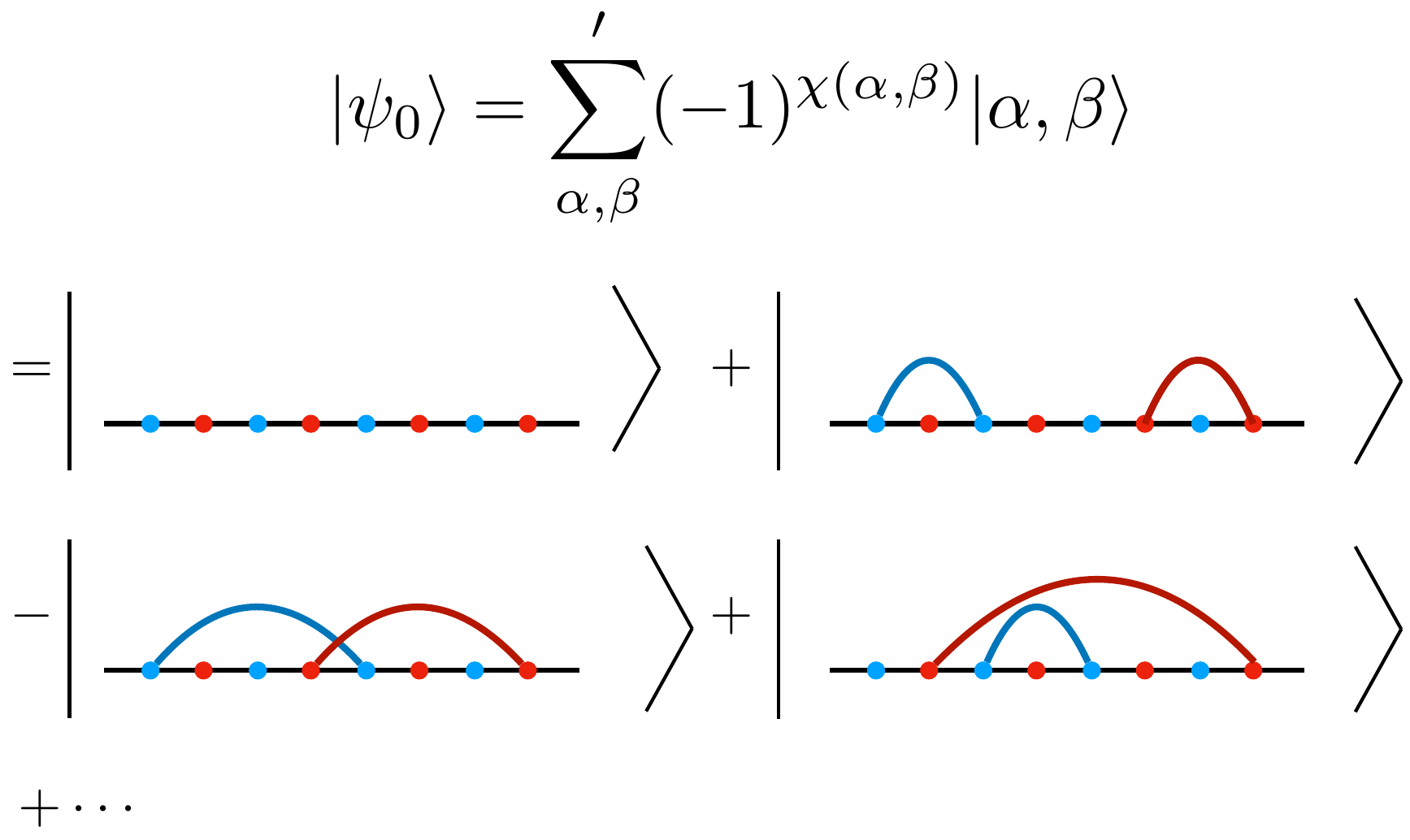}
		\end{subfigure}
		\caption{1d fixed-point $\mathbb{Z}_2 \cross \mathbb{Z}_2$ SPT (with parent Hamiltonian  $H_0 = -\sum_{i}Z_{a,i} X_{b,i} Z_{a,i+1 }  -\sum_{i}Z_{b,i} X_{a,i+1} Z_{b,i +1 } $)  as a condensate of two types of strings with a braiding sign structure in the Pauli-X basis. A product state $\ket{\alpha, \beta }$ in X basis with  $\prod_i \alpha_i = \prod_i \beta=1 $ corresponds to a configuration of blue strings and red strings whose endpoints  indicate $\alpha_i=-1$ and $\beta_i=-1$ respectively. The braiding number $\chi(\alpha,\beta)$ counts the number of times that two types of strings intersect. } 
		\label{fig:1dspt}  
	\end{figure}

	Due to the symmetry, $\alpha_i=-1 $ must come in pairs, which can be regarded as the two endpoints of a string.  Correspondingly, various $\alpha$ configurations can be understood as various $A$-strings in $A$ sublattice, and similarly, $\beta$ configurations can be represented by $B$-strings in $B$ sublattice. Within the string representations, $\chi$ is the number of times that $A$-strings intersect (braid) with $B$-strings.  When  $\ket{\psi_0}$ is a fixed-point cluster SPT with $ZXZ$ stabilizers, $\psi(\alpha,\beta)$ is a constant independent of $\alpha, \beta$ (see Fig.\ref{fig:1dspt}).

	Given Eq.\ref{eq:spt_psi_0}, measuring Pauli-X on $A$ sublattice projects $A$ sublattice to a particular $\alpha$ configuration: $P_{\alpha}\ket{\psi_0}		 = \ket{\alpha}  \otimes \sum'_{\beta} (-1)^{\chi(\alpha,\beta)} \psi(\alpha, \beta) \ket{\beta}$.  After the measurement,  applying  the unitary feedback $U_{\alpha} (= \prod_{ i  }  X_{b,i}^{ \frac{1- \prod_{ j=1, 2, \cdots }^i \alpha_j  }{2}}$)  removes  the braiding phase: $U_{\alpha}P_{\alpha}\ket{\psi_0}	 	 =  	\ket{\alpha}\otimes  \sum'_{\beta} \psi(\alpha, \beta) \ket{\beta}$.  It follows that the output state on $B$ sublattice reads
	\begin{equation}
		\begin{split}
			\rho_B  &=  \tr_A \left( \sum_{  \alpha } U_{\alpha} P_{\alpha } \ket{\psi_0}\bra{\psi_0} P_{\alpha} U^{\dagger}_\alpha \right) \\
			&=   \sum'_{\beta,\beta'} \ket{\beta} \bra{\beta'}   \sum'_{\alpha}    \psi(\alpha, \beta) \psi^*(\alpha, \beta'). 
		\end{split}
	\end{equation}
	$\rho_B$ is simply the reduced density matrix on $B$ of the following pure state: 
	
	\begin{equation}\label{eq:psi}
		\ket{  \psi}=  \sum'_{\alpha,\beta}\psi(\alpha, \beta) \ket{\alpha,\beta}. 
	\end{equation}
	One  consequence is that the  entropy of the mixed state $\rho_B$ results  from the entanglement entropy between $A$ and $B$ in $\ket{\psi}$, i.e. it is the residual quantum fluctuation when  removing the braiding phase in the initial input SPT $\ket{\psi_0}$.

	Comparing Eq.\ref{eq:spt_psi_0} and Eq.\ref{eq:psi}, it is obvious that $\ket{\psi} = U\ket{\psi_0}$ with  $U$  is a diagonal matrix whose entries encode the braiding phase:
	
	\begin{equation}\label{eq:U_map}
		U = \sum'_{\alpha,\beta} \ket{\alpha,\beta} \bra{\alpha,\beta}   (-1)^{\chi(\alpha,\beta)},  
	\end{equation}
	Although  $U$ is not unitary in the entire Hilbert space, it is unitary in the symmetric subspace specified by $\prod_{i} X_{a,i}= \prod_{i} X_{b,i}= 1$.	Note that this form of  unitary has appeared in the literature \cite{ringel_2015_hidden} as a way to reveal the hidden long-range order of certain SPTs. We also note that in the symmetric subspace, $U$ is the same as the controlled unitary $ \sum_{\alpha } P_{\alpha} U_{\alpha }  = \prod_i X_{b,i}^{ \frac{1  -  \prod_{j=1}^i{ X_{a,j }   } }{2}  }$ discussed  in the operator-based approach.

	With $U= \sum'_{\alpha,\beta}  (-1)^{\chi(\alpha,\beta)} \ket{\alpha,\beta} \bra{\alpha,\beta}$, one can derive the parent Hamiltonian  of $\ket{\psi}$ from $H  =   U H_0 U^{\dagger} $, where the following symmetry constraint is further imposed on $H$:     $\prod_{i} X_{a,i}= \prod_{i} X_{b,i}= 1$\footnote{More specifically, based on the condition that  $P\ket{\psi_0} = \ket{ \psi_0}$ and $ P\ket{\psi} = \ket{ \psi} = U \ket{\psi_0}$, where $P$ is a projector to the subspace with $\prod_{i} X_{a,i}= \prod_{i} X_{b,i}= 1$, the eigenequation $H_0\ket{\psi_0 } = E_g \ket{\psi_0}$ implies   $P  U H U^{\dagger}  P \ket{\psi } = E_g \ket{\psi_0}$, so the parent Hamiltonian of $\ket{\psi}$ is $P  U H U^{\dagger}  P $.}.

	Now we present the derivation of operator mapping:
	\begin{equation} 
		\begin{split}
			&UZ_{b,i} Z_{b,i+1} U^{\dagger} \\
			&=\sum'_{\alpha',\beta',\alpha,\beta}  \ket{\alpha',\beta'} \bra{\alpha,\beta}\\ 
			& \bra{\alpha',\beta'}(-1)^{\chi(\alpha',\beta')}  Z_{b,i} Z_{b,i+1} (-1)^{\chi(\alpha,\beta)}
			\ket{\alpha,\beta}.  
		\end{split}
	\end{equation}
	While $Z_{b,i} Z_{b,i+1}$ has no effect on $\ket{\alpha}$ (i.e. $\alpha'= \alpha$), this operator makes the transition from $\beta$ to $\beta'$, where   $\beta$ and $\beta'$ only differ in the site $(b,i)$ and site $(b,i+1)$. 	It is not hard to find that the product of the braiding phase $(-1)^{\chi(\alpha',\beta')}   (-1)^{\chi(\alpha,\beta)}$, i.e. the braiding sign change induced by $Z_{b,i} Z_{b,i+1}$ is simply equal to $\alpha_{i+1}$. In other words, $\bra{\alpha',\beta'}(-1)^{\chi(\alpha',\beta')}  Z_{b,i} Z_{b,i+1} (-1)^{\chi(\alpha,\beta)}
	\ket{\alpha,\beta}$
	\begin{equation}
		\begin{split}
			& = \bra{\alpha',\beta'}Z_{b,i} \alpha_{i+1} Z_{b,i+1} 
			\ket{\alpha,\beta  }\\
			&= \bra{\alpha',\beta'}Z_{b,i}  X_{a,i+1} Z_{b,i+1} 
			\ket{\alpha,\beta }.
		\end{split}
	\end{equation}
	Therefore one finds $Z_{b,i} Z_{b,i+1}	  \to   Z_{b,i}X_{a,i+1} Z_{b,i+1}$.  With a similar calculation, one  finds $	Z_{a,i} Z_{a,i+1}	  \to   Z_{a,i}X_{b,i} Z_{a,i+1}$.

	\begin{figure}[h]
		\centering
		\begin{subfigure}{0.5\textwidth}
			\includegraphics[width=\textwidth]{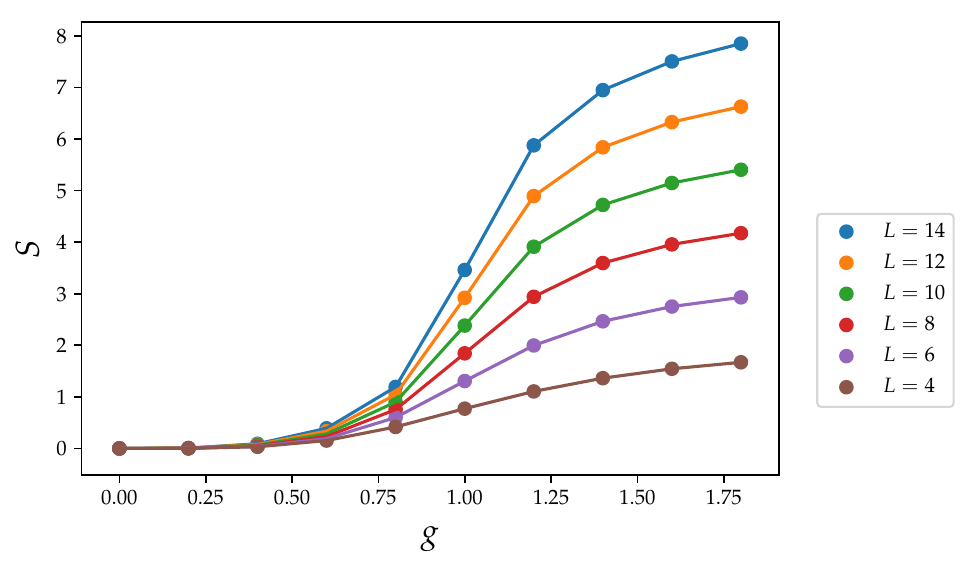}
		\end{subfigure}
		\begin{subfigure}{0.5\textwidth}
			\includegraphics[width=\textwidth]{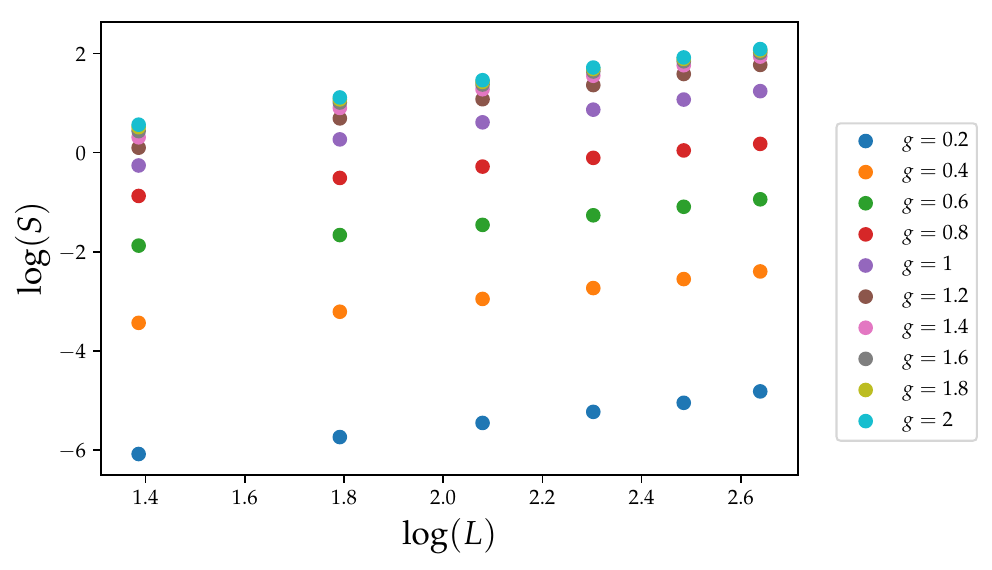}
		\end{subfigure}
		\caption{von Neumann entropy $S$ of the mixed state $\rho_B (= \tr_A \ket{\psi}\bra{\psi})$ with $\ket{\psi}$ being the ground state of $H$ in Eq.\ref{eq:1d_spt_symmetric_measure}. Upper panel: $S$ as a function of the tuning parameter $g$. Lower panel: the scaling of $S$ with the system size $L$ of sublattice $B$. We find $S$ scales linearly with $L$ for any non-zero $g$.}
		\label{fig:mixed_logS}
		
	\end{figure}

	\subsection{Numerical data on volume-law entropy}\label{sec:volume_law_entropy}
	Here we report the data on von Neumann entropy of the  mixed state $\rho_B$ from the ground state of $H$ (Eq.\ref{eq:1d_spt_symmetric_measure}). The entropy is defined as $S(\rho_B) = - \tr_B \big[\rho_B \ln(\rho_B)\big]$. The ED calculation for various system sizes is shown in Fig.\ref{fig:mixed_logS}. Our data suggest that $S(\rho_B)$ scales linearly with $L$, i.e. volume-law scaling, for any non-zero $g$. In particular, this means that at the critical point $g=1$, $B$ sublattice is a mixed state with volume-law classical entropy coexisting with critical long-range entanglement.

	\subsection{Other non-fixed-point SPT as input}\label{sec:general_perturbation}
	
	Here we consider  various types of symmetric perturbation on the fixed-point $\mathbb{Z}_2 \times \mathbb{Z}_2$ SPT, and discuss the structure of the corresponding output state. 
	
	\noindent \underline{\textit{On-site X perturbation}}: here the input is the ground state of the following Hamiltonian: 
	\begin{equation}\label{eq:input_H}
		\begin{split}
			H_0  = & -\sum_{i}Z_{a,i} X_{b,i} Z_{a,i+1 }  -\sum_{i}Z_{b,i} X_{a,i+1} Z_{b,i+1 } \\
			&		    -g  \sum_{i}  (X_{a,i} +X_{b,i}).
		\end{split}
	\end{equation} 
	
	The ground state of $H_0$ belongs to $\mathbb{Z}_2 \times \mathbb{Z}_2$ SPT for $g<1$, and belongs to a trivial phase for $g>1$\cite{Plenio_2004_spt,Bartlett_2009_spt}. Using the transformation rule in  Eq.\ref{eq:U_transform}, one obtains a Hamiltonian of  two decoupled transverse-field Ising chain: $H=H_A +H_B$, where  $ H_A = - \sum_{i}Z_{a,i}   Z_{a,i+1} -g  \sum_{i} X_{a,i}  $ and  $ H_B = - \sum_{i}Z_{b,i}   Z_{b,i+1} -g  \sum_{i} X_{b,i}  $.  The state $\rho_B$ from  the measurement-feedback channel is the reduced density matrix of the ground state of $H$ by tracing out $A$. Since $A$ and $B$ are decoupled, $\rho_B$ is in fact a pure state, namely, the ground state of the  transverse-field Ising chain, whose long-range order exists for $\abs{g}<1$ (i.e. the regime where the input state is an SPT). In particular,  $\rho_B$ being  pure means that all  post-measurement state trajectories are transformed to the same state on $B$ with the unitary feedback.\\

	
	\noindent \underline{\textit{Two-body ZZ perturbation}}: here the input is the ground state of the Hamiltonian with independently tunable two-body $ZZ$ perturbation on $A$ and $B$ sublattices: 
	
	\begin{figure}
		\centering
		\begin{subfigure}{0.49\textwidth}
			\includegraphics[width=\textwidth]{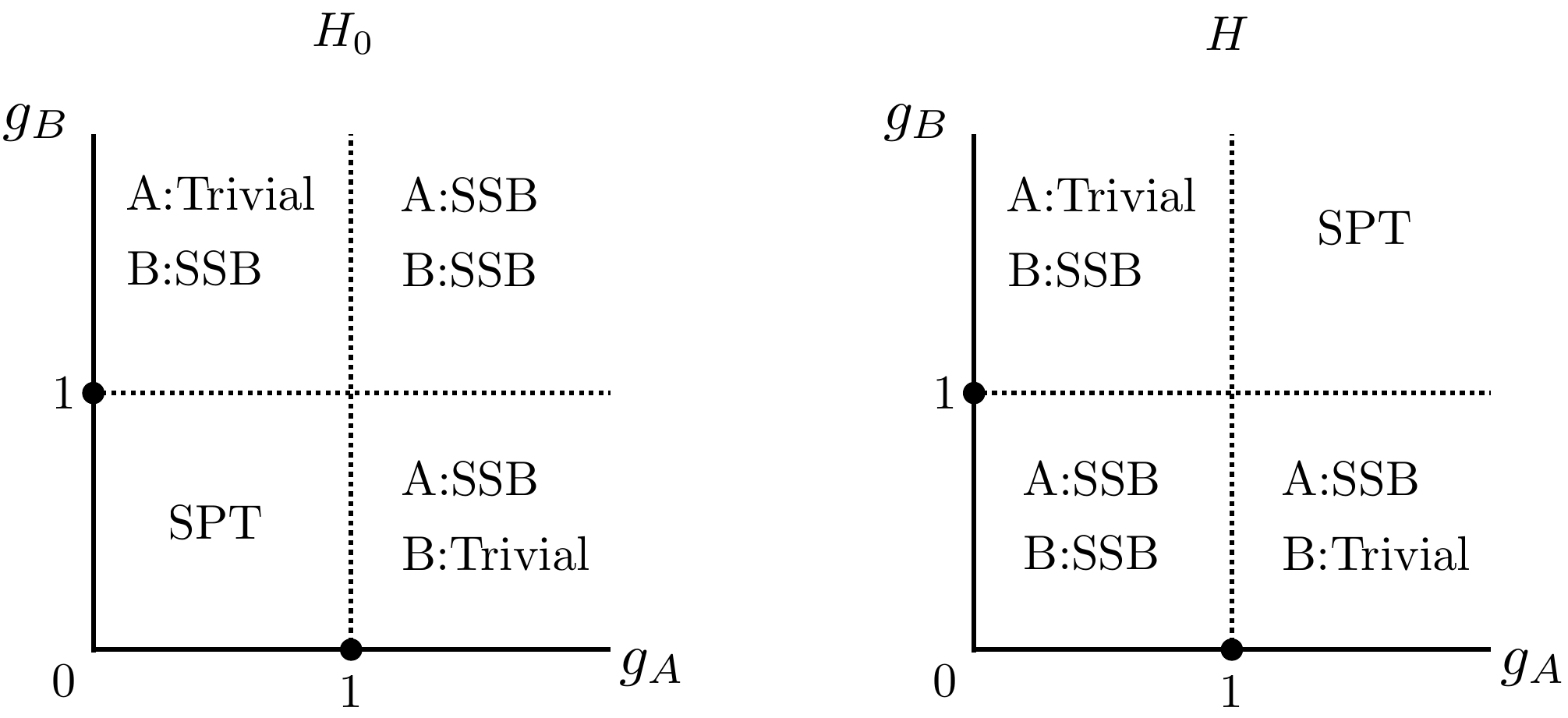}
		\end{subfigure}
		\caption{Phase diagram of $H_0$ and $H$ defined in Eq.\ref{eq:1d_tunable} and Eq.\ref{eq:1d_tunable_transform}. Taking the ground state of $H_0$ as an input, the measurement-feedback channel prepares a mixed state on $B$ sublattice, which is the reduced density matrix of the ground state of $H$. 	} 
		\label{fig:1d_phase_diagram}  
	\end{figure}
	
	\begin{equation}\label{eq:1d_tunable}	
		\begin{split}
			H_0  = & -\sum_{i}Z_{a,i} X_{b,i} Z_{a,i+1 }  -\sum_{i}Z_{b,i} X_{a,i+1} Z_{b,i +1 } \\
			&		    -g_A  \sum_{i}  Z_{a,i} Z_{a,i+1 }   -g_B  \sum_{i}  Z_{b,i}Z_{b,i+1 }
		\end{split}
	\end{equation} 
	The measurement-feedback channel will prepare a state $\rho_B$ on $B$ sublattice with the parent Hamiltonian:
	
	\begin{equation}\label{eq:1d_tunable_transform}
		\begin{split}
			H = & -\sum_{i}Z_{a,i} Z_{a,i+1 }  -\sum_{i}Z_{b,i}  Z_{b,i +1 } \\
			&		    -g_A  \sum_{i}  Z_{a,i}X_{b,i}    Z_{a,i+1 }   -g_B  \sum_{i}  Z_{b,i}X_{a,i+1} Z_{b,i+1 }.
		\end{split}
	\end{equation} 
	$H_0$ and $H$ are dual to each other, and their phase diagrams are presented in Fig.\ref{fig:1d_phase_diagram}. Interestingly, one notices that with measurement-feedback channel acting on a gapless state, we can prepare a non-critical mixed state with long-range order.  To see this, we consider $g_B=1$ and $g_A<1$. In this case, $H_0$ exhibits the gapless SPT order \cite{gapless_spt_2017_vasseur} since the $\mathbb{Z}_2$ charges on $A$ are used to decorate the domain walls in $B$  sublattice that is tuned to a critical point between a trivial phase and a $\mathbb{Z}_2$ SSB phase. It follows that by measuring  the gapped degrees of freedom on $A$ followed by unitary feedback, one obtains  mixed state $\rho_B$ on $B$ sublattice that exhibits a $\mathbb{Z}_2$ SSB order.

	More generally, one may consider an input state of the form $\ket{ \psi_0}= U\ket{\psi_A}_A \ket{\psi_B }_B$ with  $U=\prod CZ_{(a,i), (b,i) }CZ_{(b,i), (a,i+1)  }$ being  a product of controlled-Z that entangles $A$ and $B$ sublattices.  The measurement-feedback protocol will always lead to a mixed state on $B$ with long-range order, as long as  $\ket{\psi_0}$ satisfies the following two conditions (in particular $\ket{\psi_0}$ does not need to be an SPT): (1) $\ket{\psi_A}$ is trivial, where domain-walls condense (i.e. the product of $X$ along an open string takes a non-zero expectation value $c$). (2) The input state on $B$ sublattice, i.e. $\ket{\psi_B}$, is invariant under the global $\mathbb{Z}_2$ symmetry: $\prod_{i \in B} X_i  \ket{\psi_B}_B=  \ket{\psi_B}_B$ (e.g. $\ket{\psi_B}$ can be trivial, spontaneously breaking the symmetry, or critical). The first condition implies that a long-range string operator $\bra{\psi_0}Z_BX_AX_A\cdots X_A Z_B \ket{\psi_0} =c$; measuring $X_A$ on $A$ sublattice followed by a unitary feedback acting on $B$ leads to a mixed state $\rho_B$ on $B$ with a long-range $ZZ$ two-point function. (2) The second condition implies that $\tr[\rho \prod_{i \in B} X_i ]=1$. As a result, combining these two conditions already guaranteed the existence of long-range order in the mixed state $\rho_B$.  Practically speaking, it is of best interest of our protocol to consider $\ket{\psi_B}_B$ as a trivial state (in which case $\ket{\psi_0}$ is an SPT), since it furnishes an example of obtaining long-range order only after implementing the protocol.

	\subsection{SPT under decoherence as input }\label{append:decoherence}
	In the main text, we discussed a measurement-feedback channel that transforms the string order of the  input pure SPT to a long-range order. Specifically, the main result is  
	
	\begin{equation}
		\tr[   \rho  Z_{b,i} Z_{b,j}    ]=  \tr [  \rho_0  Z_{b,i}  \left( \prod_{k=i+1 }^{j}  	 X_{a,k} \right)  Z_{b,j}] 
	\end{equation}
	where $\rho_0  = \ket{\psi_0} \bra{\psi_0}$ is the input state. In fact, with a straightforward derivation, one finds the equation holds true for arbitrary input  density matrix $\rho_0$. Namely, $\rho_0$ needs not be a pure state. As an application, the  input state $\rho_0$ may be a mixed state obtained by subjecting an SPT to certain decoherence channels. If the string order is robust under the channel, the output state of our protocol will remain long-range ordered. Below we will first discuss the decohered SPTs and the corresponding emergent long-range order. We will then discuss the fate of those decohered SPTs when tuned to criticality.

	First, we consider a dephasing channel in Pauli-X basis. A local channel is given by $\mathcal{E}_i: \rho\to (1-p) \rho + p X_i \rho X_i = \sum_{\sigma=0,1} p_\sigma K^{\dagger}_{\sigma} \rho K_{\sigma} $ with $K_\sigma = X_i^{\sigma_i}$, and $p_{\sigma}=p, 1-p$ for $\sigma=1, 0$ respectively. This can be understood as applying $X_i$ operator with probability $p$, so $p=1/2$ corresponds to maximal dephasing. We consider the local channels on every lattice site, so the overall dephasing channel $\mathcal{E}$ is the composition of the local noise channels: $\mathcal{E}=\mathcal{E}_1 \circ \mathcal{E}_2 \circ \cdots\circ  \mathcal{E}_{2L}$, and 
	
	\begin{equation}
		\mathcal{E}[\rho ]  = \sum_{  \boldsymbol{\sigma} }    P( \boldsymbol{\sigma}   )  K^{ \dagger}_{  \boldsymbol{\sigma}  }  \rho  K_{  \boldsymbol{\sigma}  },   
	\end{equation}
	where $\boldsymbol{\sigma}  = \{  \sigma_i \}$,   $  P( \boldsymbol{\sigma}   ) = \prod_{i=1}^{2L} p_{\sigma_i}$, and $  K_{  \boldsymbol{\sigma}  } = \prod_{i=1}^{2L} X^{\sigma_i}_{i} $.  Given a pure SPT, $\rho_{\text{SPT}} = \ket{\text{SPT}} \bra{\text{SPT}}$, the dephasing channel leads to $\rho_0  = \mathcal{E}[\rho_{\text{SPT}}]$, and an operator $O$ with respect to  $\rho_0$ becomes 
	
	\begin{equation}
		\expval{O}_{\rho_0}  =  \sum_{\boldsymbol{\sigma}  }	  P(\boldsymbol{\sigma} ) \bra{  \text{SPT}} K_{  \boldsymbol{\sigma}  }  O K^{\dagger}_{  \boldsymbol{\sigma}  }  \ket{\text{SPT}}. 
	\end{equation}
	One immediate consequence is that if $[O, 	K_{\boldsymbol{\sigma}} ] =0$, then  $\expval{O}_{\rho_0}  = \bra{  \text{SPT}} O  \ket{\text{SPT}}$, i.e. the expectation value of $O$ remains invariant  under the dephasing channel. Therefore, $\expval{\prod_{i=1}^L X_{b,i}    }_{\rho_0}=1$ as in $\ket{\text{SPT}}$. On the other hand, for the string operator $\mathcal{S} \equiv Z_{b,i}  \left( \prod_{k=i+1 }^{j}   X_{a,k} \right)  Z_{b,j}$, since only $Z_{b,i}$ and $Z_{b,j}$ may anticommute with the noise operator $K_{\boldsymbol{\sigma}}$, $\expval{ \mathcal{S} }_{\rho_0} $ can be simplified as 
	\begin{equation} 
		\sum_{\sigma_{b,i}  \sigma_{b,j}}  p_{\sigma_{b,i}   }  p_{\sigma_{b,j}}   \bra{  \text{SPT}}  X_{b,i}^{\sigma_{b,i}}     X_{b,j}^{\sigma_{b,j}} \mathcal{S} X_{b,i}^{\sigma_{b,i}}     X_{b,j}^{\sigma_{b,j}}          \ket{\text{SPT}}. 
	\end{equation}
	
	Since the string operator  $\mathcal{S}$ conjugated by $X_{b,i}^{\sigma_{b,i}}     X_{b,j}^{\sigma_{b,j}} $ acquires a $+1$, $-1$ sign when  $\sigma_{b,i} = \sigma_{b,j}$, $\sigma_{b,i} \neq  \sigma_{b,j } $ respectively, one finds 
	
	\begin{equation}
		\begin{split}
			&			\expval{ \mathcal{S} }_{\rho_0} \\
			&= \left[ \text{Prob.} (\sigma_{b,i}	=\sigma_{b,j}    )  -  \text{Prob.} (\sigma_{b,i}	\neq \sigma_{b,j}    )   \right] \bra{  \text{SPT}}  \mathcal{S}     \ket{\text{SPT}}\\
			& = (1-2p)^2  \bra{  \text{SPT}}  \mathcal{S}     \ket{\text{SPT}}. 
		\end{split}
	\end{equation}	 
	Therefore, under the X-dephasing channel, the string order survives for any non-maximal dephasing (i.e. $p<1/2$). Given this dephased SPT mixed state, one can then use the measurement-feedback channel to obtain a mixed state with GHZ-like long-range order.

	If one instead considers the symmetry-breaking dephasing channel based on Pauli-Z noise, with a similar calculation, one finds the resulting dephased state cannot have both $Z_B X_A \cdots X_A Z_B \sim O(1)$ and $\prod_{i} X_{b,i}=1$. Specifically, due to the Pauli-Z noise on $A$ sublattice, one finds $Z_B X_A \cdots X_A Z_B \sim (1-2p )^{\abs{S}}$, i.e.  exponentially decaying with the length of the string $\abs{S}$ for any non-zero dephasing. On the other hand, the Pauli-Z noise on $B$ sublattice results in  $\prod_{i} X_{b,i} = (1-2p )^{L}$, i.e.  exponentially decaying with sublattice size of $B$ for any non-zero dephasing. As a result,  the measurement-feedback channel cannot lead to a mixed state with the GHZ-like long-range order based on such dephased state.

\begin{figure}[h]
		\centering
		\begin{subfigure}{0.5\textwidth}
			\includegraphics[width=\textwidth]{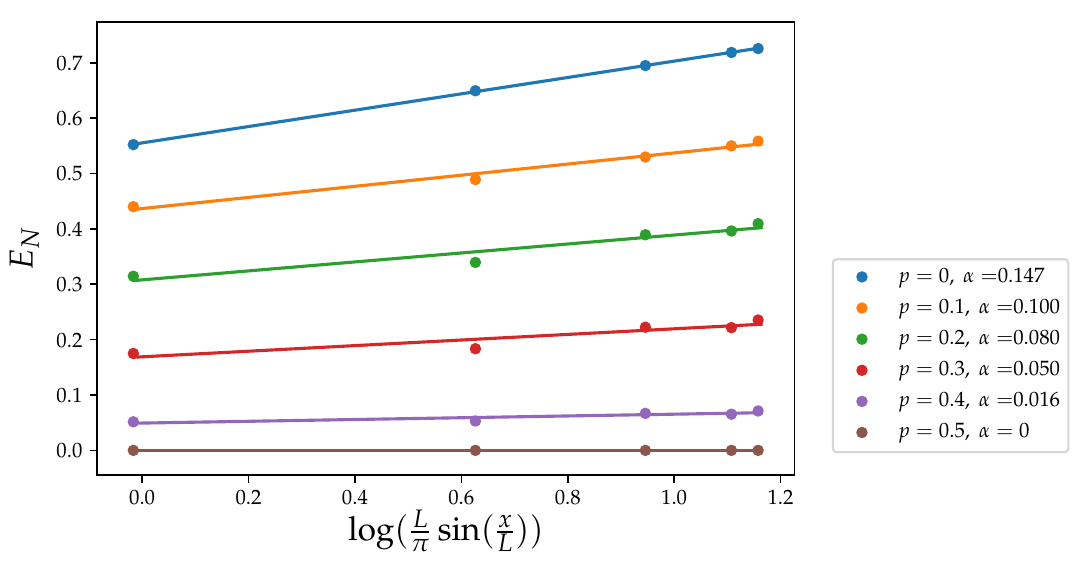}
		\end{subfigure}
		\begin{subfigure}{0.5\textwidth}
			\includegraphics[width=\textwidth]{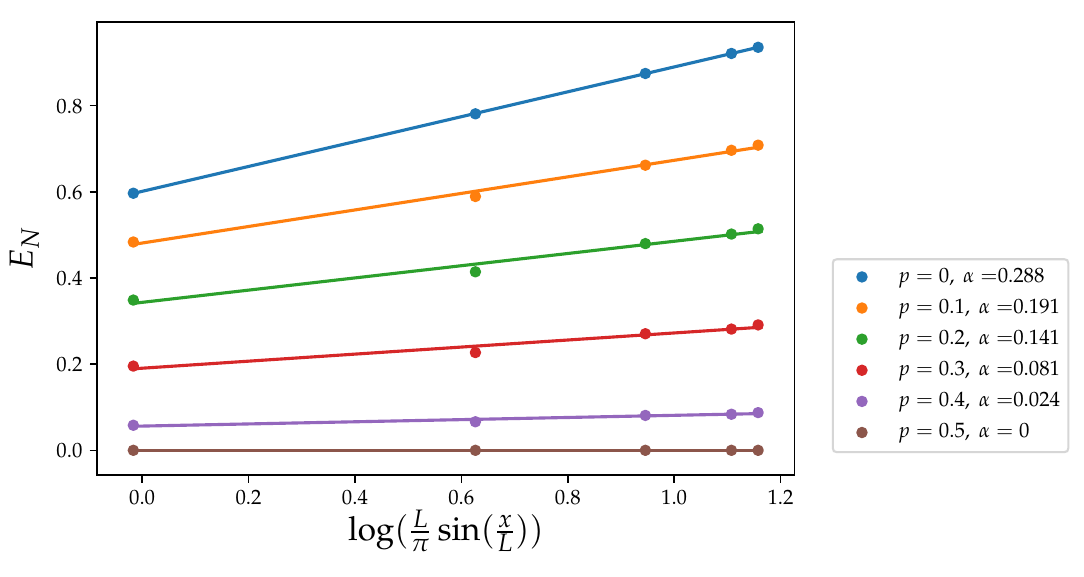}
		\end{subfigure}
		
		\caption{Entanglement negativity $E_N$ between two complementary segments of size $x$ and $L-x$ on $B$ sublattice for the mixed state $\rho_B$. The input state is a mixed state $\rho_0$ obtained by acting the input SPT $\psi_0$ with a decoherence channel. Upper panel: negativity of $\rho_B$ obtained by the measurement-feedback channel. Lower panel: negativity averaged over mixed-state trajectories obtained by measuring $A$ sublattice for the input dephased SPT $\rho_0$.}
		\label{fig:dephasing_nega}
		
	\end{figure}

	\noindent\underline{\textit{At criticality}}: when $\ket{\text{SPT}}$ is tuned to a critical point (e.g. $g=1$ in Eq.\ref{eq:1d_spt_symmetric_deform}), the string operator exhibits critical order  $Z_{b,i}  ( \prod_{k=i+1 }^{j}   X_{a,k} )  Z_{b,j}  \sim  \frac{1}{\abs{i-j}^{\eta}  }$. Under the Pauli-X dephasing, with a similar analysis as above, one obtains a mixed state $\rho_0$ with  $  \expval{Z_{b,i}  ( \prod_{k=i+1 }^{j}   X_{a,k} )  Z_{b,j}    }_{\rho_0} =(1-2p)^2  \expval{  Z_{b,i}  ( \prod_{k=i+1 }^{j}   X_{a,k} )  Z_{b,j}    }_{\text{SPT}}  \sim   \frac{1}{\abs{i-j}^{\eta}  } $. Therefore, the measurement-feedback channel discussed in Sec.\ref{sec:1d_bspt} gives rise to a mixed state $\rho_B$ with the critical correlation  $Z_{b,i}   Z_{b,j} \sim   \frac{1}{\abs{i-j}^{\eta}  }$. On the other hand,  the disorder operator $X_{b,i}X_{b,i+1}...X_{b,j}  $ commutes with both the dephasing channel and the measurement-feedback channel, and therefore, $  \expval{ X_{b,i}X_{b,i+1}...X_{b,j}   }_{\rho_0}  =   \expval{X_{b,i}X_{b,i+1}...X_{b,j}}_{\text{SPT}} \sim  \frac{1}{\abs{i-j}^{2\eta}  }   $. The algebraic order in these operators suggests that $\rho_B$ exhibits critical scaling  of entanglement. This is verified with  our  ED calculation, where the bipartite entanglement negativity follows the universal scaling form as in a 1+1D CFT (see Fig.\ref{fig:dephasing_nega} Upper panel):
	
	\begin{equation}
		E_N(\alpha, \beta) = \alpha \ln (\frac{L}{\pi}\sin(\frac{\pi x}{L})) + \beta.	
	\end{equation}
	In particular, the prefactor $\alpha$ decreases as increasing the strength of dephasing.

	Another natural question is, given an input dephased SPT $\rho_0$, what are the typical entanglement structures of the post-measurement (mixed) state trajectories right after measuring $A$ sublattice? To address this question, we  numerically compute the entanglement negativity averaged over these trajectories. As indicated by Fig.\ref{fig:dephasing_nega} lower panel, this quantity also exhibits the critical scaling in entanglement, indicating the pattern of critical entanglement in $\rho_B$ may be attributed to the post-measurement state trajectories that constitute the ensemble. Note that for a given dephasing strength $p$, the trajectory-average negativity is always smaller than the negativity in $\rho_B$. This is consistent with the intuition that some of the long-distance entanglement is diminished by  classical fluctuations in the ensemble of those trajectories.

	\section{Mixed state with topological order from measuring 2d SPT}\label{append:2d_spt}
	
	Here we discuss a finite-depth protocol that converts higher dimensional SPTs to mixed-state topological orders. In particular, we will consider the input state as a 2d SPT protected by the $\mathbb{Z}_2$ 0-form $\times$ $\mathbb{Z}_2$ 1-form symmetry, in which case the resulting  mixed state exhibits a $Z_2$ topological order. 
	
	Consider a 2d lattice with every vertex $v$ and every edge $e$ accommodating a qubit, the fixed-point SPT Hamiltonian reads
	\begin{equation}
		H_0=- \sum_v  X_v \prod_{e\ni v } Z_e - \sum_{e}  X_e \prod_{v \in e} Z_v.
	\end{equation}	
	The first term is a product of a Pauli-X on vertex $v$ and four neighboring Pauli-Zs on edges, and the second term is a product of a Pauli-X on edge $e$ and two  neighboring Pauli-Zs on vertices. This Hamiltonian can be obtained from a trivial paramagnet $- \sum_v  X_v   -\sum_e X_e$ by a depth-1 unitary circuit $U_{CZ}=\prod_{\expval{v,e}} CZ_{v,e}$, i.e. a product of controlled-Z gates, each of which acts on a pair of neighboring vertex and edge. The ground state is the 2d cluster state, a non-trivial SPT protected by $\mathbb{Z}_2$ 0-form $\times$ $\mathbb{Z}_2$ 1-form symmetry, where the $\mathbb{Z}_2$ 0-form symmetry is given  by $\prod_v X_v$ on vertices and the $\mathbb{Z}_2$ 1-form symmetry is  given by $\prod_{e\in \gamma} X_e$ acting on edges along any closed loop $\gamma$ (including non-contractible ones).

	The non-trivial SPT order can be diagnosed by the  membrane operator	 $M_{\tilde{\gamma}}= \prod_{e\in \tilde{\gamma}}  Z_e  \prod_{v \in A_{\tilde{\gamma}}  } X_v$, where $\tilde{\gamma}$ is a loop in the dual lattice, and $v \in A_{\tilde{\gamma}}  $ denotes all vertices (in the primary lattice) in the area enclosed by the loop $\tilde{\gamma}$. For instance, 
 \begin{equation}
		M_{\tilde{\gamma}}= \includegraphics[width=2.7cm,valign=c]{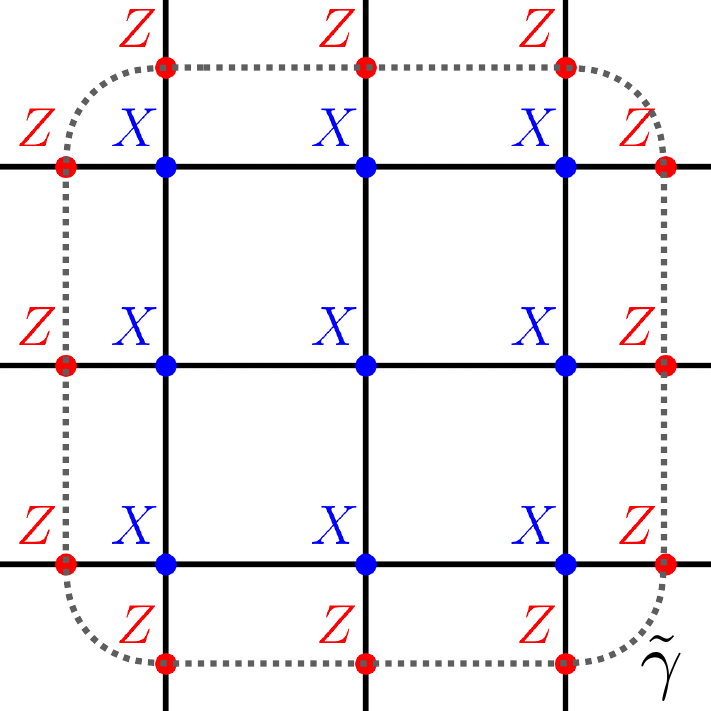}
	\end{equation}

 For the fixed-point SPT, $\expval{M_{\tilde{\gamma}}}=1$ for any loops $\tilde{\gamma}$. Away from the fixed-point limit, $M_{\tilde{\gamma}}$ exhibits a long-range correlation in the form of a perimeter law: $ \expval{ M_{\tilde{\gamma}} }   \sim e^{ - c \abs{\tilde{\gamma}} }$. This follows from the decorated domain wall description of this type of SPT; starting with a trivial paramagnet of Ising spins on vertices, the $\mathbb{Z}_2$ domain walls condense, implying  the perimeter law of a domain-wall creation operator $\prod_{v \in A_{\tilde{\gamma}}  } X_v \sim  e^{ - c \abs{\tilde{\gamma}} }$, where $\tilde{\gamma}$ is a closed loop in the dual lattice \footnote{This can also be understood in a dual picture, where the domain-wall creation operator in  the  trivial phase of Ising model correspond to a Wilson loop in the deconfined phase of the $2+1$D $\mathbb{Z}_2$ gauge theory, thereby following a perimeter law.}. We then introduce  $\mathbb{Z}_2$ charge on edges to decorate the domain walls by applying $U_{CZ}$. This leads to an SPT with  $M_{\tilde{\gamma}} = \prod_{e\in \tilde{\gamma}}  Z_e  \prod_{v \in A_{\tilde{\gamma}}  } X_v \sim e^{ - c \abs{\tilde{\gamma}} }$ because $M_{\tilde{\gamma}}$ is obtained by conjugating the domain wall operator $\prod_{v \in A_{\tilde{\gamma}}  } X_v $ with $U_{CZ}$.

	Now we can utilize the presence of long-range membrane order to devise a protocol for preparing a mixed state with $\mathbb{Z}_2$ topological order. Given an SPT $\ket{\psi_0}$, one measures $X_v$ on all vertices with outcome $\alpha= \{ \alpha_v\}$. For each post-measurement state trajectory with outcome $\alpha$, one applies a corresponding local unitary $U_\alpha$ acting on edges. This 2-step protocol leads to the ensemble $\rho  = \sum_{\alpha}  U_\alpha P_\alpha\rho_0  P_\alpha U_\alpha^{\dagger}$. Using a unitary $U_{\alpha}$ such that $U_{\alpha}^{\dagger} \prod_{e\in \tilde{\gamma}}  Z_e U_{\alpha}    =   \prod_{e\in \tilde{\gamma}}    Z_e   \left(  \prod_{v \in A_{\tilde{\gamma}}  } \alpha_v\right)$, (i.e. the dual loop operator acquires a sign that depends on the measurement outcome enclosed by the loop),  the expectation value of loop operators $\prod_{e\in \tilde{\gamma}}  Z_e$ in $\rho$ will exactly equal the expectation value of the membrane operator  $M_{\tilde{\gamma}}= \prod_{e\in \tilde{\gamma}}  Z_e  \prod_{v \in A_{\tilde{\gamma}}  } X_v$ with respect to the initial input SPT, which exhibits a perimeter-law scaling. In addition, since the symmetry sector is fixed with $\prod_{e\in \gamma} X_e=1$, the perimeter law scaling indicates a $\mathbb{Z}_2$ topological order on edges in the output mixed state, as in the deconfined phase of a 2+1D $\mathbb{Z}_2$ gauge theory.

	Here we discuss the choice of $U_{\alpha}$. Due to the global $\mathbb{Z}_2$ symmetry on vertices in the input SPT $\ket{\psi_0}$, the measurement outcomes on vertices  satisfy $\prod_v \alpha_v =1$, i.e. the $-1$ outcomes will occur in pairs. It turns out $U_{\alpha}$ is consisting of  string operators of Pauli-Xs on edges; each  string is deformable and connects two $\alpha_v = -1 $ outcomes on vertices. Physically, the $\alpha_v=-1$  outcomes may be regarded as $e$ particles in the $\mathbb{Z}_2$ topological order, and $U_{\alpha}$ is the string operator that annihilates those anyon excitations.

	%
	%
	%
	%

	\begin{figure}
		\centering
		\begin{subfigure}{0.38\textwidth}
			\includegraphics[width=\textwidth]{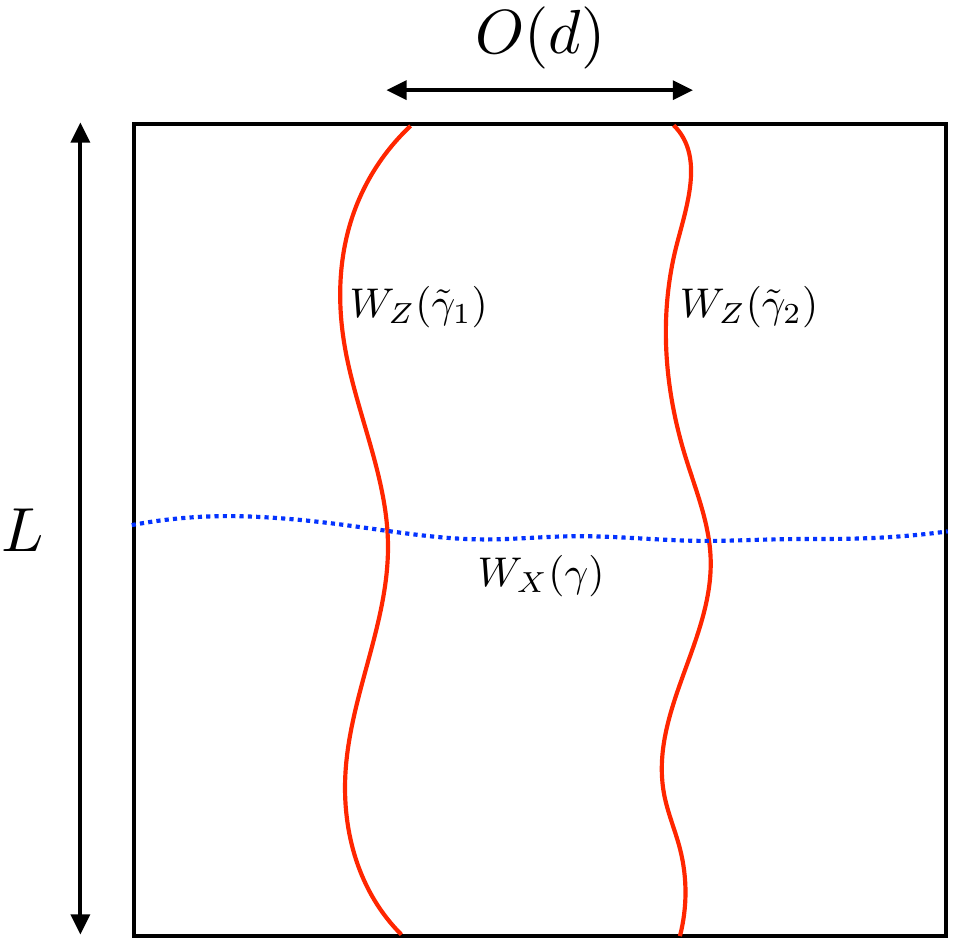}
		\end{subfigure}
		\caption{On a 2d lattice of size $L\times L$ on the surface of a 2-torus (by imposing periodic boundary conditions), the perimeter law decay of the Wilson loops in the dual lattice and the 1-form symmetry in the primary lattice together imply a non-trivial mixed state.}
		\label{fig:loops}  
	\end{figure}

\textit{Nontrivialness of the mixed state}: now we show that the 1-form symmetry $\prod_{e\in \gamma}X_e=1$ together with  the perimeter law of the Wilson loop in the dual lattice $\prod_{e\in \tilde{\gamma}} Z_e  \sim e^{ -c \abs{\tilde{\gamma}}  }   $ implies that the mixed state on edges ($\rho_B$) is non-trivial; it cannot be a mixture of short-range entangled pure states. This can be proved by contradiction, as discussed in Sec.\ref{sec:general_structure}. We first assume $\rho_B= \sum_n p_n \ket{n} \bra{n}$, where each $\ket{n}$ is a trivial short-range entangled state that can be connected to a product state using a finite-depth local unitary. We define two Wilson loop operators $W_Z(\tilde{\gamma}_i )= \prod_{e\in \tilde{\gamma}_i} Z_e $ for $i=1, 2$, where $\tilde{\gamma}_1, \tilde{\gamma}_2$ are  non-contractible loops around the vertical direction, and they are horizontally separated by a scale $O(d)$, see Fig.\ref{fig:loops}. For each $\ket{n}$, the connected correlator is $\bra{n} W_Z(\tilde{\gamma}_1)W_Z(\tilde{\gamma}_2) {\ket{n}}_c   = \bra{n} W_Z(\tilde{\gamma}_1)W_Z(\tilde{\gamma}_2) {\ket{n}}  -  \bra{n} W_Z(\tilde{\gamma}_1) {\ket{n}}  \bra{n} W_Z(\tilde{\gamma}_1) {\ket{n}}$. Due to the 1-form symmetry $W_X(\gamma)= \prod_{e\in \gamma} X_e=1$ with $\gamma$ winding around torus horizontally, $\bra{n} W_Z(\tilde{\gamma}_i){\ket{n}} =0$ for both $i=1,2$. On the other hand, the connected correlator in the trivial state $\ket{n}$ decays exponentially with $d$ \footnote{For example, in the ground state of a $\mathbb{Z}_2$ gauge theory, such an operator follows an area-law decay as $e^{-uLd}$ with $u$ being an $O(1)$ constant.}, and correspondingly  
 $\bra{n}W_Z(\tilde{\gamma}_1)W_Z(\tilde{\gamma}_2) {\ket{n}}$ decays with the separation $d$ as well. This in turn implies the $W_Z(\tilde{\gamma}_1)W_Z(\tilde{\gamma}_2)$ decays with $d$ in the ensemble of $\ket{n}$, which contradicts the perimeter law scaling $W_Z(\tilde{\gamma}_1)W_Z(\tilde{\gamma}_2)  \sim e^{-\alpha L }$, which does not decay with $d$ in the output mixed state. Therefore, the assumption that $\rho_B$ is a mixture of trivial states must be false.

	\textbf{Structure of mixed state topological order}:
	Here we discuss the structure of the topologically ordered mixed state using a wave-function perspective. This  allows us to derive a duality transformation that facilitates the  characterization of the phase diagram of the mixed state as varying the input state.
	
\begin{figure}[t]
		\centering
		\begin{subfigure}{0.5\textwidth}
			\includegraphics[width=\textwidth]{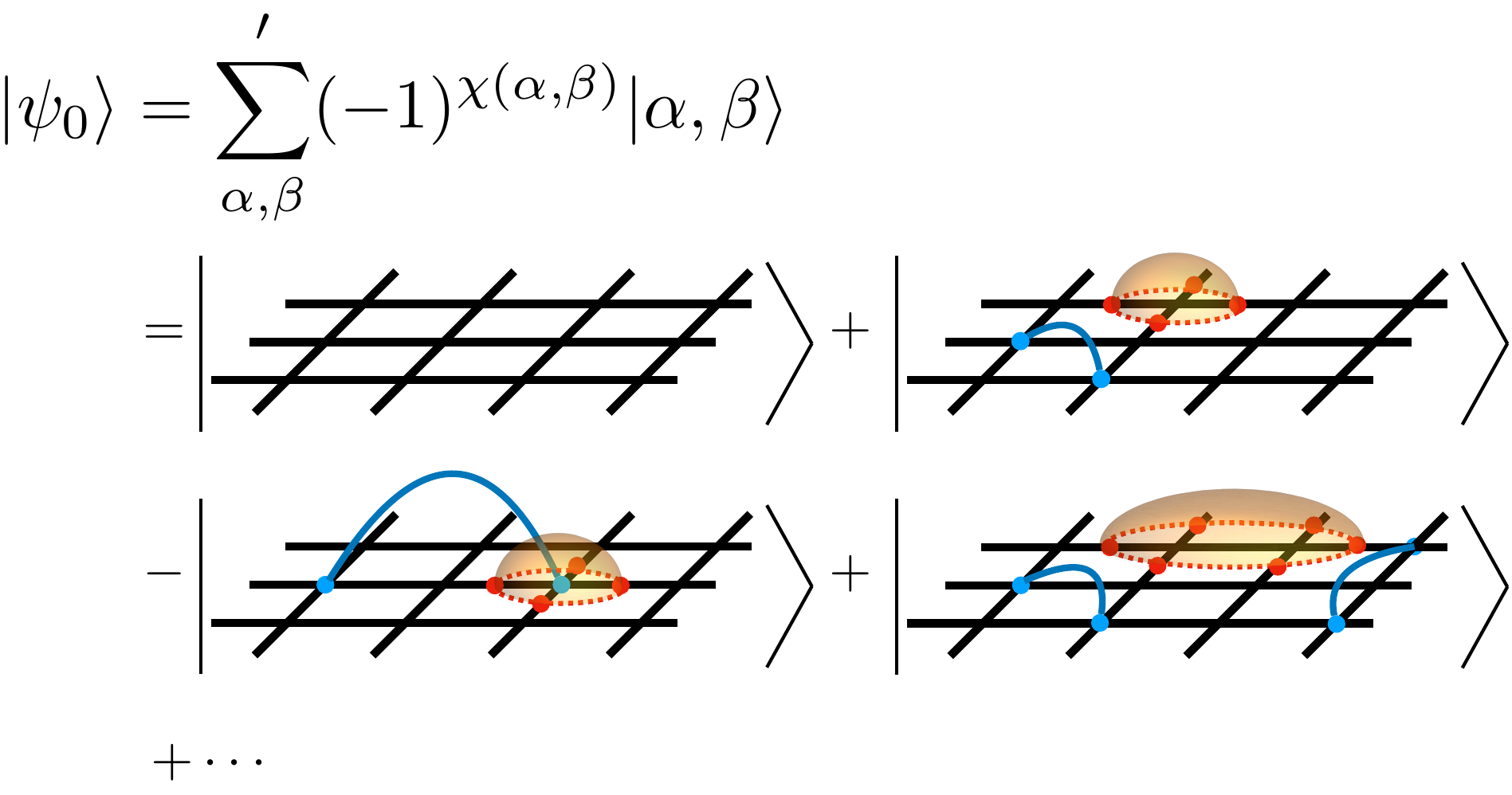}
		\end{subfigure}
		\caption{Wave function of the  $\mathbb{Z}_2\times\mathbb{Z}_2 $ fixed-point SPT, i.e. the ground state of  $H_0=- \sum_v  X_v \prod_{e\ni v } Z_e - \sum_{e}  X_e \prod_{v \in e} Z_v$. The fixed-point state is understood as a condensate of open blue strings and open red membranes with a non-trivial braiding sign structure in Pauli-X basis. The wave function $(-1)^{\chi(\alpha,\beta)}$ takes the value 1 (-1) corresponding to the even (odd)
			number of times that blue strings pierce through red membranes. }
		\label{fig:2dspt}
		
	\end{figure}

	First, we express the input state  in Pauli-X basis: 
	\begin{equation}
		\ket{\psi_0}   =\sum'_{\alpha,\beta} (-1)^{\chi(\alpha,\beta)} \psi(\alpha,\beta)  \ket{\alpha,\beta}
	\end{equation}
	$\alpha=\{\alpha_v  \}$ and $\beta=\{ \beta_e  \}$ denote the configurations in $X$ basis for qubits on vertices and edges, respectively. $\sum_{\alpha,\beta}'$ denotes summing all symmetry-allowed $\alpha, \beta$ configurations: the global symmetry on vertices implies $\alpha_v=-1$ occurs in pairs, so every allowed $\alpha$ configuration corresponds to the configuration of open strings whose end points label the location of $\alpha_v=-1$. On the other hand, the 1-form symmetry on edges indicates the edge with $\beta_e =-1$ must form a closed loop in the dual lattice, which equivalently can be regarded as the boundary of an open membrane.  $\chi(\alpha,\beta)$ is the number of times that the $\alpha$-strings pierce through the $\beta$-membranes. When $\ket{\psi_0}$ is a fixed-point SPT, $\psi(\alpha,\beta) $ is independent  of $\alpha, \beta$ (see Fig.\ref{fig:2dspt}).

	With the input state $\ket{\psi_0}$, measuring vertices ($A$ sublattice) projects to a particular $\alpha$ configuration, and the follow-up unitary correction $U_{\alpha}$ removes the braiding phase. One then finds the measurement-feedback protocol leads to a mixed state  $\rho_B$ on the edges  ($B$ sublattice),  which is the reduced density matrix (by tracing out $A$ sublattice) of the following state:
	
	\begin{equation}
		\ket{\psi}   =\sum'_{\alpha,\beta} \psi(\alpha,\beta)  \ket{\alpha,\beta}.
	\end{equation}
	Therefore, the input state and the purification of the output $\rho_B$ can be connected through $\ket{\psi} =  U \ket{\psi_0}$, where $U= \sum'_{\alpha,\beta} \ket{\alpha,\beta} \bra{\alpha,\beta}   (-1)^{\chi(\alpha,\beta)} $ is unitary in the symmetric subspace.  Under the conjugation of $U$, operators transform as

	\begin{equation}\label{eq:2dspt_transform}
		\begin{split}
			X_{v}	  \to   X_{v},	&    \quad  X_e	  \to   X_e,        \\
			\prod_{e\ni v } Z_e	  \to   X_v \prod_{e\ni v } Z_e       ,	& \quad   \prod_{v \in e} Z_v \to  X_e \prod_{v \in e} Z_v     .
		\end{split}
	\end{equation}
This  allows us to derive the parent Hamiltonian $H$ of $\ket{\psi}$ from the Hamiltonian $H_0 $ of the  input state via $H= U H_0 U^{\dagger}$. Also, note that  $\ket{\psi}$ lives in the subspace given by 0-form symmetry $\prod_v  X_v =1 $ and $\mathbb{Z}_2$ 1-form symmetry $\prod_{e\in \gamma} X_e =1$ (acting on edges along any closed loop $\gamma$), these two constraints need to be further imposed in $H$.

	As an application,  we may consider perturbing the fixed-point SPT by onsite Pauli-Xs: $H_0=- \sum_v  X_v \prod_{e\ni v } Z_e - \sum_{e}  X_e \prod_{v \in e} Z_v - g \sum_e X_e-  g\sum_v X_v$, the corresponding $H$ is  the decoupled Ising model on vertices and 2d toric code, both of which are subject to the onsite transverse field: $H  =  - \sum_v   \prod_{e\ni v } Z_e - \sum_{e}   \prod_{v \in e} Z_v    - g \sum_e X_e-  g\sum_v X_v  $, where the constraints $\prod_v  X_v =1 $ and $\prod_{e\in \gamma} X_e =1$ are further imposed.  Therefore, the measurement-feedback protocol leads to a pure state $\rho_B$ on edges with $\mathbb{Z}_2$ topological order. This also indicates all possible post-measurement states by measuring the input state SPT $\ket{\psi_0}$ are deterministically converted to the same pure state with topological order.

	To discuss a non-trivial mixed state topological order on edges, we consider the following form of input Hamiltonian: 
	
	\begin{equation}\label{eq:2d_perturbed_H0}
		\begin{split}
			H_0=&- \sum_v  X_v \prod_{e\ni v } Z_e - \sum_{e}  X_e \prod_{v \in e} Z_v\\
			&		  -g  \sum_v   \prod_{e\ni v } Z_e - g \sum_{e}   \prod_{v \in e} Z_v.
		\end{split}
	\end{equation} 
	Using the transformation rule in Eq.\ref{eq:2dspt_transform}, one finds the corresponding $H$ 
	
	\begin{equation}\label{eq:2d_perturbed_H}
		\begin{split}
			H=&- \sum_v   \prod_{e\ni v } Z_e - \sum_{e}   \prod_{v \in e} Z_v\\
			&		  -g  \sum_vX_v   \prod_{e\ni v } Z_e - g \sum_{e} X_e  \prod_{v \in e} Z_v, 
		\end{split}
	\end{equation} 
	where  the 0-form symmetry ($\prod_{v} X_v =1$) and 1-form symmetry ($\prod_{e \in \gamma} X_e=1$) are further imposed. To understand the phase diagram for the mixed state $\rho_B$ defined on edges, we transform $H$ by $ U_{CZ} =\prod_{\expval{v,e}} {CZ}_{v,e}$. This leads to

	\begin{equation}
		- \sum_v   \prod_{e\ni v } Z_e - \sum_{e}   \prod_{v \in e} Z_v   -g  \sum_vX_v  - g \sum_{e} X_e, 
	\end{equation} 
	i.e., a transverse-field Ising model on vertices and transverse-field toric code on edges. Their phase diagrams are well understood: the global Ising symmetry breaking order persists to $g=g_{\text{ising}} \approx 3.04438 $ \cite{mc_ising_2002}, above which the state on vertices become trivial.  On the other hand, using the (Kramer-Wannier) duality between transverse-field Ising and transverse-field toric code, the topological order persists up to $ g=  g_{\text{toric}} = 1/g_{\text{ising}} \approx  0.32847$,  above which the state on edges become trivial. As a result, starting with the ground state $\ket{\psi_0}$ of Eq.\ref{eq:2d_perturbed_H0} for $g<g_{\text{toric}}$ (the regime where $\ket{\psi_0}$ is an SPT), the measurement-feedback channel gives a  mixed state $\rho_B$ on edges with $\mathbb{Z}_2$ topological order. Similar to the discussion in 1d, one expects $\rho_B$ to have volume-law entropy due to the coupling between $A, B$ sublattices in Eq.\ref{eq:2d_perturbed_H}.

	Finally, we note that mixed-state topological order encoded in $\rho_B$ may be diagnosed by entanglement negativity, an entanglement measure for mixed states. In particular, one expects a long-range entanglement structure that manifests  in  the  universal subleading contribution in negativity as discussed in Ref.\cite{Lu_topo_nega_2020,lu2022_lre,fan2023mixed}.

	\section{Fermionic mixed state with  Kitaev topological order in 1d}\label{append:kitaev_1d}
	
	Here we discuss the finite-depth preparation of a mixed state in 1d that exhibits a topological order  of the Kitave Majorana chain \cite{kitaev_chain_2001,Alicea_2010_Majorana} by taking  a $\mathbb{Z}_2 \times \mathbb{Z}_2^f$ SPT $\ket{\psi_0}$ \cite{fermion_spt_2014_wen,floquet_spt_2016_sondhi,Ashvin_2018_commuing,Verresen_kitaev_2021,ashvin_2021_measurement} as an input. Here we closely follow Ref.\cite{ashvin_2021_measurement} to construct the fixed-point SPT, and then we will consider certain symmetric perturbation and apply measurement and unitary feedback to prepare a non-trivial fermionic topological mixed state.

\begin{figure}
		\centering
		\begin{subfigure}{0.38\textwidth}
			\includegraphics[width=\textwidth]{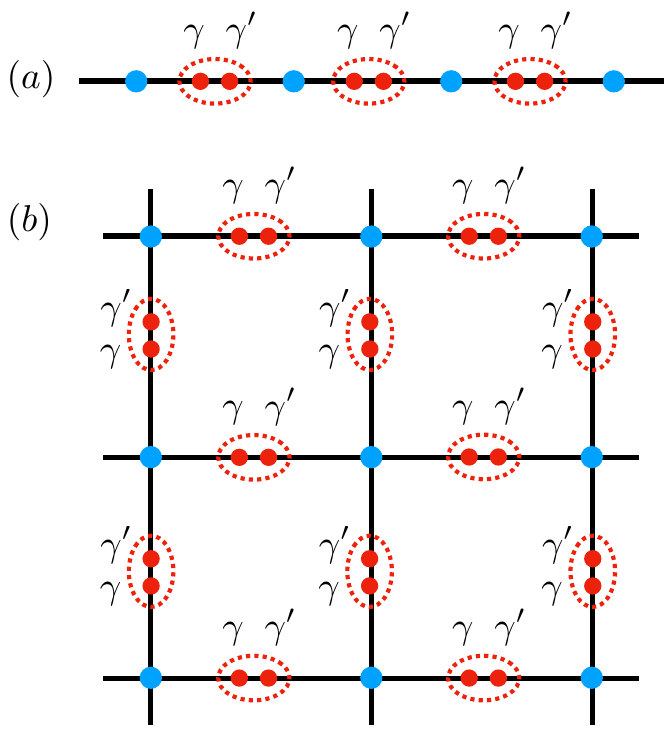}
		\end{subfigure}
		\caption{ (a) 1d lattice with every  odd site accommodating a qubit and every even site accommodating two Majorana fermions. Given an initial 0-form $\mathbb{Z}_2 \times \mathbb{Z}_2^f$ SPT, measurement and unitary feedback lead to a fermionic mixed state with the topological phase of the Kitaev chain. (b)  2d lattice with every vertex  accommodating a qubit and every edge  accommodating two Majorana fermions. Given an initial $\mathbb{Z}_2$ 0-form $ \times \mathbb{Z}_2^f$ 1-form SPT, measurement and unitary feedback leads to a fermionic mixed state with $\mathbb{Z}_2$ topological order.   }
		\label{fig:fermion_spt}  
	\end{figure}

	To start,  we consider a 1d lattice of size $2L$ initialized in a product state, see Fig.\ref{fig:fermion_spt}(a):  every odd site accommodates a qubit initiated in the $\ket{+}$ state, and  every even site accommodates a fermion initiated in the empty state, i.e. parity operator $P=-i \gamma\gamma' =  1-2 c^{\dagger} c =1$ with  left Majorana  and right Majorana  defined as $\gamma= c+ c^{\dagger}, \gamma' = -i (c-c^{\dagger})$. Define a Majorana hopping operator $S_{2n} = i \gamma'_{2n-2} \gamma_{2n}$, we  define a controlled gate 
	
	\begin{equation}
		CS_{2n-1 }	 =   \ket{\uparrow} \bra{\uparrow} +    \ket{\downarrow} \bra{\downarrow}  S_{2n}; 
	\end{equation}
	the Majorana hops from the site $2n-2$ to site $2n$ only when the qubit at $2n-1$ is in the spin-down state. Using $CS$ gate, one introduces a depth-1 local unitary circuit
	\begin{equation}\label{eq:fermion_Uspt}
		U_{\text{SPT}}  =  \prod_{n=1}^LCS_{2n-1 }.  	 
	\end{equation}
	Applying $ U_{\text{SPT}} $ on the initial product state will then lead to the   $\mathbb{Z}_2 \times \mathbb{Z}_2^f$ fixed-point SPT $\ket{\psi_0}$. Equivalently, $\ket{\psi_0  } $ is uniquely specified by the stabilizers 
	
	\begin{equation}
		\begin{split}
			U_{\text{SPT}}  X_{2n-1}	U_{\text{SPT}}^{\dagger}    &=   i \gamma_{2n-2}' X_{2n-1} \gamma_{2n}\\ 
			U_{\text{SPT}} P_{2n	}	U_{\text{SPT}}^{\dagger}    &=  Z_{2n-1} P_{2n} Z_{2n+1}. 
		\end{split}
	\end{equation}
	The SPT $\ket{\psi_0}$ is protected by the $\mathbb{Z}_2 \times \mathbb{Z}_2^f$ symmetry, where $\mathbb{Z}_2$ action is given by the product of $X_n$ on odd sites, and $\mathbb{Z}^f_2$ action is given by the product of $P_{n}$ on even sites.  The state has long-range string order diagnosed by  $i \gamma_{2n-2}' X_{2n-1} \gamma_{2n}  i \gamma_{2n}' X_{2n+1} \gamma_{2n+2}  \cdots i \gamma_{2m-2}' X_{2m-1} \gamma_{2m}   =1$ (this is simply the product of stabilizers). Being a non-trivial SPT, the string order is robust under weak symmetric perturbations, and it approaches a finite constant $c$ with $0<c<1$ in the limit $\abs{m-n} \to \infty$.

	Given a non-fixed point SPT, as in the discussion for 1d $\mathbb{Z}_2 \times \mathbb{Z}_2$ bosonic SPT, we measure Pauli-Xs on all spins on odd sites and record the outcome $\alpha=\{ \alpha_i | \text{odd } i\}$. For each post-measurement state, one applies unitary feedback, i.e. a product of on-site fermionic parity operators:
	
	\begin{equation}\label{eq:fermion_unitary}
		U_\alpha = \prod_{ n=1 }^{L}  P_{2n}^{ \frac{1- \prod_{ m=1, 2, \cdots }^n \alpha_{2m-1}  }{2}},
	\end{equation}
	namely, a parity operator at site $2n$ is applied when there is an odd number of $-1$ outcomes among $\alpha_1, \alpha_3, ... , \alpha_{2n-1}$.
	With this,  one obtains a mixed state of Majorana fermions with the long-range order $ i \gamma_{2n-2}'  \gamma_{2n}  i \gamma_{2n}'  \gamma_{2n+2}  \cdots i \gamma_{2m-2}'  \gamma_{2m}   =c$, which is a universal feature in the topological phase of a 1d Kitaev chain \cite{kitaev_chain_2001}.

	\textit{Wave function perspective -} It is useful to discuss the above result from the wave function perspective. As in the  $\mathbb{Z}_2 \times \mathbb{Z}_2$ fixed-point bosonic SPT, the $\mathbb{Z}_2 \times \mathbb{Z}^f_2$ fixed-point SPT can also be understood as the condensate of strings with a braiding structure. Specifically, the fixed-point $\ket{\psi_0}$ can be written as 
	
	\begin{equation}
		\ket{\psi_0}   =\sum'_{\alpha,\beta} (-1)^{\chi(\alpha,\beta)}  \ket{\alpha,\beta},
	\end{equation}
	where $\alpha =\{  \alpha_i \in  \{  \pm 1  \}  \}$ denotes the product state in Pauli-X basis for qubits and $\beta =\{  \beta_i  \in  \{  \pm 1  \}  \}$ denotes the product state basis with definite fermion parity on even sites \footnote{The basis state $\ket{\beta}$ corresponds to the $(c^{\dagger}_2)^{\frac{1-\beta_2}{2}} (c^{\dagger}_4)^{\frac{1-\beta_4}{2}} \cdots \ket{0}   $, where $\ket{0 }$ is the fermionic vacuum.}. The $\mathbb{Z}_2$ symmetry and $\mathbb{Z}_2^{f}$ symmetry implies $\prod_{\text{odd }i }  \alpha_i =1  $ and  $\prod_{\text{even }i }   \beta_i =1  $ respectively. As $\alpha_i$ comes in pairs, an $\alpha$  configuration may be represented by open strings ($A$-strings) whose boundary points label $\alpha_i=-1$. Similarly one can define $B$-strings to represent product state basis $\beta$ for fermions.  $\chi(\alpha,\beta )$ is the number of times that $A$-strings and $B$-strings intersect.

	Away from the fixed-point, one can write $	\ket{\psi_0}   =\sum'_{\alpha,\beta} (-1)^{\chi(\alpha,\beta)}   \psi(\alpha,\beta  )  \ket{\alpha,\beta}$ and the corresponding measurement-feedback channel gives the  mixed states of fermion $\rho_B$, which is the reduced density matrix of $	\ket{\psi}   =\sum'_{\alpha,\beta} \psi(\alpha,\beta  )  \ket{\alpha,\beta}$ by tracing out qubits on odd sites.

$\ket{\psi}$ and $\ket{\psi_0}$ can be connected by the unitary $U= \sum'_{\alpha,\beta} \ket{\alpha,\beta} \bra{\alpha,\beta}   (-1)^{\chi(\alpha,\beta)} $\footnote{$U$ is a unitary in the $\mathbb{Z}_2 \times\mathbb{Z}_2^f$ symmetric subspace.}, which has the action that 
	
	\begin{equation}\label{eq:fermion_spt_map}
		\begin{split}
			&		X_{2n-1}	  \to  	X_{2n-1},	    \quad  P_{2n}	  \to    P_{2n},        \\
			&	Z_{2n-1} Z_{2n+1} \to    	Z_{2n-1} P_{2n} Z_{2n+1}   , \\
			& \quad  \gamma'_{2n-2} \gamma_{2n} \to \gamma'_{2n-2}X_{2n-1} \gamma_{2n}.
		\end{split}
	\end{equation}
	
	As an application, we consider the initial Hamiltonian $H_0   =  -   \sum_n  (i   \gamma_{2n-2}' X_{2n-1} \gamma_{2n}    +  Z_{2n-1} P_{2n} Z_{2n+1} )  - g  \sum_n   (i   \gamma_{2n-2}' \gamma_{2n}    +  Z_{2n-1}Z_{2n+1} )$. To understand the phase diagram, we may use the depth-1 unitary circuit (Eq.\ref{eq:fermion_Uspt}) and find $ UH_0U^{\dagger}  = -   \sum_n  ( X_{2n-1}  +  P_{2n})  - g  \sum_n   (i   \gamma_{2n-2}' \gamma_{2n}    +  Z_{2n-1}Z_{2n+1} )$, which is a transverse-field Ising chain on odd sites, and fermionic Kitaev chain on even sites. Consequently, for $\abs{g}<1$, both sublattices are trivial, while for $\abs{g}>1$, odd sites exhibit a $\mathbb{Z}_2$ symmetry-breaking order and even sites belong to the topological phase of Kitaev chain. This implies the ground state of $H_0$ exhibits the $\mathbb{Z}_2\times \mathbb{Z}_2^f$ SPT order for $\abs{g}<1$, and for $\abs{g}>1$, odd sites exhibit $\mathbb{Z}_2$ symmetry-breaking order and even sites belong to the fermionic topological phase.

	With the measurement-feedback channel, one obtains a state $\rho_B$ of fermions, which is the reduced density matrix of the ground state $\ket{\psi}$ of $H$: 
	
	\begin{equation}
		\begin{split}
			H   = & -   \sum_n  (i   \gamma_{2n-2}'  \gamma_{2n}    +  Z_{2n-1}Z_{2n+1} ) \\
			& - g  \sum_n   (i   \gamma_{2n-2}'X_{2n-1}  \gamma_{2n}    +  Z_{2n-1} P_{2n}   Z_{2n+1} ).
		\end{split}
	\end{equation}
	$H$	is exactly dual to $H_0$. As a result, the measurement-feedback channel acting on the pure state of the SPT phase (Kitaev topological  phase) leads to  the Kitaev topological  phase (trivial phase) in the mixed state defined on even sites.  Note that for $\ket{\psi_0}$ belonging to the Kitaev topological phase, $\ket{\psi}$ is a non-trivial  SPT, but tracing out odd sites leads to  a trivial  reduced density matrix on even sites. 
	
	\section{Fermionic mixed state with topological order in 2d}\label{append:fermion_2d_topo}
	One  may generalize the discussion above to higher space dimensions, which allows for the existence of (non-invertible)  topological order in fermionic systems\footnote{This is in contrast to invertible fermionic topological order in Kitaev chain, which does not host fractionalized excitations in the bulk, and can be trivialized by stacking another copy of Kitaev chain.}. 
	
	Consider a 2d lattice with every vertex  accommodating a qubit and every edge  accommodating two Majorana fermions (see Fig.\ref{fig:fermion_spt}(b)), we can construct a fixed-point SPT with $\mathbb{Z}_2~$0-form$\times \mathbb{Z}_2^f~$1-form symmetry as follows. We  introduce Majorana hopping operators $ S_{v,\rightarrow}$ which take the form $ i \gamma' \gamma$, where $\gamma', \gamma$ are  the Majoranas in the left and right of the vertex $v$. Similarly, the operators $ S_{v,\uparrow} $ take the form $ i \gamma' \gamma$,  where $\gamma', \gamma$ are  the Majoranas  right above and below the vertex $v$. Using the Majorana hopping operators, one defines a controlled gate
	\begin{equation}
		CS_v	 =   \ket{\uparrow} \bra{\uparrow} +    \ket{\downarrow} \bra{\downarrow}   S_{v,\rightarrow} S_{v,\uparrow}.  
	\end{equation}
	Starting with a product state, where qubits are in $\ket{+}$ and fermions are in product states with definite fermion parity $P=1$, simultaneously applying $CS_v$ on all vertices leads to the SPT $\ket{\psi_0}$, which is uniquely specified by the stabilizers $ Z P_e Z $ (the product of a fermion parity on the edge $e$ and two Pauli-Z on two vertices on the boundary of $e$), and  $ X_{2n-1} \gamma'\gamma'\gamma \gamma$ (i.e. the product of and a Pauli-X at the vertex $v$ and four Majoranas around $v$ counter-clockwise). The SPT is protected by a $\mathbb{Z}_2^f~$1-form symmetry ($\prod_{e\in C} P_e $,  i.e. the product of fermion parity along any loops) and $\mathbb{Z}_2~$0-form symmetry ($\prod_v X_v$).  In particular, the non-trivialness of the SPT manifests in the membrane operator  $M_{\tilde{C}}= \prod_{e\in \tilde{C}} \tilde{\gamma}   \prod_{v \in A_{\tilde{C}}  } X_v$ =1, where $\prod_{e\in \tilde{C}} \tilde{\gamma} $ is the product of Majoranas ($\tilde{\gamma}$ could be $\gamma$ or $\gamma'$) along a loop in the dual lattice, and  $\prod_{v \in A_{\tilde{C}}  } X_v$ is the product of $X_v$ in the area enclosed by the loop $\tilde{C}$. For instance, \begin{equation}
		M_{\tilde{C}}= \includegraphics[width=2.7cm,valign=c]{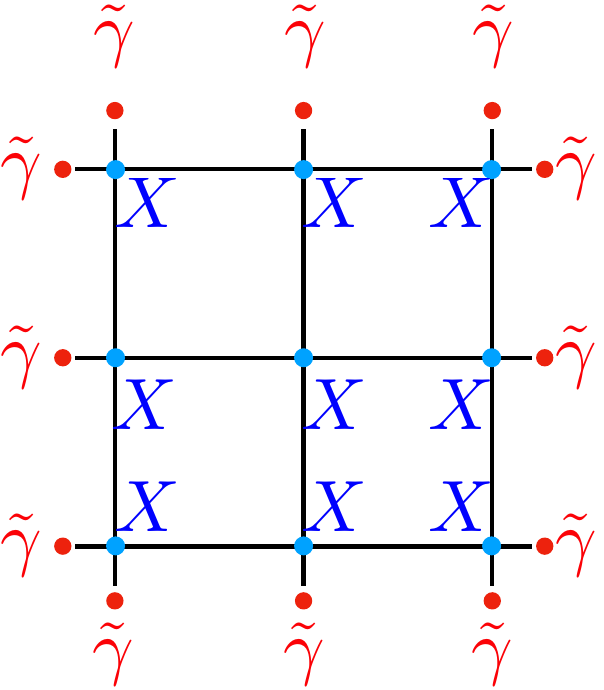}.
	\end{equation}

For non-fixed-point SPT, the membrane operator will survive in the form of a perimeter law  $\expval{M_{\tilde{C}}} \sim e^{- \mu\abs{\tilde{C}}}$. With the protocol of first measuring  $X_v$ on every vertex, followed by a unitary correction, one obtains an output mixed state $\rho$ with $\tr [ \rho \prod_{e\in \tilde{C}}  \tilde{\gamma}] \sim  e^{- \mu\abs{\tilde{C}}}$. This  perimeter law of the loop in the dual lattice together with  the 1-form symmetry $\prod_{e\in C} P_e =1 $, which inherits from the input SPT, in  the primary lattice indicates a topological order consisting of Majorana fermions.  When the input SPT is a fixed-point state, the output reduces to a pure state, which is exactly a ground state of the Majorana surface code discussed in Ref.\cite{vijay_2016_surface_code}.

\section{Purification of the mixed state from measuring 1d free-fermions}\label{append:purification_1d_fermions}
	In Sec.\ref{sec:1d_fermion} we discussed a measurement-feedback protocol that leads to a mixed state $\rho$ with an enhanced spin-spin correlation. Here we discuss the structure of $\rho$, and in particular, by unveiling a novel braiding structure between ``spin'' and ``charge'' in the initial input state  $\ket{\psi_0}$,  $\rho$ can be purified as a ground state of a local Hamiltonian $H$ which we derive below.
	
	To begin with, instead of using the conventional occupation number basis, we will consider the basis that manifests the structure  of ``spin'' and ``charge''.  For simplicity, let's first consider the $i$'th lattice site as an example. This single site is ssociated to a $4$-dim Hilbert space spanned by $\ket{0}, \ket{\uparrow}  = c_{i,\uparrow}^\dagger \ket{0}  , \ket{\downarrow}  = c_{i,\downarrow}^\dagger \ket{0} ,  \ket{ \uparrow\downarrow}   =  c_{i,\uparrow}^\dagger c_{i,\downarrow}^\dagger  \ket{0}$. In the 2-dim subspace with single-fermion occupation number, i.e. spanned by $\ket{\uparrow}, \ket{\downarrow}$, one can define two orthonormal states $ \frac{1}{\sqrt{2}} (\ket{\uparrow}  \pm  \ket{\downarrow})$ as the symmetric/anti-symmetric superposition of spin-up and spin down, i.e. the eigenstates of the spin-flip operator $\mathcal{S}_i^x = 2 S_i^x$.  Therefore, denoting the fermion occupation number as $ n_i=n_{i,\uparrow}+n_{i,\downarrow} $, one can define the complete basis states $\ket{ n_i=0  }$, $\ket{n_i =2  }$, $\ket{n_i =1, \beta_i =1   }$, $\ket{n_i=1, \beta_i =-1   }$, where $\beta=  \pm 1 $ corresponds to the eigenvalues of the spin-flip operator $\mathcal{S}_i^x$. As a more compact notation, the basis may be defined as  $\ket{n_i, \beta_i }$ with $n_i \in \{0,1,2  \}$ and $\beta_i \in \{\pm 1\}$, and importantly, $\beta_i$ degree of freedom only exists when $n_i=1$. Generalizing the discussion to all lattice sites, one constructs the basis\footnote{Such basis can be written as the linear combination of the fermion occupation number basis $(c^{\dagger}_{1,\uparrow})^{n_{1,\uparrow}}(c^{\dagger}_{1,\downarrow})^{n_{1,\downarrow}} (c^{\dagger}_{2,\uparrow})^{n_{2,\uparrow}}(c^{\dagger}_{2,\downarrow})^{n_{2,\downarrow}} \cdots \ket{0} $ with $\ket{0}$ being the fermionic vacuum.} $   \ket{ n  ,   \beta    }$   with  $ n \equiv \{ n_i \} ,  \beta  \equiv\{  \beta_i  \}  $, and the input fermionic pure state may be written as 
	
	\begin{equation}\label{eq:fermion_braiding}
		\ket{  \psi_0} = \sum'_{n,\beta}(-1)^{\chi(n,\beta)} \psi(n,\beta ) \ket{n  ,   \beta }.
	\end{equation}
	Since $\beta_i$ only exists when $n_i=1$,  the $\beta$ configuration must be consistent with fermion number configuration $n$. Alternatively, this consistency condition can be imposed on the wave function so that $\psi(n,\beta)=0$ when $\beta$ is not consistent with $n$. In addition to the above consistency condition, for the summation $\sum'_{n,\beta}$, the allowed $n, \beta$ must respect the  global $\mathbb{Z}_2\times \mathbb{Z}_2$ symmetry as a consequence  of the  equal number of up-spins and down-spins in $\ket{\psi_0}$. The first $\mathbb{Z}_2$ corresponds the fermion parity symmetry: the total fermion number $N$ must be even since $N=N_{\uparrow}+ N_{\downarrow} = 2 N_{\uparrow}$. This implies configuration $n$ must satisfy $\prod_{i}(-1)^{n_i}=1$, so the number of lattice sites with a single occupation number (odd fermion parity) must be even.  The second $\mathbb{Z}_2$ corresponds to the global spin flip given by $\prod_i\mathcal{S}_i^x$, so the product of $\beta_i$ (at the site with $n_i=1$) must be one. This means $\beta_i=-1$ comes in pairs. Since both $n_i=1$ (odd fermion parity) and $\beta_i=-1$ come in pairs, with the ordering $n_1, \beta_1, n_2, \beta_2, \cdots$ along a 1d line, one can use strings to label the configurations of $n_i=1, \beta_i=-1$, and define a braiding sign structure between charge $n$ and spin $\beta$, as in the bosonic $\mathbb{Z}_2 \times \mathbb{Z}_2$ SPT discussed in Appendix.\ref{append:1dspt_duality}.

	Given Eq.\ref{eq:fermion_braiding}, measuring global fermion occupation number gives $P_n\ket{  \psi_0} = \sum'_{\beta}(-1)^{\chi(n,\beta)} \psi(n,\beta ) \ket{n, \beta }$. As one can check, the unitary feedback $U_n = \prod_{i=1}^{L} \left(\mathcal{S}_i^{x} \right)^{ \sum_{j\leq i} n_j }$ (Eq.\ref{eq:U}) removes the braiding sign: $U_n P_n\ket{  \psi_0} = \sum'_{\beta}\psi(n,\beta ) \ket{n, \beta }$, and therefore, the resulting mixed state  is 
	\begin{equation}
		\begin{split}
			\rho &=  \sum'_n   U_n P_n\ket{  \psi_0} \bra{  \psi_0}  P_n U_n^{\dagger}  \\
			&=  \sum'_{\beta,\beta'} \left[ \sum'_{n}  \psi(n,\beta ) \psi(n,\beta') \right] \ket{ n,\beta }  \bra{ n,\beta' }. 
		\end{split}
	\end{equation}
	Since only the spin sector carries long-range correlation, one may consider the reduced density matrix  in the spin sector by tracing out the charge sector, which leads to $\rho_s = \sum'_{\beta,\beta'} \left[ \sum'_{n}  \psi(n,\beta ) \psi(n,\beta') \right] \ket{\beta }  \bra{\beta' }$. This is nothing but the reduced density matrix of
	
	\begin{equation}\label{eq:1dfermion_psi}
		\ket{  \psi} = \sum'_{n,\beta} \psi(n,\beta ) \ket{n  ,   \beta }.
	\end{equation}
	To better characterize the structure of $\ket{\psi}$, one can derive its parent Hamiltonian. This is achieved by finding the unitary $U$ such that $\ket{\psi} = U\ket{\psi_0}$. Then the parent Hamiltonian $H$ of $\ket{\psi}$ can be obtained by   $H=UH_0 U^{\dagger}$, where $H_0$ is the parent Hamiltonian of the spinful free fermion state $\ket{\psi_0}$, namely, $H_0 =  - \sum_{ i,\sigma} ( c_{i+1,\sigma}^{\dagger} c_{i,\sigma} + \text{h.c.})$. By comparing  $\ket{\psi_0}$ (Eq.\ref{eq:fermion_braiding}) and $\ket{\psi}$ (Eq.\ref{eq:1dfermion_psi}), $U$ may be chosen as 
	\begin{equation}\label{eq:u_transform_fermions}
		U =  \sum'_{n,\beta }\ket{n  ,   \beta } \bra{n, \beta} (-1)^{\chi(n,\beta)}.
	\end{equation}
	$U$ is diagonal in the basis $\ket{n,\beta}$ with the diagonal entries encoding the braiding phases. While $U$ is not a unitary in the entire Hilbert space, it is a unitary in the restricted Hilbert space with the $\mathbb{Z}_2 \times\mathbb{Z}_2$ symmetry. Alternatively, $U$ may be written as  $U= \prod_{i=1}^{L} \left(\mathcal{S}_i^{x} \right)^{ \sum_{j\leq i} \hat{n}_j }$, which is the same as Eq.\ref{eq:u_transform_fermions} in the symmetric subspace. We note that $\hat{n}_j$ is a number operator instead of a $c$-number.  $U$  allows us to derive the parent Hamiltonian of $\ket{\psi}$ through  $H=UH_0U^{\dagger}$:

	\begin{equation}\label{eq:1d_fermion_parent}
		\begin{split}
			H=-&  \sum_{i,\sigma}   \Bigl[    c_{i+1,\sigma}^{\dagger} c_{i,\sigma} P_{n_i + n_{i+1} \neq 2 } \\
			&+  c_{i+1,\sigma}^{\dagger} c_{i,-\sigma} P_{n_i + n_{i+1}  =2 } P_{n_i=1} \\
			&- c_{i+1,\sigma}^{\dagger} c_{i,-\sigma} P_{n_i + n_{i+1}  =2 } P_{n_i=2}       \Bigr]  + \text{h.c.},
		\end{split}
	\end{equation}
	where $P$ is a projector with a subscript that labels the subspace it projects to, and the constraint $\prod_i (-1)^{n_i } = \prod_i \mathcal{S}_i^x =1 $ is further imposed on $H$.

	Below we present the derivation of  $H$: consider the term $c_{i+1,\uparrow}^{\dagger} c_{i,\uparrow} $, it transforms as $Uc_{i+1,\uparrow}^{\dagger} c_{i,\uparrow} U^{\dagger}$. To proceed, we compute the matrix entries of this matrix:
	
	\begin{equation}
		\begin{split}
			&\bra{n',\beta'}Uc_{i+1,\uparrow}^{\dagger} c_{i,\uparrow} U^{\dagger}\ket{n,\beta}\\
			&=\bra{n',\beta'} (-1)^{\chi(n',\beta')}c_{i+1,\uparrow}^{\dagger} c_{i,\uparrow}(-1)^{\chi(n,\beta)} \ket{n,\beta}.
		\end{split}
	\end{equation}
	Since the fermion hops from $i$-th site to $i+1$-th site, there are only four possible cases for obtaining non-vanishing values: $(n_i,n_{i+1}) =  (2,1)$,  $(n_i,n_{i+1}) =  (1,0)$,  $(n_i,n_{i+1}) =  (2,0)$,  $(n_i,n_{i+1}) =  (1,1)$. Below we separately discuss these cases. \\
	
	\noindent \underline{Case (a)} $(n_i,n_{i+1}) =  (2,1)$: in this case, one finds the fermion hopping does not change the braiding phase, namely, $(-1)^{\chi(n',\beta')}   (-1)^{\chi(n',\beta')}  =1$. Therefore, $\bra{n',\beta'} (-1)^{\chi(n',\beta')}c_{i+1,\uparrow}^{\dagger} c_{i,\uparrow}(-1)^{\chi(n,\beta)} \ket{n,\beta}  = \bra{n',\beta'}c_{i+1,\uparrow}^{\dagger} c_{i,\uparrow} \ket{n,\beta}  $. Alternatively, one can replace the braiding phase $(-1)^{\chi (n,\beta)} $ with the operator  $\prod_{i=1}^{L} \left(S_i^{x} \right)^{ \sum_{j\leq i} \hat{n}_j }$, and go through the algebra to derive the same result. \\

	\noindent \underline{Case (b)} $(n_i,n_{i+1}) =  (1,0)$: as in case (a), in this case, one finds the fermion hopping does not change the braiding phase as well. Therefore, $\bra{n',\beta'} (-1)^{\chi(n',\beta')}c_{i+1,\uparrow}^{\dagger} c_{i,\uparrow}(-1)^{\chi(n,\beta)} \ket{n,\beta}  = \bra{n',\beta'}c_{i+1,\uparrow}^{\dagger} c_{i,\uparrow} \ket{n,\beta}  $.\\
	
	\noindent \underline{Case (c)} $(n_i,n_{i+1}) =  (2,0)$: in this case, after the fermion hopping, both $i$-th site and $i+1$-th site contain a single fermion. It follows that the braiding sign will change when  $\beta_i'=-1$. Therefore, $\bra{n',\beta'} (-1)^{\chi(n',\beta')}c_{i+1,\uparrow}^{\dagger} c_{i,\uparrow}(-1)^{\chi(n,\beta)} \ket{n,\beta}  = \bra{n',\beta'} \beta_i'c_{i+1,\uparrow}^{\dagger} c_{i,\uparrow} \ket{n,\beta} = \bra{n',\beta'}  \mathcal{S}_i^xc_{i+1,\uparrow}^{\dagger} c_{i,\uparrow} \ket{n,\beta} $. Writing $\mathcal{S}_i^x  = c_{i,\uparrow}^{\dagger} c_{i,\downarrow} +c_{i,\downarrow}^{\dagger} c_{i,\uparrow}$, the result can be simplified as  $\bra{n',\beta'} (-1)^{\chi(n',\beta')}c_{i+1,\uparrow}^{\dagger} c_{i,\uparrow}(-1)^{\chi(n,\beta)} \ket{n,\beta}  = \bra{n',\beta'} (-c_{i+1,\uparrow}^{\dagger} c_{i,\downarrow}) \ket{n,\beta}  $.\\

	\noindent \underline{Case (d)} $(n_i,n_{i+1}) =  (1,1)$: in this case,
	the braiding sign will change if before hopping, $\beta_i=-1$. Therefore, $\bra{n',\beta'} (-1)^{\chi(n',\beta')}c_{i+1,\uparrow}^{\dagger} c_{i,\uparrow}(-1)^{\chi(n,\beta)} \ket{n,\beta}  = \bra{n',\beta'} c_{i+1,\uparrow}^{\dagger} c_{i,\uparrow} \beta_i\ket{n,\beta} = \bra{n',\beta'} c_{i+1,\uparrow}^{\dagger} c_{i,\uparrow} \mathcal{S}_i^x  \ket{n,\beta} $. Using the representation $\mathcal{S}_i^x  = c_{i,\uparrow}^{\dagger} c_{i,\downarrow} +c_{i,\downarrow}^{\dagger} c_{i,\uparrow}$, one finds $\bra{n',\beta'} (-1)^{\chi(n',\beta')}c_{i+1,\uparrow}^{\dagger} c_{i,\uparrow}(-1)^{\chi(n,\beta)} \ket{n,\beta}  = \bra{n',\beta'} c_{i+1,\uparrow}^{\dagger} c_{i,\downarrow}\ket{n,\beta}  $.\\
	
	By combining the discussion on these four cases, one obtains the parent Hamiltonian in Eq.\ref{eq:1d_fermion_parent}.

\end{document}